\documentclass[prd,showpacs,nofootinbib,preprint,10 pt]{revtex4}
\usepackage{amsmath,graphicx,color,epsfig}
\usepackage{appendix}
\usepackage{slashed}
\usepackage{multirow}
\usepackage{soul}
\usepackage{pifont}
\usepackage{amsfonts}
\usepackage{bbding}
\usepackage{placeins}
\usepackage{float}

\begin{document}
\begin{titlepage}
\begin{center}
	\setlength {\baselineskip}{0.3in}
	{\bf\Large\boldmath Probing new physics effects in $\Lambda_b \to \Lambda (\to p\pi^-)\ell^{+}\ell^{-}$ decay via model independent approach }\\[15mm]
	\setlength {\baselineskip}{0.3in}
	{\large Aqsa Nasrullah$^{1,}$ \footnote[1]{aqsanasrullah54@gmail.com}, Ishtiaq Ahmed$^{2}$, M. Jamil Aslam$^{1}$, Z. Asghar$^{3,4}$, Saba Shafaq$^{5}$\\[5mm]
		~{\it $^{1}$Department of Physics, Quaid-i-Azam University, Islamabad 45320, Pakistan. \\
			$^{2}$National Centre for Physics, Quaid-i-Azam University, Islamabad 45320, Pakistan.\\
			$^{3}$ Center for Mathematical Sciences, Pakistan Institute of Engineering and Applied Sciences, Nilore, Islamabad 45650, Pakistan \\
			$^{4}$ Department of Physics and applied sciences, Pakistan Institute of Engineering and Applied Sciences, Nilore, Islamabad 45650, Pakistan \\
			$^{5}$ Department of Physics, International Islamic University, Islamabad 44000, Pakistan.}}\\[5mm]
		%
		
		%
		%

		{\bf Abstract}\\[5mm]
\end{center}

\setlength{\baselineskip}{0.2in}
The New Physics (NP) effects are studied in the rare baryonic decay $\Lambda_b \to \Lambda (\to p\pi^-)\ell^{+}\ell^{-}$, with unpolarized $\Lambda_b$ using most general model independent approach by introducing new axial(vector), (pseudo)scalar and tensor operators in the weak effective Hamiltonian corresponding to $b\to s$ transitions. Recently, for $\Lambda_b \to \Lambda (\to p\pi^-)\mu^{+}\mu^{-}$ decay the LHCb collaboration has measured the branching ratio $(d\mathcal{B}/ds)$, lepton- and hadron-side forward-backward asymmetries, denoted by $A_{FB}^{\ell}$ and $A_{FB}^{\Lambda}$, respectively, and the longitudinal polarization fraction $F_L$ both in the low- and high-recoil regions. To see whether the new $VA$, $SP$ and $T$ couplings can accommodate the available experimental data of these observables, first we have examined their influence on these observables and later we have checked the imprints of these new couplings on a number of interesting but yet not measured observables; namely the combined lepton-hadron forward-backward asymmetry  $(A_{FB}^{\ell\Lambda})$, transverse polarization fraction $(F_T)$, asymmetry parameters $\alpha_i's$ and some other angular observables, extracted from certain foldings. It is found that compared to the $VA$ the $SP$ couplings favor experimental data for all the four observables but still no individual coupling is able to accommodate all of the available data simultaneously. To achieve this goal, the pairs of new WCs are taken to check their range that simultaneously satisfy constraints of $B$-Physics and available LHCb data on $d\mathcal{B}/ds$, $F_L$, $A_{FB}^{\ell}$ and $A_{FB}^{\Lambda}$ in several bins for the decay channel under consideration. We find that most of the available data could be accommodated by the different pairs of $VA$ and $SP$ WCs giving more severe constraints on the parametric space of these WCs that is still satisfied with the $B$-physics data.
\end{titlepage}

\section{Introduction}\label{sec1}
The rare exclusive $B \to K^{*},\phi (\rho)\ell^{+}\ell^{-}$ decays, governed by the quark level transitions $b\rightarrow s(d)\gamma $ or $b\rightarrow s(d)\ell^+\ell^-$, have been the focus of the theoretical and experimental studies for some time. The prime reason for them to be at the spotlight is due to their potential to test the predictions of Standard Model (SM) for different observables such as branching fractions, angular distributions and the lepton flavor universality. Though the interest in such decays dates back to the era of B-factories, the recent triggering point was the measurements at the LHCb, most prominently the $P_{5}^{\prime}$ anomaly \cite{DescotesGenon:2012zf, Descotes-Genon:2013vna} in $B
\rightarrow K^*\mu^+ \mu^-$ decay \cite{Aaij:2013qta, Aaij:2015oid}. Later, the interest in these rare decays was enhanced when the deviations from the SM predictions were observed in the measurements of the lepton flavor universality $R_K$ and
$R_{K^*}$($R_{K^{(*)}}=d\mathcal{B}/ds(B \rightarrow K^{(*)}\mu^+
\mu^-)/d\mathcal{B}/ds(B \rightarrow K^{(*)}e^+ e^-)$) \cite{Aaij:2014ora,Bordone:2016gaq,Aaij:2017vbb}.
Similarly,
 the deviations are also observed in the differential decay rate of
 $B_s \rightarrow \phi \mu^+ \mu^-$ \cite{Aaij:2013aln,Horgan:2013pva,Aaij:2015esa}, the branching ratios of $B \rightarrow D^* \tau \nu$ \cite{Fajfer:2012vx} and
 $B \rightarrow K^{(*)}\mu^+\mu^-$ \cite{Aaij:2014pli}. Moreover, the BaBar Collaboration measured the lepton flavor universality
  violation \cite{Huschle:2015rga,Aaij:2017uff} in $R_D$ and
  $R_{D^{*}}$ ($R_{D^{(*)}}=d\mathcal{B}/ds(B \rightarrow D^{(*)}\tau\nu_{\tau})/d\mathcal{B}/ds(B \rightarrow D^{(*)}\ell\nu_{\ell})$ where $\ell=e,\mu$).
 These mismatch between the SM predictions and experimental measurements are rather significant ($2\sigma - 3.4\sigma$) \cite{Lees:2012xj}, providing clear hints of the presence of some new couplings along with that of the SM ones. Due to these facts, the four-body decay
  $B \rightarrow K^*(\to K\pi) \ell^+ \ell^-$ has been extensively studied in literature
\cite{Kruger:1997jk,Aliev:1999gp,Aliev:1999re,Kruger:2005ep,Altmannshofer:2008dz,Becirevic:2011bp,Bobeth:2010wg,Faessler:2002ut,Matias:2012qz,Mahmoudi:2014mja}. More precise experimental studies are already part of the programs for the LHC upgrade \cite{LHCb-19} and Belle-II \cite{Belle-II}, and there is no doubt  that these decays will help us to to see if these anomalies are due to physics beyond the SM or it involve some QCD physics. Supposing that these anomalies persist in future data, similar kind of
deviations would also expected to be seen in the baryonic partners of
these rare $B-$meson decays, especially in $\Lambda_b \to \Lambda\left(\to
p \pi^{-}\right) \ell^{+}\ell^{-}$ decays.

The advantage of the baryon decay $\Lambda_b \to \Lambda\left(\to
p \pi^{-}\right) \ell^{+}\ell^{-}$ over $B \rightarrow K^*(\to K\pi) \ell^+ \ell^-$ is that even the
initial state baryons $\Lambda_b$ are unpolarized, the final state baryon $\Lambda$ spin can be
used to understand the helicity structure of weak effective Hamiltonian
\cite{Mannel:1997xy,Hiller:2007ur,Wang:2008sm,Chen:2001ki,Faessler:2002ut,Gutsche:2013pp,Gutsche:2013oea}. Furthermore, similar to the $B \rightarrow K^*(\to K\pi) \ell^+ \ell^-$ decay, it also provides a large number of angular observables and is sensitive to all the Dirac structures present in the weak Hamiltonian. As the number of angular observables increases due to the
polarization of $\Lambda_b$ baryon, it makes this decay to be more
prolific to test new physics (NP)
\cite{Bhattacharya:2019der}. Due to these distinctive features, the
radiative and semileptonic $\Lambda_b$ decays have been well studied in the literature
\cite{Korner:1994nh,Huang:1998ek,Hiller:2001zj,Mott:2011cx, Gutsche:2013pp,Gutsche:2013oea,Detmold:2016pkz,Roy:2017dum,Das:2018iap}.
To look for the imprints of NP in $\Lambda_b$ decays, there are some
dedicated studies of this decay in different NP models, namely in 2HDM \cite{Aliev:1999ap}, $Z^\prime$ model
\cite{Nasrullah:2018puc}, Randall-Sundrum model with custodial
protection \cite{Nasrullah:2018vky}, Left-right
models \cite{Chen:2001zc}, Supersymmetric theories \cite{Aslam:2008hp} and
using a most general model independent weak effective Hamiltonian \cite{Aliev:2002ww,Aliev:2002tr}. In model independent approach the study of $\Lambda_b \rightarrow \Lambda \ell^+ \ell^-$ decay
is confined to the analysis of branching ratio ($d\mathcal{B}/ds$) and lepton-side
forward-backward asymmetry ($A_{FB}^{\ell}$). However, an analysis of full set of angular observables for
this decay have already in other approaches, see e.g., \cite{Chen:2001zc,
Nasrullah:2018puc, Hu:2017qxj} showing that these are quite sensitive to the parameters of different NP models.
Motivated by these studies, the present work focuses the analysis of $\Lambda_b \rightarrow \Lambda \ell^+ \ell^-$ decay in the model independent approach. Particularly, we study the $\Lambda_b \rightarrow \Lambda (\to p
\pi) \ell^+ \ell^-$ decay, with unpolarized $\Lambda_b$, using a most
general effective Hamiltonian involving new (axial)vector
$(VA)$ with combination $V+A$, pseudo(scalar) $(SP)$ and the tensor $(T)$ operators.
Being an
exclusive decay process, some of the physical observables are not clean due to the
uncertainties arsing from the form factors (FF). However, some high level
precision lattice QCD calculations of FFs are available for
$\Lambda_b \to \Lambda$ \cite{Detmold:2016pkz} transition and by using
the Bourrely-Caprini-Lellouch parametrization their profile in the
full momentum transfer square $(q^2=s)$ is obtained
in \cite{Bourrely:2008za}. The lattice results are not only consistent
with the recent QCD light- cone sum rule calculations
\cite{Wang:2015ndk} but also have much smaller uncertainty in most
of the kinematic range. 

It is a well established fact that in contrast to the $B$-decays, the
QCD factorization is not fully developed for the $b$-baryon decays, therefore, we will not include the non-factorizable contributions
in the present study. It is also important to emphasis that a similar
study is performed for the few observables, namely, $d\mathcal{B}/ds$, $A_{FB}^{\ell}$ and $F_L$ in ref.
\cite{Das:2018sms}. However, in comparison to this the present study is quite extensive, especially in two ways: the
first and the most important is that we have taken the four-folded
decay rate distribution which helps us to explore some more interesting
physical observables which depend both on the angle of cascade decay
$\theta_{\Lambda}$ and on the angle $\phi$. The second is to see the
lepton mass effects on the observables, we keep the terms involving
the mass of the final state lepton which are previously ignored in
\cite{Das:2018sms} and hence helping us to extend our analysis to the case when we have $\tau$'s in the final state. As a first step, by using the constraints
on the NP Wilson coefficients (WCs) given in \cite{Das:2018sms}, we reproduced their
results of $d\mathcal{B}/ds$, $A_{FB}^{\ell}$ and $F_L$  in  $\Lambda_b \rightarrow \Lambda (\to p
\pi) \mu^+ \mu^-$ decay. Later, we
see the imprints of these new operators on the longitudinal
asymmetry $\alpha_L$, the transverse asymmetry $\alpha_U$ and the
observables named as $\mathcal{P}_i'$'s that are derived from
different foldings and hence have minimal dependence on the form factor. We have already mentioned that non-factorizable contributions which are quite challenging to calculate in $\Lambda_b \to \Lambda \ell^{+}\ell^{-}$ decay might question the analysis of different observables in the large-recoil region; i.e., for dilepton invariant masses $s < m^2_{J/\psi}$, therefore, to see the imprints of NP we have discussed these observables separately in the low-recoil bin too.

The study performed here is organized as follows: Sect. \ref{Heff}
discusses the effective Hamiltonian of the SM and its extension to take care of the NP operators arising due to the model independent approach. 
In the same section, the matrix elements of $\Lambda_b \to \Lambda$ for different possible currents are expressed in terms of the FFs. Also, a brief discussion on the
formalism of cascade decay $\Lambda \rightarrow p\pi^-$ is given at the end of this section. The four folded angular distribution and the expressions of physical observables for different NP operators are 
given in Sect. \ref{angular}. The discussion of the impact of new
$VA$, $SP$ and $T$ couplings on different physical observables has been done in section \ref{discussion}, where we also discuss the lepton mass effects in $\Lambda_b \rightarrow \Lambda (\to p
\pi) \ell^+ \ell^-$ decay. In the
same section, we present the simultaneous plots of observables for which experimental data is available to see if we could find the values of the pairs of NP WCs that can 
satisfy the experimental data for more than one physical observables. At the end of Sect. \ref{discussion}, impact of lepton mass effects on different observables is briefly explored for $\Lambda (\to p
\pi) \tau^+ \tau^-$. Finally, the main findings of this study are concluded in Sect. \ref{conclusion}. At the end, the Appendix provides the details of the calculation of different helicity fractions for hadron and leptons for all possible operators.

\section{Effective Hamiltonian \label{Heff}}

In the SM the exclusive $\Lambda_b(P_{\Lambda_b}) \rightarrow \Lambda(P_{\Lambda}) \ell^+ (q_1) \ell^-(q_2)$ decay is governed by the the four fermion operators build out of $V-A$ currents for the quark level $b\to s$ transitions and the vector/axial vector leptonic currents. In the most general model independent approach extend the operator basis to new (pseudo)scalar and tensor operators, and the $V+A$ combination for the $b \to s$ transitions. The relevant Hamiltonian in this case is \cite{Das:2018sms}:
\begin{eqnarray}
\mathcal{H}_{eff} &=& - \frac{G_F \alpha_e}{\sqrt{2}\pi} V_{tb} V_{ts}^* \bigg[ \left(C_9^{eff} \bar{s}\gamma^{\mu}P_L b -\frac{2m_b}{s} C_7^{eff} \bar{s} iq_{\nu} \sigma^{\mu\nu}P_R b +C_V \bar{s}\gamma^{\mu}P_L b +C_V^{\prime} \bar{s}\gamma^{\mu}P_R b\right)\bar{\ell}\gamma_{\mu} \ell \notag \\
&&+\left(C_{10}\bar{s}\gamma^{\mu}P_L b+ C_A \bar{s}\gamma^{\mu}P_L b+C_A^{\prime}\bar{s}\gamma^{\mu}P_R b\right) \bar{\ell}\gamma_{\mu}\gamma_5 \ell +\left(C_S^{\prime} \bar{s} P_L b +C_S \bar P_R b\right) \bar{\ell}\ell \notag \\
&&+\left(C_P^{\prime} \bar{s} P_L b+ C_P \bar{s} P_R b\right)\bar{\ell}\gamma_5 \ell+C_T \left(\bar{s} \sigma^{\mu\nu} b\right) \bar{\ell} \sigma_{\mu\nu}\ell+C_{T_5} \left(\bar{s} \sigma^{\mu\nu} b\right) \bar{\ell} \sigma_{\mu\nu}\gamma_5\ell \bigg]\label{hamilt}
\end{eqnarray}
where $G_F$ is Fermi-constant, $\alpha_e$ is fine structure constant, $V_{tb} V_{ts}^*$ are the corresponding elements of CKM matrix, $s\equiv q^2$ is dilepton mass squared \cite{Faessler:2002ut,Gutsche:2013pp,Gutsche:2013oea} and $P_{R,L}=\frac{1\pm \gamma_5}{2}$. In SM, the WCs $C^{(\prime)}_{V},\;C^{(\prime)}_{A},\; C^{(\prime)}_{S},\; C^{(\prime)}_{P},\;C_{T}\;$ and $C_{T_5}$ are equal to zero and the explicit form of the SM WCs $C_{7,9}^{eff}$ that has been used in this study are given in \cite{Nasrullah:2018vky}.

Collecting the vector and axial-vector operators from Eq. (\ref{hamilt}), the relevant $VA$ part of the Hamiltonian is
\begin{equation}
\mathcal{H}_{VA}=- \frac{G_F \alpha_e}{\sqrt{2}\pi} V_{tb} V_{ts}^* \bigg[\left(H_V^{\mu} \widetilde{C}_{9}^+ -H_A^{\mu} \widetilde{C}_{9}^- -\frac{2m_b}{s} C_7^{eff} (H_T^{\mu} +H_{T_5}^{\mu})\right)\bar{\ell}\gamma_{\mu} \ell+\left( H_V^{\mu}\widetilde{C}_{10}^+ -H_A^{\mu} \widetilde{C}_{10}^- \right)\bar{\ell}\gamma_{\mu}\gamma_5 \ell\bigg].\label{hva1}
\end{equation}
where $H_{V}^{\mu}=\frac{1}{2}\left(\bar{s}\gamma^{\mu}b\right)$ and $H_{A}^{\mu} = \frac{1}{2}\left(\bar{s}\gamma^{\mu}\gamma_{5}b\right)$ with $\widetilde{C}_{9}^+ = C_9^{eff}+C_V+C_V^{\prime}$, $\widetilde{C}_{9}^- = C_9^{eff}+C_V-C_V^{\prime}$, $\widetilde{C}_{10}^+ = C_{10}^{eff}+C_A+C_A^{\prime}$ and $\widetilde{C}_{10}^- = C_{10}^{eff}+C_A-C_A^{\prime}$.

Writing $C_{S,P}^{\pm}=C_{S,P}\pm C_{S,P}^{\prime}$, the  scalar-pseudoscalar $(SP)$ part of weak effective Hamiltonian is
\begin{equation}
\mathcal{H}_{SP}=[C_S^+ H_S+ C_S^-H_P] \overline{\ell} \ell +[C_P^+ H_S+C_P^- H_P] \overline{\ell} \gamma_5 \ell
\end{equation}
with $H_S = \bar{s}b$ and $H_P = \bar{s}\gamma_{5}b$.
Likewise, we can write the tensor $(T)$ part from Eq. (\ref{hamilt}) as
\begin{equation}
\mathcal{H}_{T^{\prime}}=-\frac{G_F \alpha_e}{\sqrt{2}\pi} V_{tb} V_{ts}^* \bigg[\left(\bar{s} \sigma^{\mu\nu} b\right)\bar{\ell} \sigma_{\mu\nu}\left(C_T+C_{T_5}\gamma_{5}\right)\ell\bigg].\label{ht}
\end{equation}

\subsection{ Matrix Elements \label{HF}}

As an exclusive process, the matrix elements for $\Lambda_b \to \Lambda$ transition for different possible currents can be parameterized in terms of FFs, $f^{V}_{t,0,\perp},\; f^{A}_{t,0,\perp},\; f^{T}_{0,\perp}$ and $f^{T_5}_{0,\perp}$ \cite{Feldmann:2011xf}. In case of vector-currents the corresponding matrix elements for $\Lambda_b \to \Lambda$ become
\begin{eqnarray}
\langle \Lambda\left(P_{\Lambda},\;s_{\Lambda}\right)\left|\bar{s}\gamma^{\mu}b\right| \Lambda_b\left(P_{\Lambda_b},\;s_{\Lambda_b}\right)\rangle&=&\bar{u}\left(P_{\Lambda_b},\;s_{\Lambda_b}\right)\bigg[f^{V}_{t}\left(s\right)\left(m_{\Lambda_b}-m_{\Lambda}\right)\frac{q^\mu}{s} +f_{0}^{V}\left(s\right)\frac{m_{\Lambda_b}+m_{\Lambda}}{s_+} \notag\\
&&\times \left(P^{\mu}_{\Lambda_b}+P^{\mu}_{\Lambda}-\frac{q^\mu}{s}\left(m^2_{\Lambda_b}-m^2_{\Lambda}\right)\right) \notag\\
&&+f_{\perp}^{V}\left(s\right)\left(\gamma^{\mu}-\frac{2m_\Lambda}{s_+}P^\mu_{\Lambda_b}-\frac{2m_{\Lambda_b}}{s_+}P^\mu_{\Lambda}\right)\bigg]u\left(P_\Lambda,\;s_\Lambda\right), \label{HelV1}
\end{eqnarray}
with $s_+ = (m_{\Lambda_b}+ m_{\Lambda})^2-s$ and $q=P_{\Lambda_b}-P_{\Lambda}$. 
Similarly, by sandwiching the axial-vector currents between $\Lambda_b$ and $\Lambda$, the corresponding matrix elements can be written as
\begin{eqnarray}
\langle \Lambda\left(P_{\Lambda},\;s_{\Lambda}\right)\left|\bar{s}\gamma^{\mu}\gamma_5 b\right| \Lambda_b\left(P_{\Lambda_b},\;s_{\Lambda_b}\right)\rangle&=&-\bar{u}\left(P_{\Lambda_b},\;s_{\Lambda_b}\right)\bigg[f^{A}_{t}\left(s\right)\left(m_{\Lambda_b}+m_{\Lambda}\right)\frac{q^\mu}{s} +f_{0}^{A}\left(s\right)\frac{m_{\Lambda_b}-m_{\Lambda}}{s_-} \notag\\
&&\times \left(P^{\mu}_{\Lambda_b}+P^{\mu}_{\Lambda}-\frac{q^\mu}{s}\left(m^2_{\Lambda_b}-m^2_{\Lambda}\right)\right) \notag\\
&&+f_{\perp}^{V}\left(s\right)\left(\gamma^{\mu}+\frac{2m_\Lambda}{s_+}P^\mu_{\Lambda_b}-\frac{2m_{\Lambda_b}}{s_-}P^\mu_{\Lambda}\right)\bigg]u\left(P_\Lambda,\;s_\Lambda\right). \label{HelA1}
\end{eqnarray}

For the dipole operators $i\bar{s}q_\nu \sigma^{\mu\nu}b$ and  $i\bar{s}q_\nu \sigma^{\mu\nu}\gamma_5 b$, the respective transition matrix elements are
\begin{eqnarray}
\langle \Lambda\left(P_{\Lambda},\;s_{\Lambda}\right)\left|\bar{s}i \sigma^{\mu\nu}q_\nu b\right| \Lambda_b\left(P_{\Lambda_b},\;s_{\Lambda_b}\right)\rangle&=&-\bar{u}\left(P_{\Lambda_b},\;s_{\Lambda_b}\right)\bigg[f_{0}^{T}\frac{s}{s_+} \left(P^{\mu}_{\Lambda_b}+P^{\mu}_{\Lambda}-\frac{q^\mu}{s}\left(m^2_{\Lambda_b}-m^2_{\Lambda}\right)\right) \notag\\
&&+f_{\perp}^{T}\left(m_{\Lambda_b}+m_{\Lambda}\right)\left(\gamma^{\mu}-\frac{2m_\Lambda}{s_+}P^\mu_{\Lambda_b}-\frac{2m_{\Lambda_b}}{s_+}P^\mu_{\Lambda}\right)\bigg]u\left(P_\Lambda,\;s_\Lambda\right), \label{HelT1}
\end{eqnarray}
and
\begin{eqnarray}
\langle \Lambda\left(P_{\Lambda},\;s_{\Lambda}\right)\left|\bar{s}i \sigma^{\mu\nu}q_\nu\gamma_5 b\right| \Lambda_b\left(P_{\Lambda_b},\;s_{\Lambda_b}\right)\rangle&=&-\bar{u}\left(P_{\Lambda_b},\;s_{\Lambda_b}\right)\gamma_5\bigg[f_{0}^{T_5}\frac{s}{s_-} \left(P^{\mu}_{\Lambda_b}+P^{\mu}_{\Lambda}-\frac{q^\mu}{s}\left(m^2_{\Lambda_b}-m^2_{\Lambda}\right)\right) \notag\\
&&+f_{\perp}^{T_5}\left(m_{\Lambda_b}-m_{\Lambda}\right)\left(\gamma^{\mu}+\frac{2m_\Lambda}{s_-}P^\mu_{\Lambda_b}-\frac{2m_{\Lambda_b}}{s_-}P^\mu_{\Lambda}\right)\bigg]u\left(P_\Lambda,\;s_\Lambda\right). \label{HelT2}
\end{eqnarray}

The matrix elements for the scalar and pseudo-scalar currents can be obtained from Eq. (\ref{HelV1}) and Eq. (\ref{HelA1}) after contracting with the di-lepton momentum transfer $q_{\mu}$ and invoking the Dirac equation. This gives
\begin{eqnarray}
\langle \Lambda\left(P_{\Lambda},\;s_{\Lambda}\right)\left|\bar{s} b\right| \Lambda_b\left(P_{\Lambda_b},\;s_{\Lambda_b}\right)\rangle &=&f^{V}_t\left(s\right)\frac{m_{\Lambda_b}-m_{\Lambda}}{m_b}\bar{u}\left(P_{\Lambda_b},\;s_{\Lambda_b}\right)u\left(P_{\Lambda},\;s_{\Lambda}\right)\label{S1}\\
\langle \Lambda\left(P_{\Lambda},\;s_{\Lambda}\right)\left|\bar{s}\gamma_5 b\right| \Lambda_b\left(P_{\Lambda_b},\;s_{\Lambda_b}\right)\rangle &=&f^{A}_t\left(s\right)\frac{m_{\Lambda_b}+m_{\Lambda}}{m_b}\bar{u}\left(P_{\Lambda_b},\;s_{\Lambda_b}\right)\gamma_5u\left(P_{\Lambda},\;s_{\Lambda}\right)\label{S2}.
\end{eqnarray}
where we have ignored the strange quark mass. One can see from above equation that these currents do not contribute any new FFs. 

In the theoretical study of the exclusive decays, FFs being the non-perturbative quantities are the major source of uncertainties and hence having a good control on their precise calculation is always a need of time. To address this, several approaches have been opted to compute them, e.g., the  quark models \cite{Cheng:1995fe,Mohanta:1999id,Mott:2011cx},  the Lattice QCD \cite{Detmold:2016pkz}, light cone sum rules (LCSR) \cite{Aliev:2010uy,Khodjamirian:2011jp} and the perturbative QCD approach \cite{He:2006ud}. In order to reduce the number of independent FFs, some effective theories are used, e.g., the Heavy quark effective theory (HQET) \cite{Hussain:1990uu,Hussain:1992rb} helps to reduce the number of independent FFs from ten to two i.e., the Isuger-wise relations $\xi_1$ and $\xi_2$. Similarly, in soft-collinear effective theory (SCET) the evaluation of the FFs \cite{Feldmann:2011xf} reduces this number to one.  In our analysis, we use the FFs calculated by using Lattice QCD for full dilepton mass square range and these can be expressed as \cite{Detmold:2016pkz}:
\begin{equation}
f(s)= \frac{a^f_0 +a^f_1 z(s)}{1-s/(m^f_{pole})^2}
\end{equation}
The inputs $a^f_0$ and $a^f_1$ are given in table V and $m^f_{pole}$ in Table III of \cite{Detmold:2016pkz} and these are summarized in Tables I and II of \cite{Nasrullah:2018puc} with the replacement of $f^{V}_{0,\;+,\;\perp} \to f^{V}_{t,\;0,\;\perp}$, $g_{0,\;+,\;\perp} \to f^{A}_{t,\;0,\;\perp}$, $h_{+,\; \perp} \to f^{T}_{0,\; \perp}$ and $h_{+,\; \perp} \to f^{T_5}_{0,\; \perp}$. The parameter $z$ is defined as \cite{Detmold:2016pkz}
\begin{equation}
z(s)= \frac{\sqrt{t_{+}-s}- \sqrt{t_{+}-t_{0}}}{\sqrt{t_{+}-s}+\sqrt{t_{+}-t_{0}}} \label{z-parameter},
\end{equation}
where $t_0=( m_{\Lambda_b}-m_\Lambda)^2$ and $t_+ = (m_B-m_K)^2$.

\subsection{Cascade Decay $\Lambda \rightarrow p\pi^-$ \label{cascade}}

The SM effective Hamiltonian for the cascade decay $\Lambda \rightarrow p\pi$ is given as
\begin{eqnarray}
H_{\Lambda}^{eff}= \frac{4 G_F}{\sqrt{2}}V_{us}V_{ud}^{*}(\bar{d}\gamma^{\mu}P_L u)(\bar{u}\gamma^{\mu}P_Ls) \label{hami-cas}.
\end{eqnarray}
Again, by using the Hamiltonian (\ref{hami-cas}) between initial state $\Lambda$ and final state $p$, the matrix elements can be expressed in term of the QCD parameters (see ref. \cite{Boer:2014kda} for details).  The non-zero helicity contributions to the total decay width for this decay are
\begin{eqnarray}
\Gamma^{\prime}(+1/2,+1/2) &=& (1+\alpha \cos{\theta_{\Lambda}})\Gamma_{\Lambda} \hspace{2cm} \Gamma^{\prime}(+1/2,-1/2)=-\alpha \sin{\theta_{\Lambda}}e^{i\phi}\Gamma_{\Lambda} \notag \\
\Gamma^{\prime}(-1/2,-1/2) &=& (1-\alpha \cos{\theta_{\Lambda}})\Gamma_{\Lambda} \hspace{2cm} \Gamma^{\prime}(-1/2,+1/2)=-\alpha \sin{\theta_{\Lambda}}e^{-i\phi}\Gamma_{\Lambda}
\end{eqnarray}
where $\Gamma_{\Lambda}$ is the decay width of $\Lambda \to p\pi$.

\section{Four Fold Angular Distribution and Physical Observables\label{angular}}
The four fold differential decay width for the four-body decay process $\Lambda_b \rightarrow \Lambda (\rightarrow p\pi)\ell^+ \ell^-$ is
\begin{eqnarray*}
\frac{d^{4}\Gamma}{ds\;d\cos{\theta_\ell}d\cos{\theta_\Lambda} d\phi}=d\Gamma_{VA}+d\Gamma_{SP}+d\Gamma_{T^{\prime}}+d\Gamma_{VA-SP}+d\Gamma_{VA-T^{\prime}}+d\Gamma_{SP-T^{\prime}}, \label{DDR}
\end{eqnarray*}
denoting
\begin{eqnarray*}
d\Gamma_{i}=\frac{d^{4}\Gamma_{i}}{ds d\cos{\theta_\ell}d\cos{\theta_\Lambda} d\phi}
\end{eqnarray*}
with $i=VA, \;SP,\; T^{\prime},\; VA-SP,\; VA-T^{\prime}$ and $SP-T^{\prime}$. Eq. (\ref{DDR}) can be written in the form of different matrix elements as
\begin{eqnarray}
\frac{d^{4}\Gamma}{ds d\cos{\theta_\ell}d\cos{\theta_\Lambda} d\phi} = \mathcal{N}\left[ \vert M_{VA}\vert^2+\vert M_{SP}\vert^2+\vert M_{T^{\prime}}\vert^2 +\left(M_{VA}M^*_{SP}+M_{VA}M^*_{T^{\prime}}+M_{SP}M^*_{T^{\prime}}+ h.c.\right) \right],
\end{eqnarray}
where the normalization constant $\mathcal{N}$ is given by
\begin{eqnarray*}
\mathcal{N}= \frac{(G_F \alpha_e V_{tb}V_{ts}^*)^2 \lambda v}{3\times 2^{11}m_{\Lambda_b}^3\pi^5},
\end{eqnarray*}
with $v=\sqrt{1-\frac{4m_{\ell}^2}{s}}$ and $\lambda= (m_{\Lambda_b}^2-m_{\Lambda}^2-s)^2+4sm_{\Lambda}^2$. The non-zero helicity components of hadron and lepton current are given in the Appendix. Here, we would like to mention that our expressions of the lepton helicity components corresponding to different currents include the lepton mass term and by setting it equal to zero, these are reduced to the simple ones given in \cite{Das:2018sms}.

As in the SM, the currents corresponding to $\Lambda_b \to \Lambda \ell^{+}\ell^{-}$ is vector and axial-vector with combination $V-A$, therefore, the contribution of $VA$ operators will only modify some of the angular coefficients appearing in the SM. However, contributions from the scalar and pseudo-scalar operators which are absent in the SM will introduce the new angular coefficients. Putting everything together the four-fold angular decay distribution for $\Lambda_b \to \Lambda (\to p \pi)\ell^{+}\ell^{-}$ becomes
\begin{eqnarray}
\frac{d^4\Gamma}{ds \ d\cos {\theta}_{\ell} \ d\cos{\theta_{\Lambda}} \ d\phi} &=& \frac{3}{8 \pi} \bigg[K_{1ss} \sin^2{\theta_{\ell}}+K_{1cc} \cos^2{\theta_{\ell}}+ K_{1c} \cos{\theta_{\ell}} + (K_{2ss} \sin^2{\theta_{\ell}}+K_{2cc} \cos^2{\theta_{\ell}}+ K_{2c} \cos{\theta_{\ell}}) \cos{\theta_{\Lambda}} \notag \\
&&+\left((K_{3sc} \cos{\theta_{\ell}}+K_{3s}) \sin{\theta_{\ell}}\right) \sin{\theta_{\Lambda}} \sin{\phi} + ((K_{4sc}\cos{\theta_{\ell}}+K_{4s}) \sin{\theta_{\ell}}) \sin{\theta_{\Lambda}} \cos{\phi}  \bigg]. \label{eq6}
\end{eqnarray}
In order to extract the different angular observables, we will follow the approach adopted in Refs. \cite{Nasrullah:2018puc} and \cite{Hu:2017qxj}, the explicit expressions of the physical observables of interest in our model independent approach are
\begin{eqnarray}
\frac{d\Gamma}{ds} &=& K_{1ss}+K_{1cc}, \hspace{1.5cm}
A_{FB}^{\ell} =\frac{3K_{1c}}{4K_{1ss}+2K_{1cc}}, \hspace{1.5cm}
A_{FB}^{\Lambda }=\frac{2K_{2ss}+K_{2cc}}{4K_{1ss}+2K_{1cc}}, \notag \\
A_{FB}^{\ell\Lambda } &=& \frac{3K_{2c}}{8K_{1ss}+4K_{1cc}}, \hspace{1.5cm}
\alpha_{U} = \frac{\widetilde{K}_{2cc}}{ K_{1cc}} ,\hspace{3.3cm}  \alpha_{L} = \frac{\widetilde{K}_{2ss}}{ K_{1ss}}, \hspace{2.4cm} \notag \\
\alpha_{\theta_{\Lambda}} &=& \frac{\widetilde{K}_{2cc}+2 \widetilde{K}_{2ss}}{{K}_{1cc}+2 {K}_{1ss}}, \hspace{1.6cm}
\alpha_{\theta_{\ell}} = \frac{K_{1cc}-K_{1ss}}{K_{1ss}},\hspace{2.1cm}
\alpha^{\prime}_{\theta_{\ell}}=\frac{{K}_{1c}}{K_{1ss}}, \notag\\
\alpha_{\phi} &=& \frac{3 \pi^2 \widetilde{K}_{4s}}{16 (K_{1cc}+2 K_{1ss})}, \hspace{1.1cm}
\alpha^{\prime}_{\phi} = \frac{3 \pi^2 \widetilde{K}_{3s}}{16 (K_{1cc}+2 K_{1ss})}, \hspace{1.4cm}
\mathcal{P}_{3} =\frac{2K_{2c}}{ \widehat{\Gamma}}, \notag\\  \mathcal{P}_{8} &=& \frac{4K_{1c}}{ \widehat{\Gamma}}, \hspace{3cm}
\mathcal{P}_{9} =\frac{4K_{2cc}}{ \widehat{\Gamma}}, \hspace{3.2cm} Y_2=\frac{3(K_{2cc}-K_{2ss})}{8\widehat{\Gamma}}, \notag \\
	Y_{3sc} &=& \frac{K_{3sc}}{2 \widehat{\Gamma}}, \hspace{3cm} Y_{4sc}=\frac{K_{4sc}}{2 \widehat{\Gamma}}, \label{observables1}
	\end{eqnarray}
where $\widetilde{K}_{i,j} = \frac{K_{i,j}}{\alpha_{\Lambda}}$ and $\widehat{\Gamma}=\frac{d\Gamma}{ds}$. The detailed expressions of $K_{i,j}$ in terms of helicity amplitudes are given in the Appendix. In Eq. (\ref{observables1}), $\alpha_{\Lambda}$ is the asymmetry parameter corresponding to the parity violating $\Lambda \to p \pi^{-}$ decay and its  experimentally measured value is $0.642\pm 0.013$ \cite{Patrignani:2016xqp}.

\section{Impact of New Couplings on Physical Observables\label{discussion}}
In this section, we will discuss the impact of the NP
couplings corresponding to $VA$, $SP$ and $T$ operators on the
different physical observables discussed above. First we start with $d\mathcal{B}/ds$, $F_L$, $A_{FB}^{\ell}$ and $A_{FB}^{\Lambda}$ for which the experimental data is available. By using the most recent constraints on NP couplings from \cite{Altmannshofer:2017fio}, the idea behind this approach is to see whether these NP couplings accommodate the currently available data \cite{Aaij:2015xza} or not. To accomplish this task, first of all we discuss the impact of individual NP couplings on the above mentioned observables and later we analyze their simultaneous impact. In doing so, we will explore all the available range of new couplings constrained by $B-$meson decays in different bins of $s$. After this, we will discuss the observables $Y_{2,\;3sc,\;4sc}$, $\mathcal{P}_{3,\; 8,\; 9}$ and $\alpha^{(\prime)}_{i}$ where, $i = \theta_\ell,\; \theta_{\Lambda},\; \xi,\; L, \; U$ which show minimum dependence on the FFs and hence are the potential candidates to 
search for NP in some on going and future experiments. In order to present our results of different physical observables, we plot them against the square of the
momentum transfer $s$ in the SM as well as in the presence of NP couplings. In these plots, we have presented our results both for the zero and non-zero lepton ($\mu$) mass. Therefore, our formalism is more general from the previous study of the same decay presented in ref. \cite{Das:2018sms}. Just to distinguish the lepton mass effects \cite{Gutsche:2015mxa}, we have also discussed the different physical observables for $\Lambda_{b}\to \Lambda(\to p\pi)\tau^{+}\tau^{-}$ decay where the mass of final state $\tau'$s is significantly large compared to the $\mu's$ case. Here, we would like to emphasis that all the plots are drawn for the central values of the FFs however, in order to quantify the uncertainties arising due to the FFs and other input parameters we have calculated these observables in different bins of $s$ and tabulated them in Tables \ref{lambda-obs-1}, \ref{lambda-obs-2} and \ref{lambda-obs-3}. Furthermore,
to see whether the NP couplings, $VA$, $SP$ and $T$ could  simultaneously accommodate all available data for the observables $d\mathcal{B}/ds$, $F_L$, $A_{FB}^{\ell}$ and $A_{FB}^{\Lambda}$ of $\Lambda_b \to \Lambda (\to p\pi^-)\mu^{+}\mu^{-}$ decay, we have plotted these observables against the new WCs.  

\begin{figure}
\begin{center}
\begin{tabular}{ll}
\includegraphics[height=4.8cm,width=8cm]{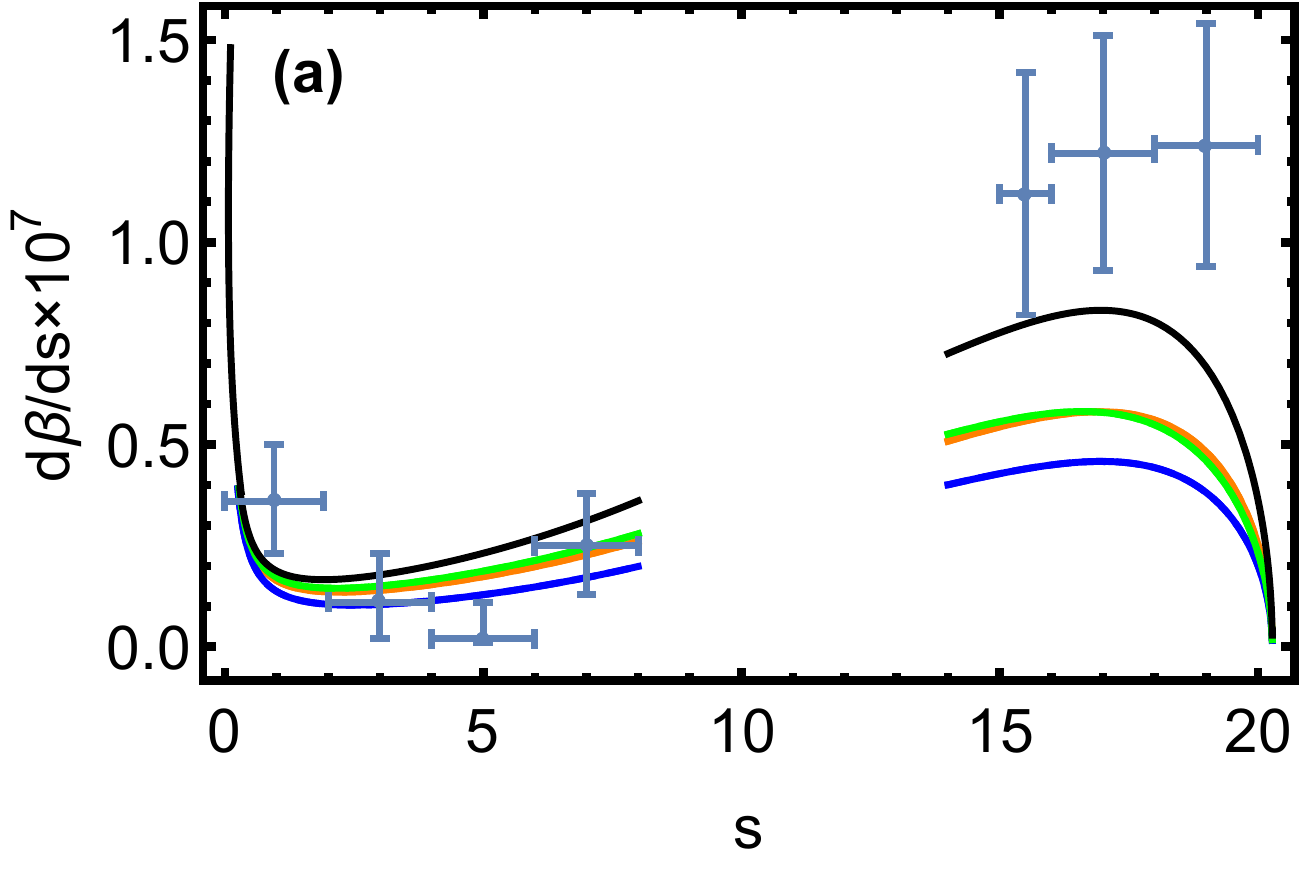}  & \ \ \includegraphics[height=4.8cm,width=8cm]{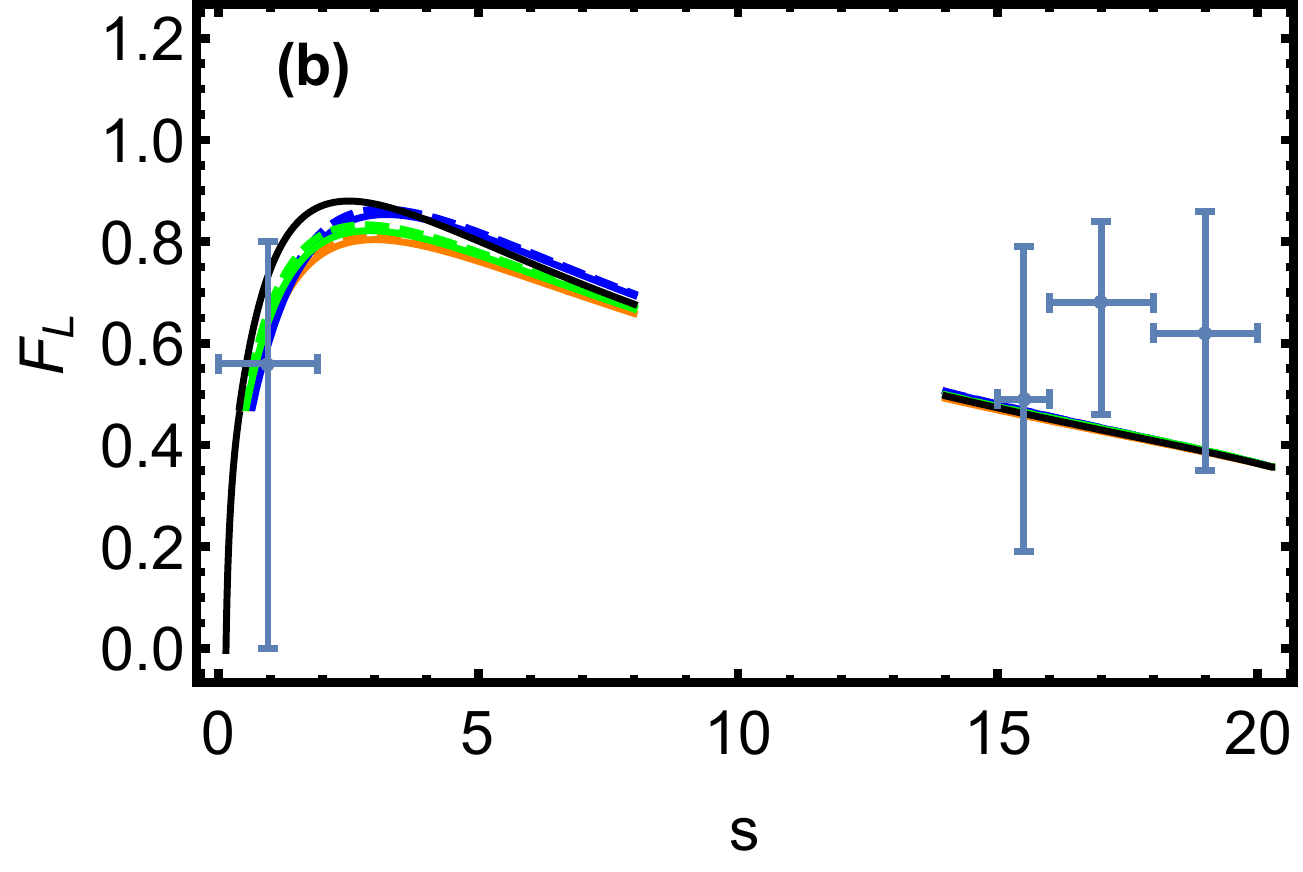} \\
\\
\includegraphics[height=4.8cm,width=8cm]{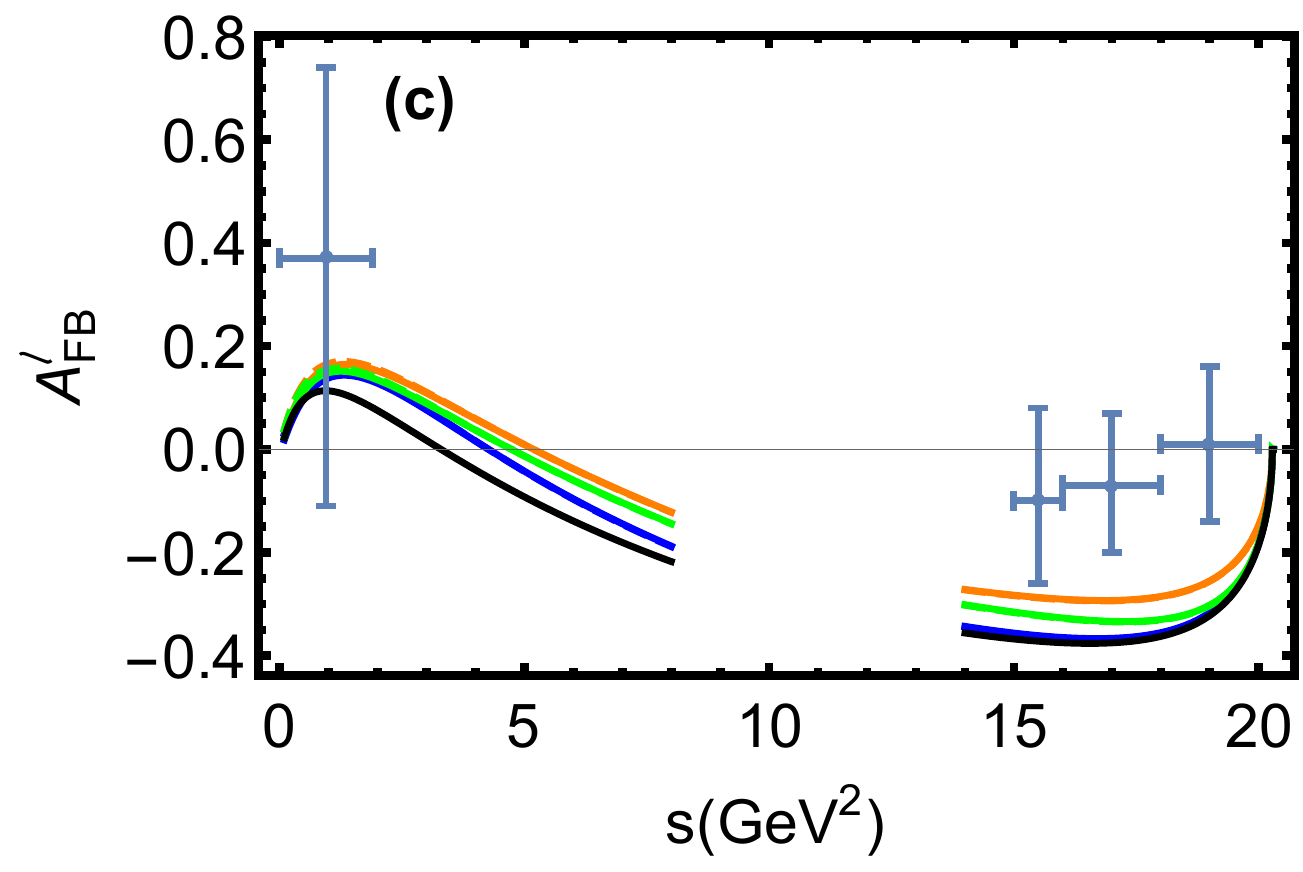}  & \ \ \includegraphics[height=4.8cm,width=8cm]{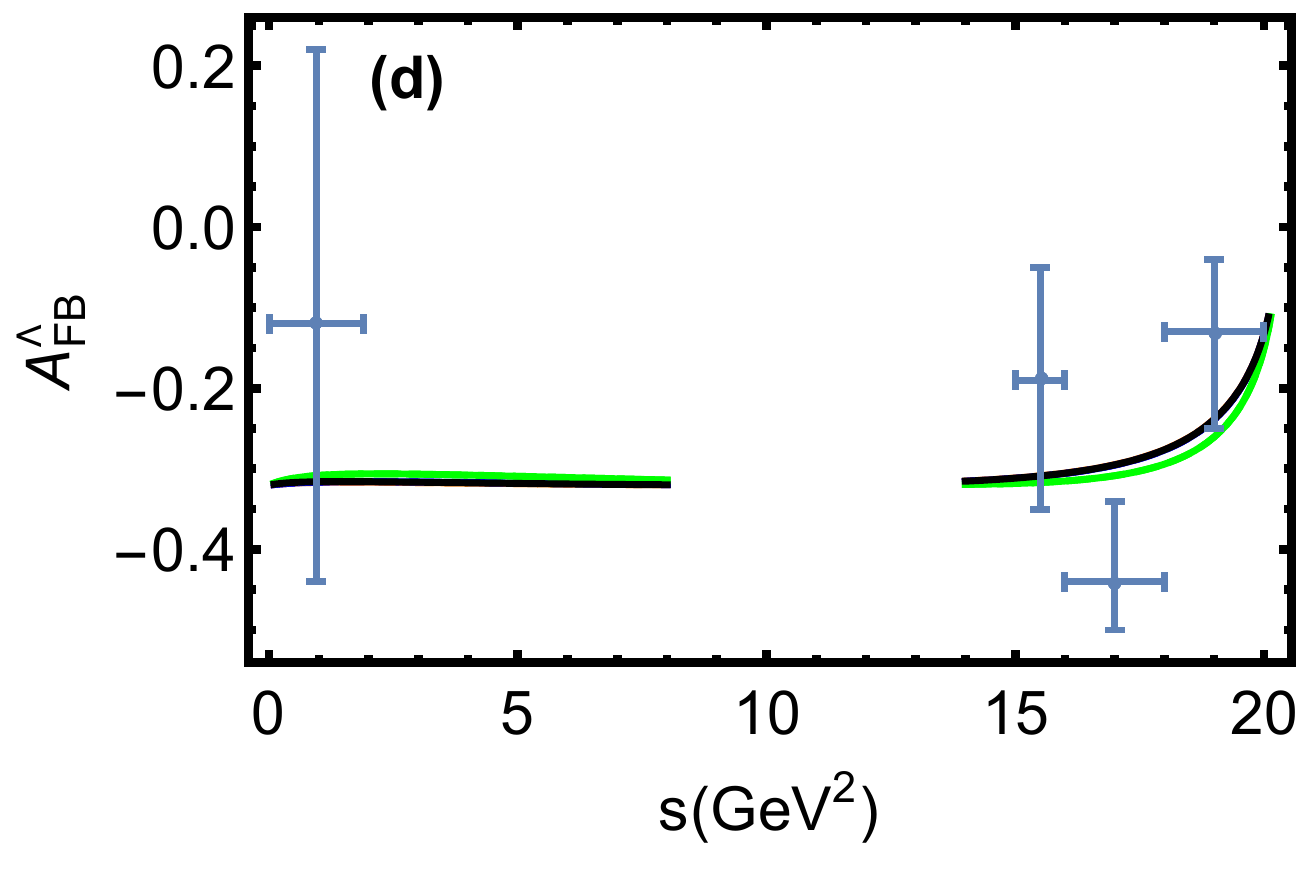} \\
\\
\includegraphics[height=4.8cm,width=8cm]{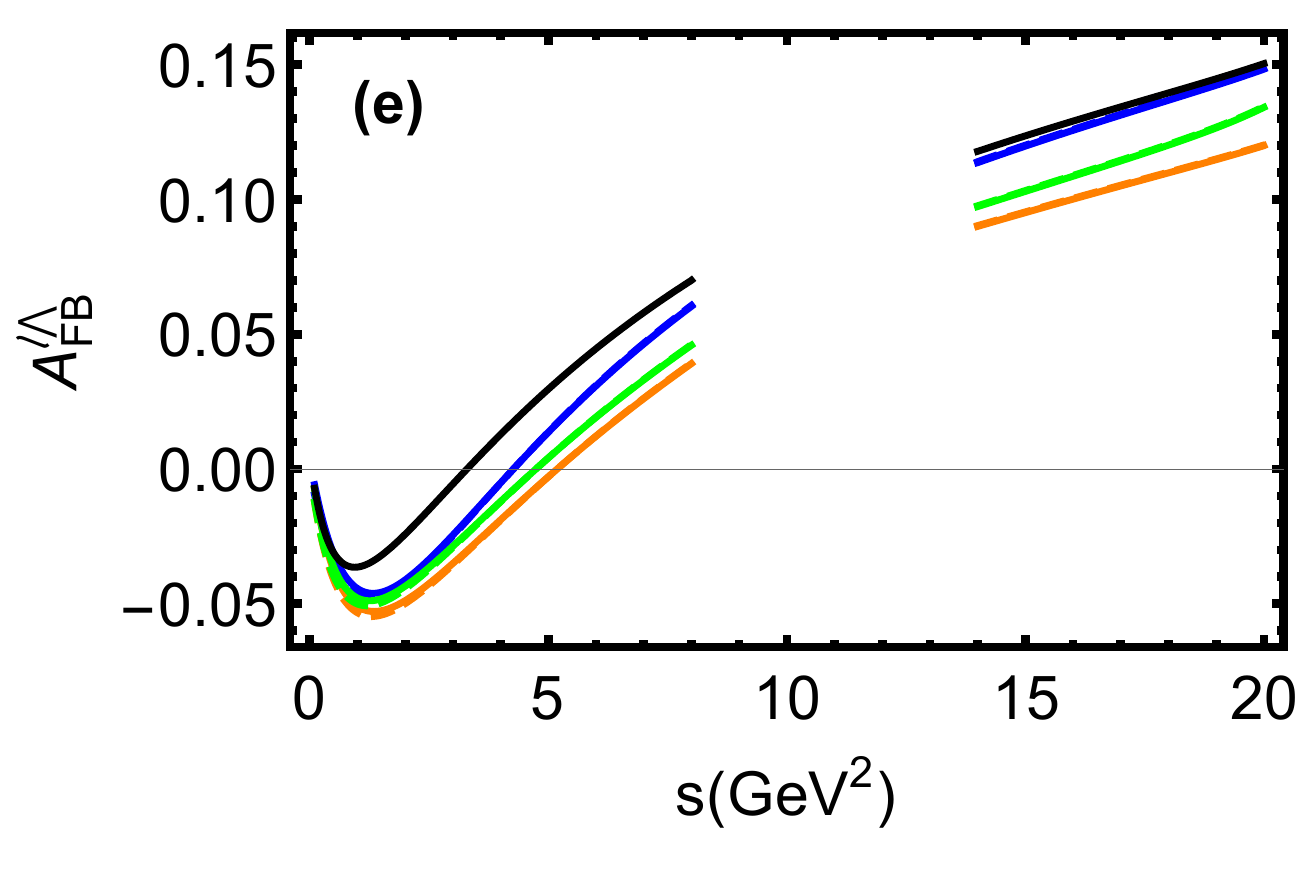}  & \ \ \includegraphics[height=4.8cm,width=8cm]{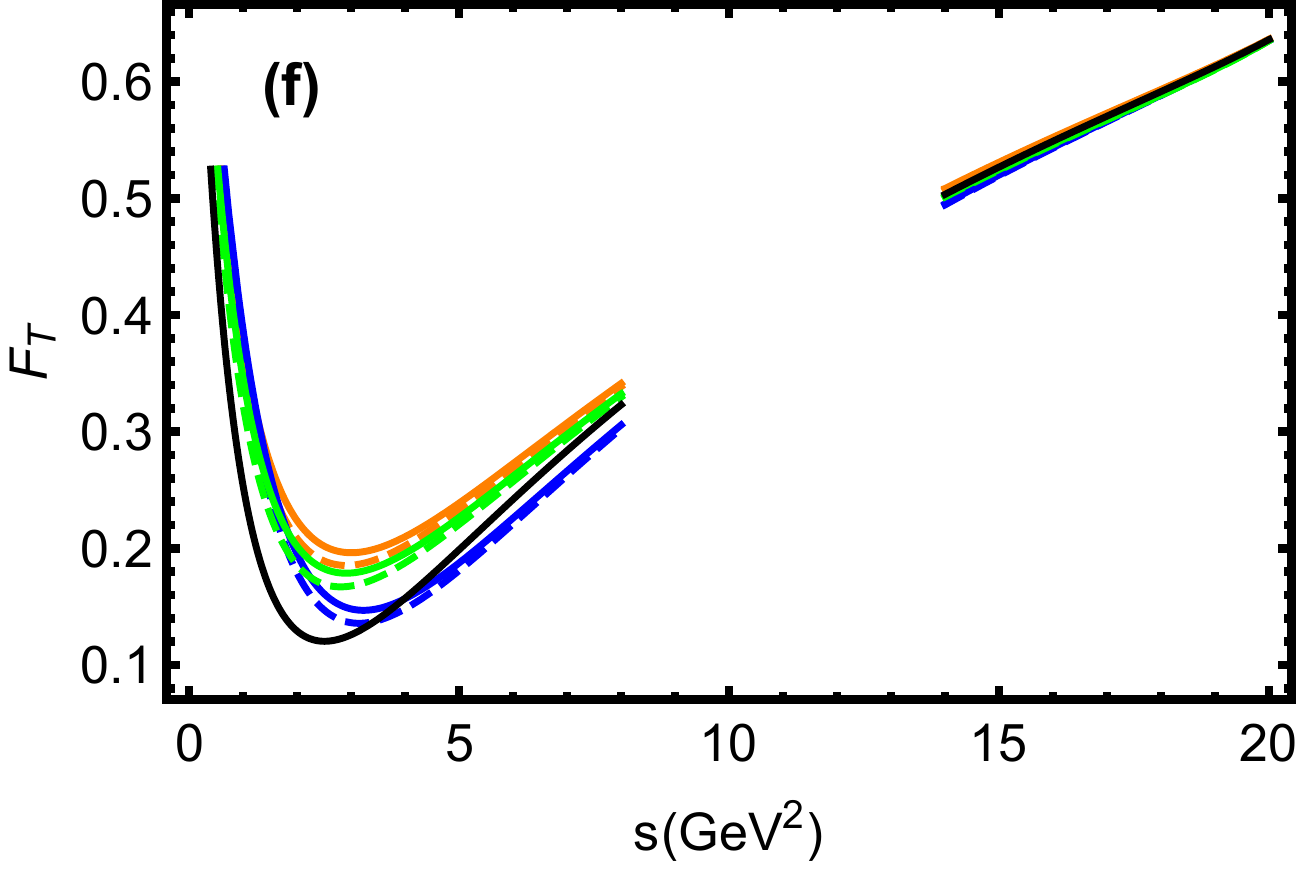} \\
\\
\includegraphics[height=4.8cm,width=8cm]{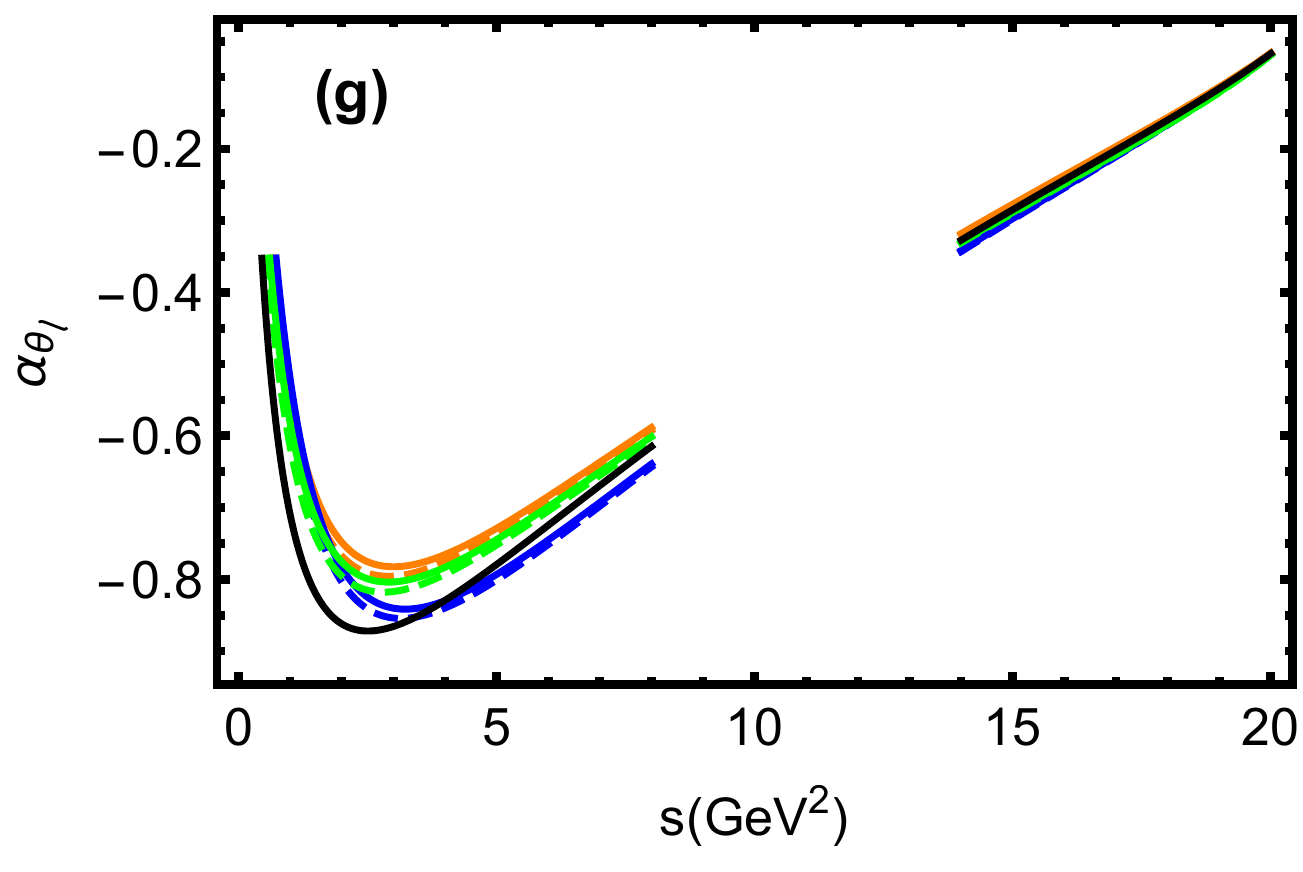}  & \ \ \includegraphics[height=4.8cm,width=8cm]{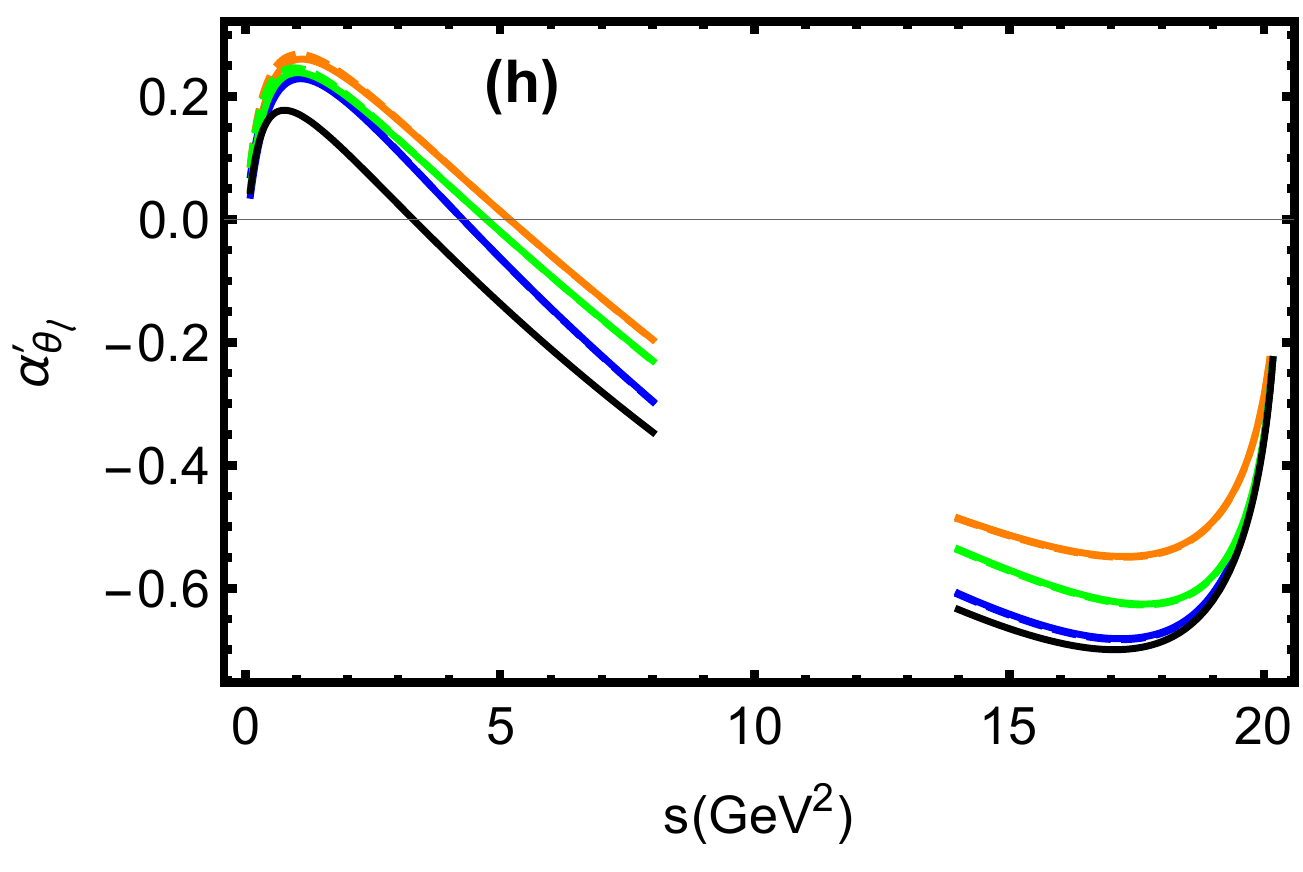}
\end{tabular}
\end{center}
\caption{Observables in the SM and in the presence of new VA couplings. The SM curves are denoted by black color. The orange curve is obtained with $C_V=-1.61$ and $C_V^{\prime}=C_A=C_A^{\prime}=0$. The blue line is for $C_V=-C_A=-1$ and $C_V^{\prime}=C_A^{\prime}=0$ and green color is for $C_V=-1.34$, $C_A^{\prime}=-0.4$ and $C_V^{\prime}=C_A=0$. The solid and dashed lines are for the massive and massless $\mu-$ cases respectively. }\label{Fig1hh}
\end{figure}

\subsection{Vector and Axial-Vector Part $(VA)$}
It is a well established fact that in order to accommodate the discrepancies between the SM predictions and the experimental measurements in different $B-$meson decays, some models with new $VA$ couplings have been proposed \cite{Dutta:2019wxo,Bernlochner:2017jxt}. As these couplings are already present in the SM, therefore, they will only modify the SM WCs leaving the operator bases to be the same. Hence, no new angular coefficient arise in this particular case. In case of the massless lepton and varying the $VA$ couplings in the range $C_V=[-1.61,-1]$, $C_V^{\prime}=0$, $C_A=1$ and $C_A^{\prime}=-0.4$ which take care of the global fit sign, the observables
$d\mathcal{B}/ds$, $A^{\ell}_{FB}$ and $F_{L}$ have already been discussed in \cite{Das:2018sms}. As a first step, we have repeated their analysis for the massless lepton case (dashed lines in all plots) in our formalism and obtained the same results. Later, the same analysis has been done by setting the non-zero
mass for our final state muons (solid lines of all colors). Fig. \ref{Fig1hh}(a) shows that by using available range of  $C^{(\prime)}_{V,A}$
couplings mentioned above, the available data of the branching
ratio could be accommodated only in three low $s$ bins ($s\in [0.1,2]$ GeV$^2$, $s\in [2,4]$ GeV$^2$ and $s\in [6,8]$ GeV$^2$) for which SM also satisfy LHCb results. In case of high $s$ region, no combination of new VA couplings satisfy experimental results as LHCb values in this region are greater than the SM results. However, due to the negative value of $C_V$, the numerical value of the branching ratio in the
presence of these couplings is smaller than the corresponding SM value in the whole $s$ region. This can also be noticed quantitatively from Table \ref{lambda-obs-1} (c.f. column 1) where we can see that in the high $s$ bins the results are suppressed significantly from the SM predictions and even further from the experimental measurements. In addition, the $\mu$-mass does not add any visible deviation for this observable.

  In Table \ref{lambda-obs-1} we can observe that the uncertainties due to FFs are quite significant in the calculation of the branching ratio in the SM as well as in any of the NP scenarios. Hence, we can look for the observables which show minimal dependence on the FFs and $A_{FB}^{\ell}$ is one of them. In case of $A_{FB}^{\ell}$ though, the NP couplings enhanced its average value but still it is small
  enough to accommodate the data (c.f. second column in Table \ref{lambda-obs-2}). As the zero position of this
  asymmetry is proportional to the vector-current coefficients $C^{\text{eff}}_{7}/C_{9}^{\text{eff}}$ therefore, the shift in the zero-position is
  expected after addition of any new vector type couplings and it can be seen in Fig. \ref{Fig1hh}(c). We hope that the future data of the
  zero-position of $A^{\ell}_{FB}$ in $\Lambda_b\to \Lambda \mu^{+}\mu^{-}$ decay will further improve the constraints on $VA$ couplings.
  Like the branching ratio, the $\mu$-mass effects are also invisible in this case too. 
  \begin{figure}
  	\begin{center}
  		\begin{tabular}{ll}
  			\includegraphics[height=4.8cm,width=8cm]{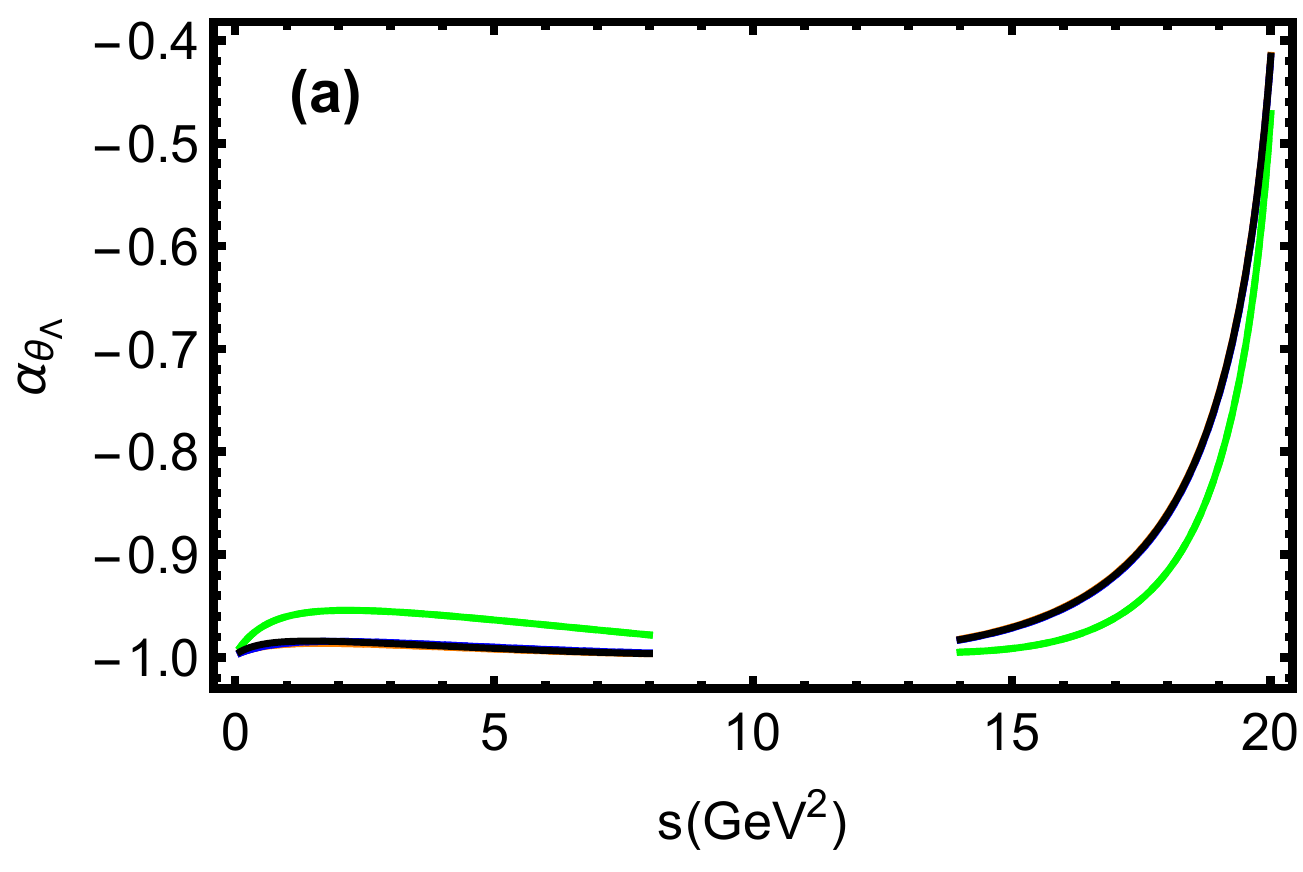}  & \ \ \includegraphics[height=4.8cm,width=8cm]{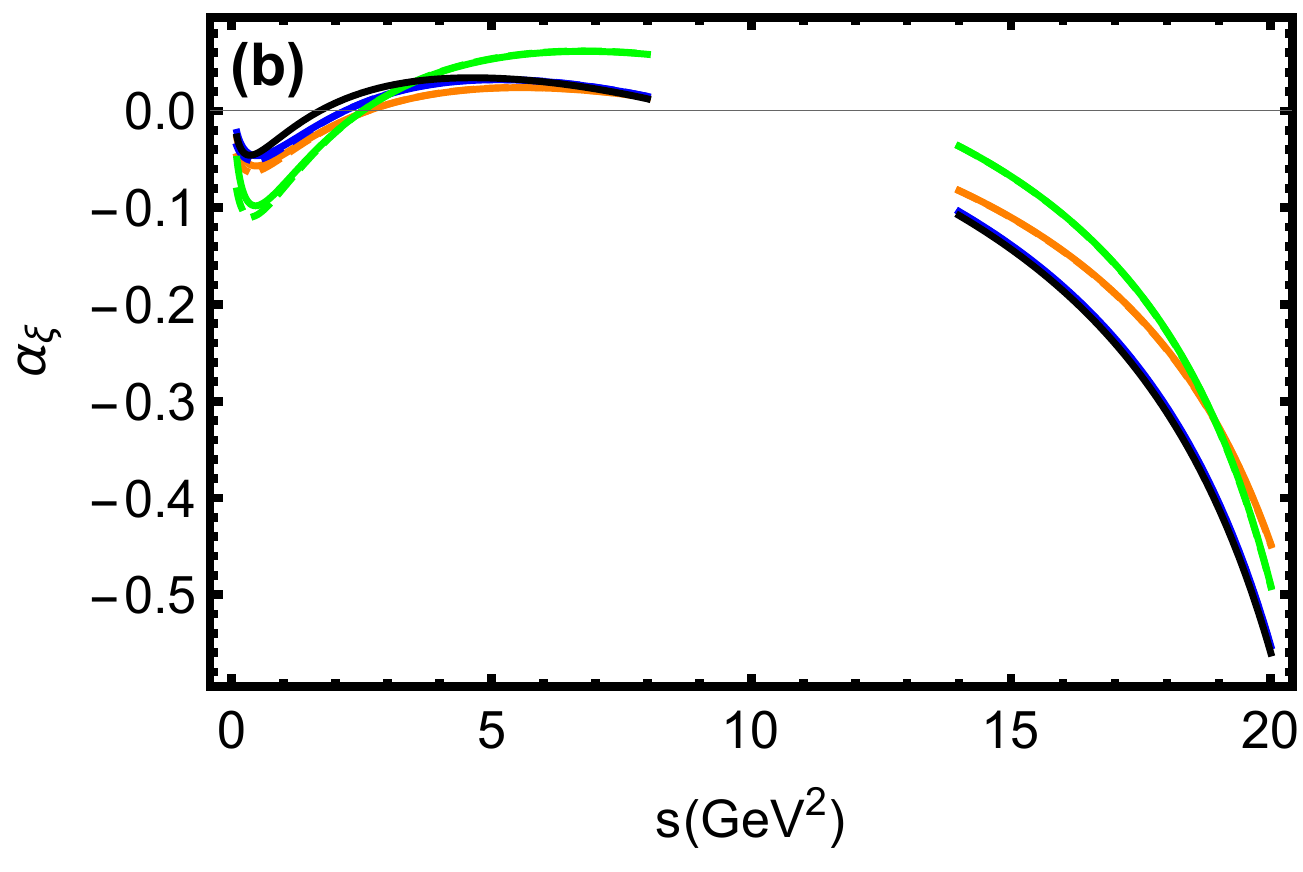} \\
  			\\
  			\includegraphics[height=4.8cm,width=8cm]{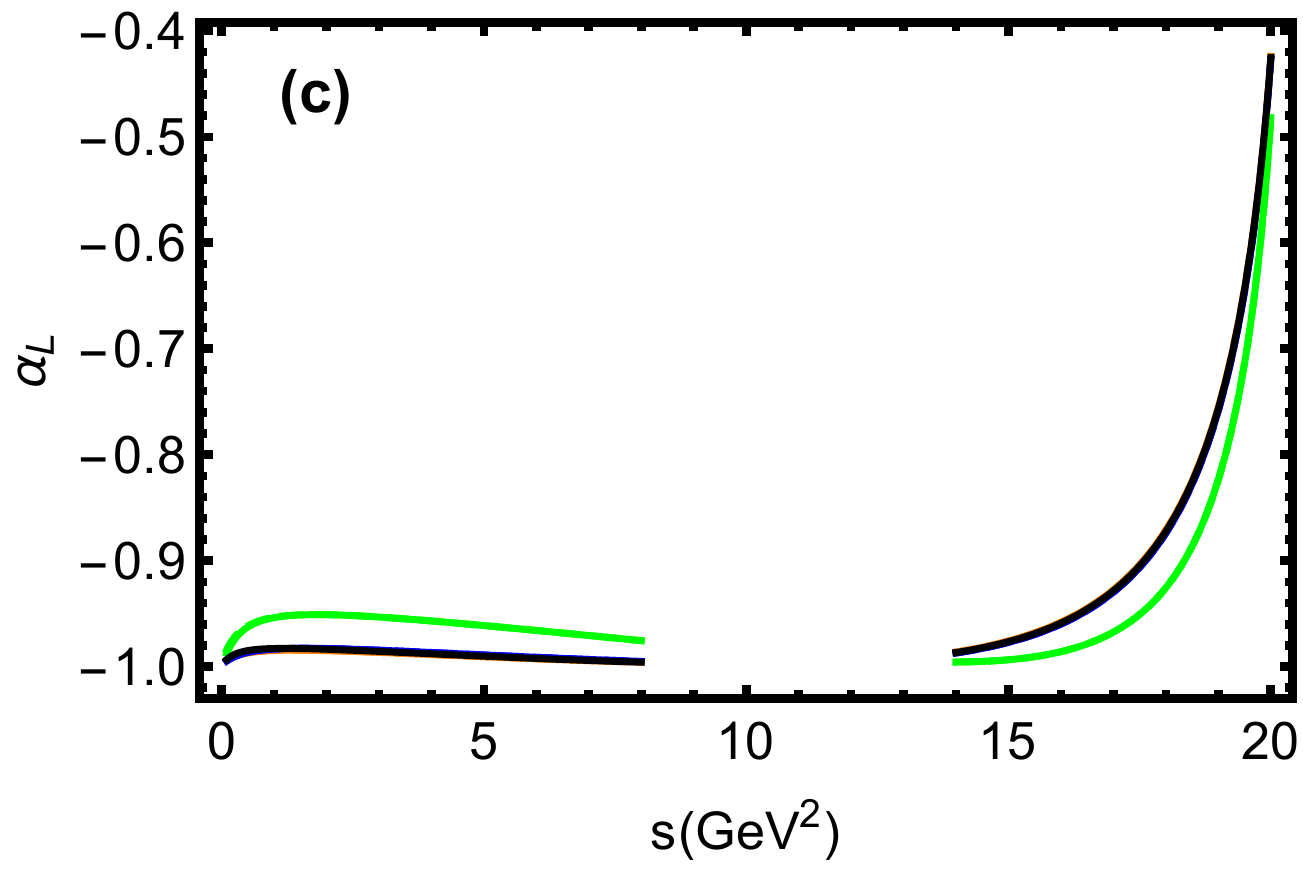}  & \ \ \includegraphics[height=4.8cm,width=8cm]{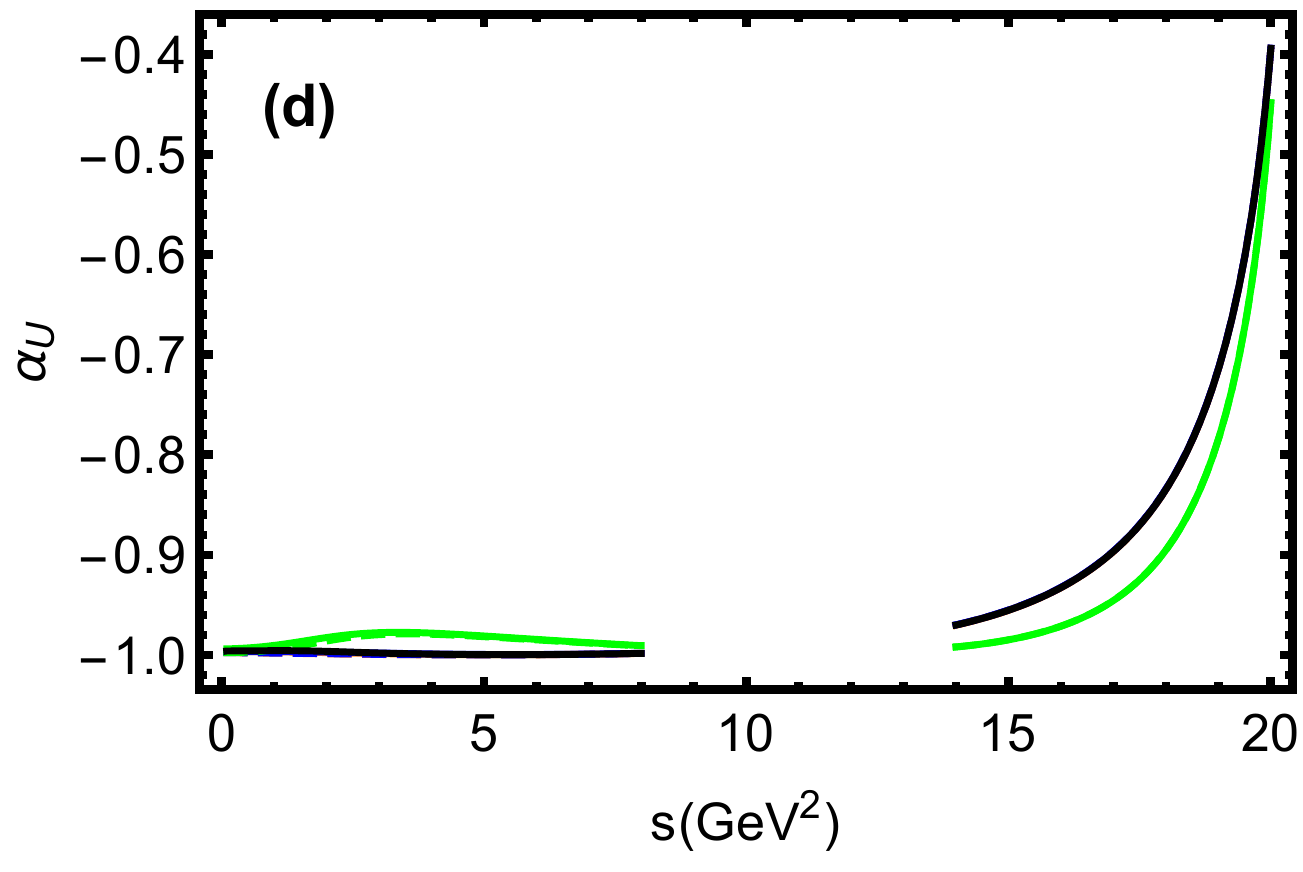} \\
  			\\
  			\includegraphics[height=4.8cm,width=8cm]{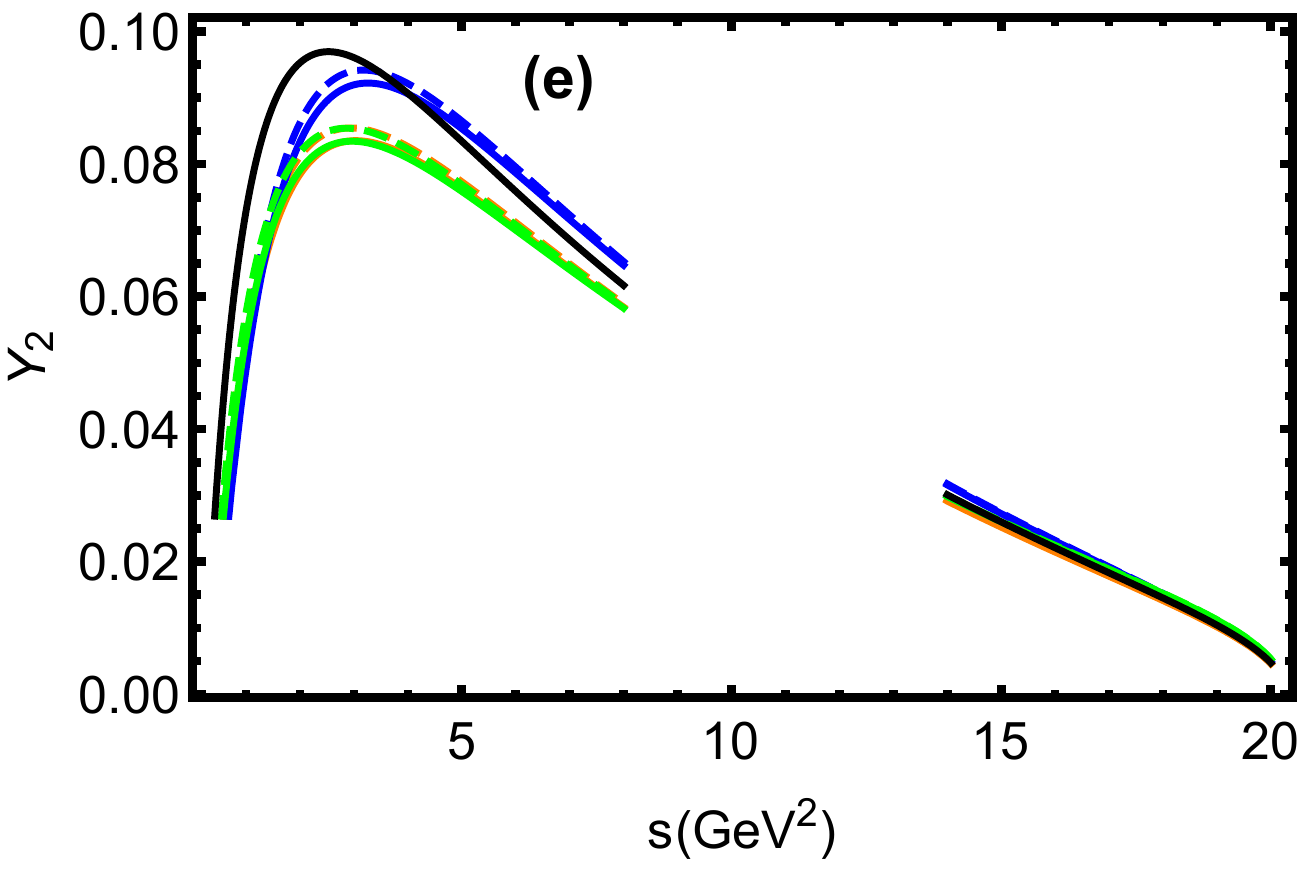} & \ \ \includegraphics[height=4.8cm,width=8cm]{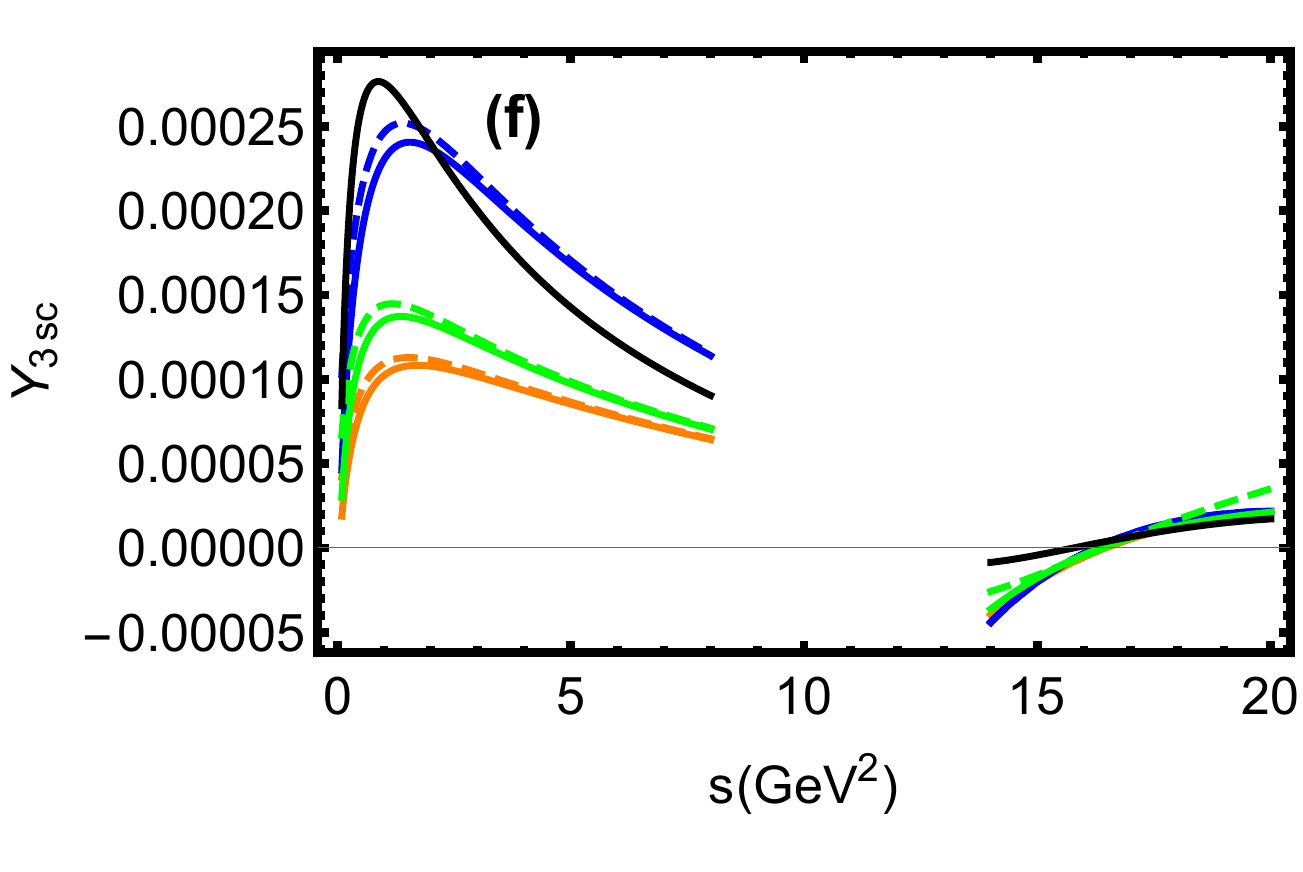} \\
  			\\ \includegraphics[height=4.8cm,width=8cm]{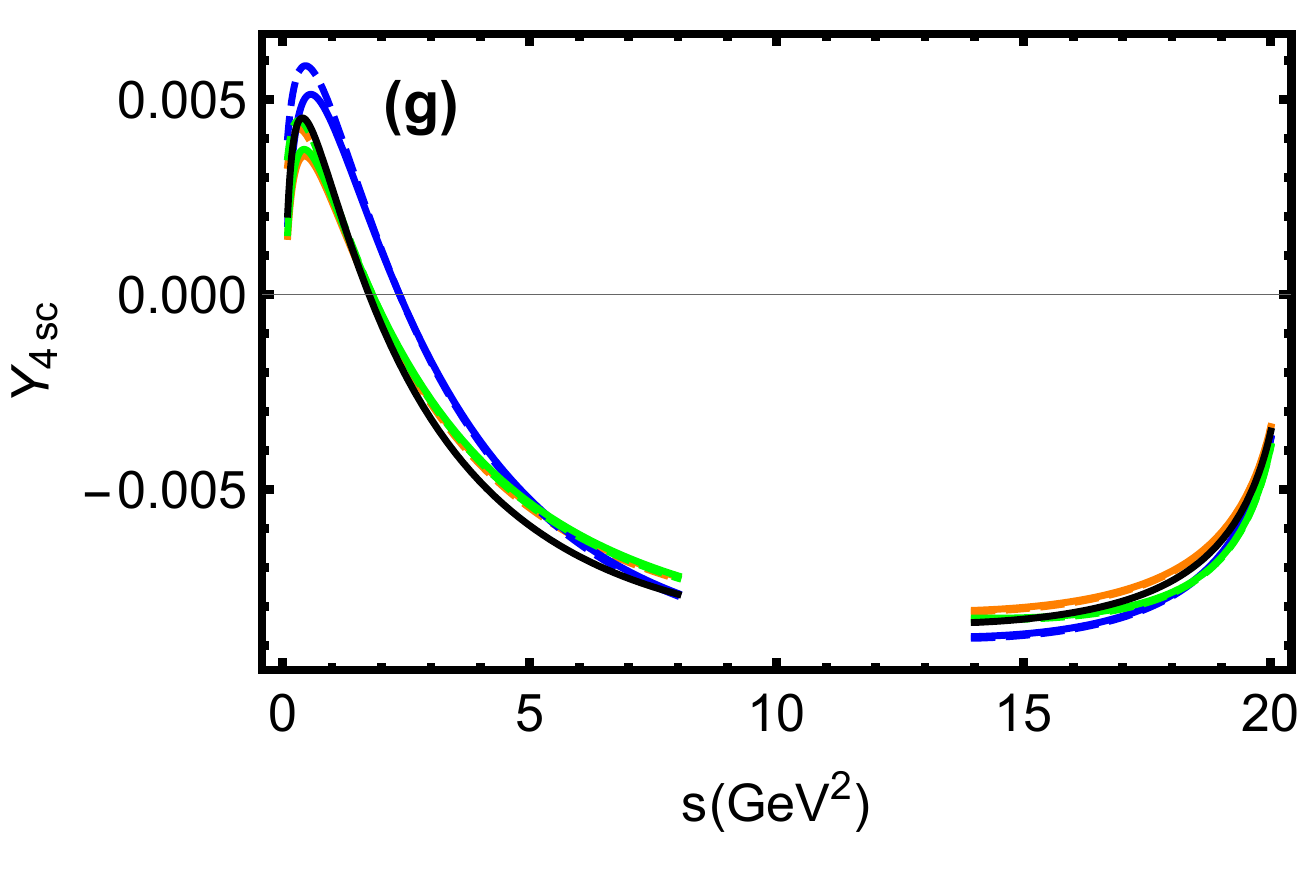} &
  		\end{tabular}
  	\end{center}
  	\caption{Observables in the SM and in the presence of new VA couplings. The description of different curves is similar to the Fig. \ref{Fig1hh} .}\label{Fig2hh}
  \end{figure}
  \begin{figure}
  	\begin{center}
  		\begin{tabular}{ll}
  			\includegraphics[height=4.8cm,width=8cm]{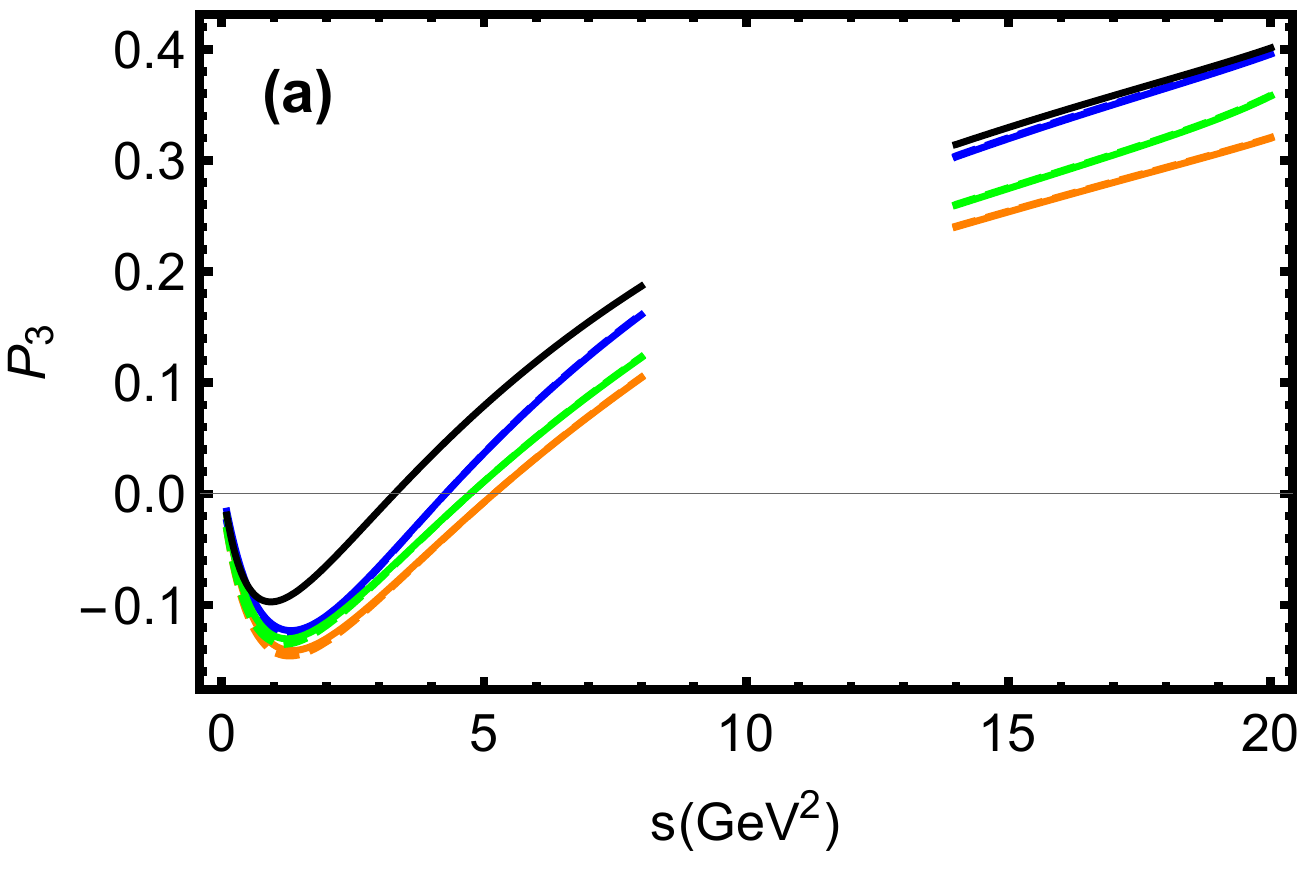}  & \ \ \includegraphics[height=4.8cm,width=8cm]{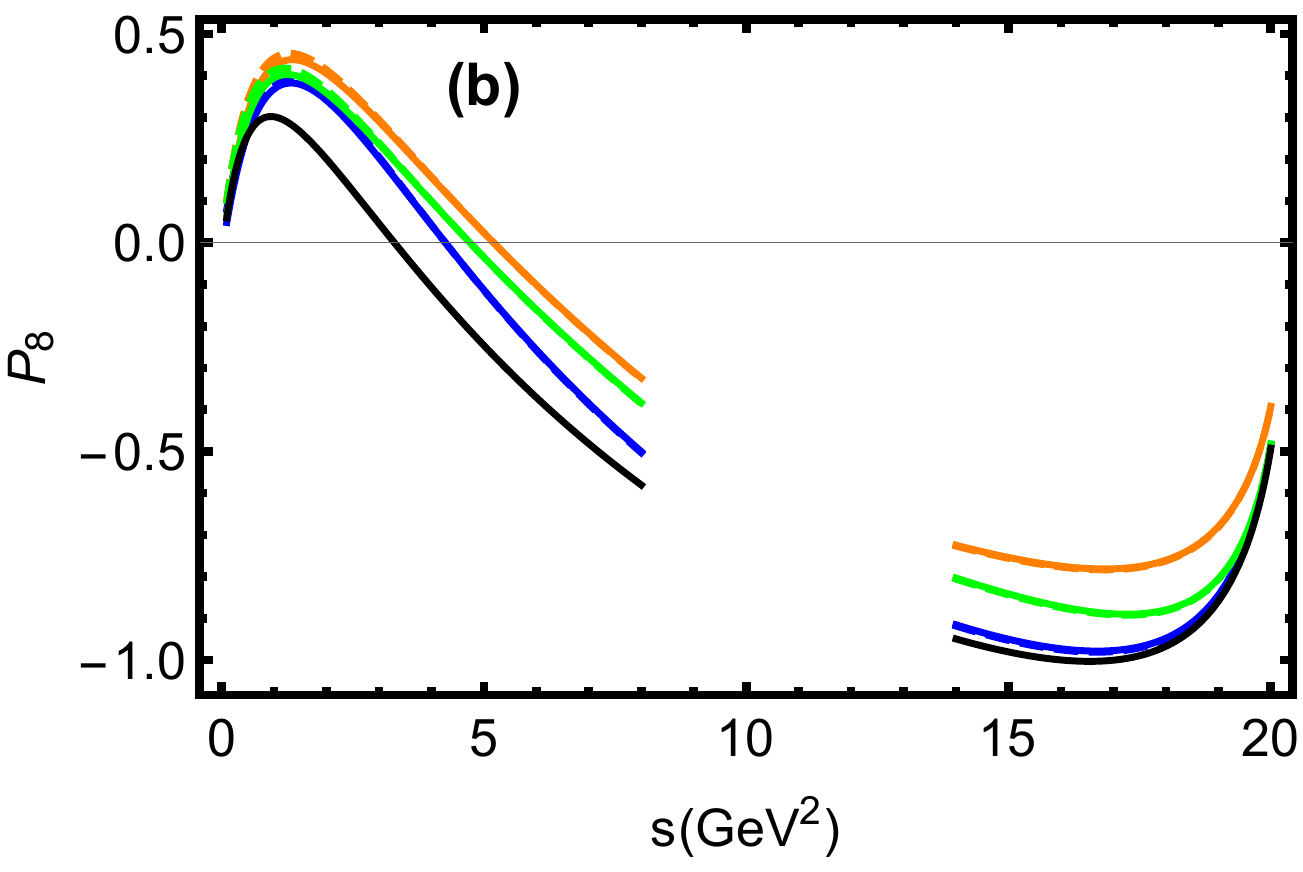} \\
  			\\
  			\includegraphics[height=4.8cm,width=8cm]{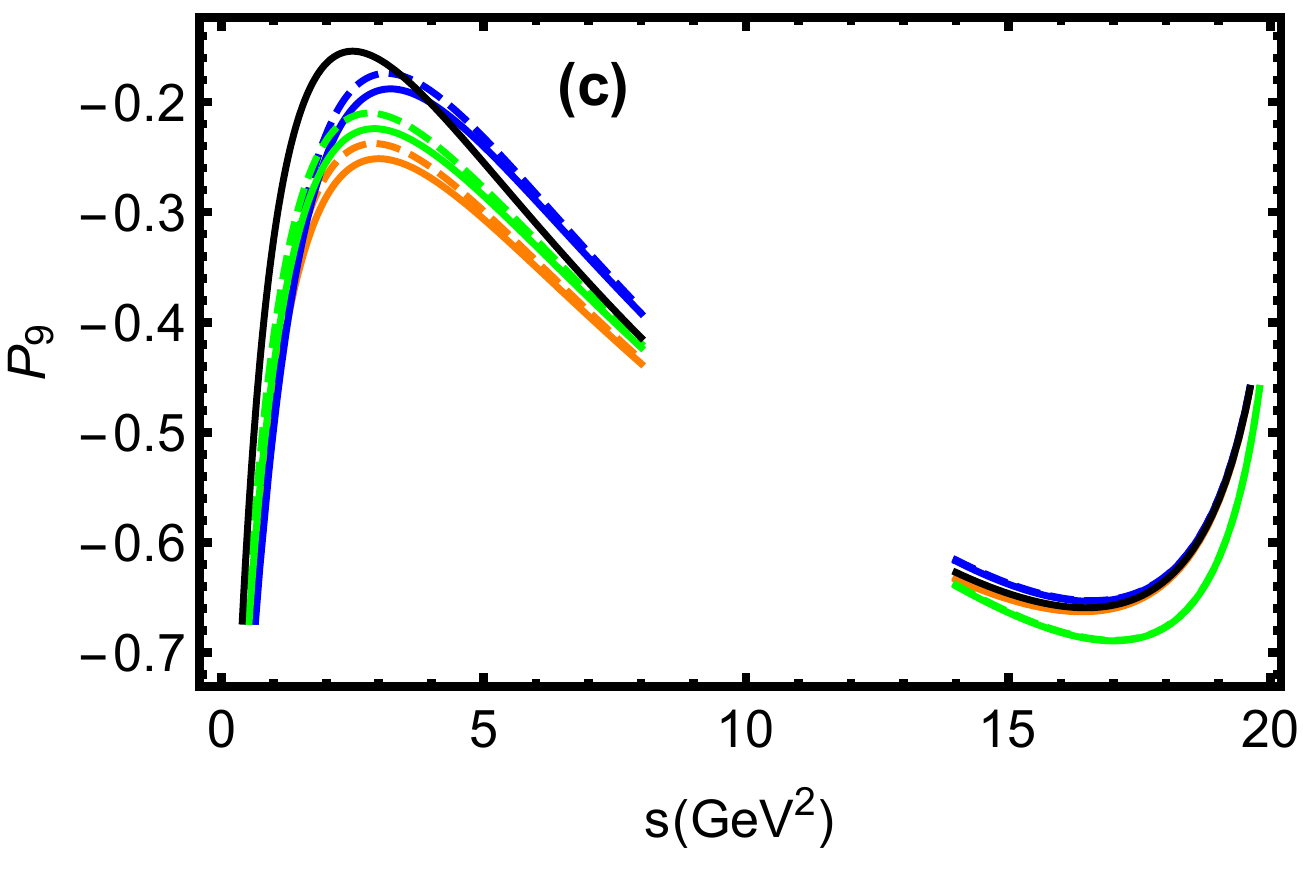} &
  		\end{tabular}
  	\end{center}
  	\caption{$\mathcal{P}_3$, $\mathcal{P}_8$ and $\mathcal{P}_9$ in the SM and in the presence of new VA couplings. The description of different curves is similar to the Fig. \ref{Fig1hh}.}\label{Fig2a}
  \end{figure}
  
  The situation is
  slightly different for $F_{L}$ (Fig. \ref{Fig1hh}(b)) and $A_{FB}^{\Lambda}$ (Fig. \ref{Fig1hh}(d)) where the above constraints on $VA$ couplings satisfy the data within errors in the measurements especially for $F_{L}$ in the $s \in [1,3]$ GeV$^2$ and $s \in [15,16]$ GeV$^2$ bins. Again, going from massless to massive $\mu-$case did not lead to any significant change. It is also worth mentioning that just like $A^{\ell}_{FB}$ the hadronic uncertainties due to FFs in $F_{L}$ and $A_{FB}^{\Lambda}$ are negligible for all low-recoil bins both for the SM and when the new $VA$ couplings are considered (c.f. Table \ref{lambda-obs-1}). This can provide a clean way to test the SM and NP models with additional $VA$ couplings.
  
  Besides the above mentioned observables we show that there are some other interesting physical observables; e.g., the combined lepton-baryon forward-backward asymmetry $A^{\ell\; \Lambda}_{FB}$, the fractions of transverse $(F_T)$ polarized dimuons, the asymmetry parameters $\alpha^{(\prime)}_{\theta_{\ell}},\;\alpha_{\theta_\Lambda},\; \alpha_{\xi},\; \alpha_L,\; \alpha_U$ and angular coefficients $Y_{2,\;3sc,\; 4sc}$ and $\mathcal{P}_{3,\; 8,\; 9}$ which are influenced by these new couplings. These observables are also interesting from the experimental point of view as they have minimum dependence on the FFs and hence these are not significantly prone by the uncertainties. Therefore, these observables will provide an optimal ground to test the SM as well as to explore the possible NP. The values of these observables are plotted against $s$ in Figs. \ref{Fig1hh}, \ref{Fig2hh} and \ref{Fig2a} for the SM and in the presence of $C^{(\prime)}_{V,A}$ couplings. Quantitatively, by considering different $VA$ scenarios their values are compared with the SM  predictions which are collected in Tables \ref{lambda-obs-1}, \ref{lambda-obs-2} and \ref{lambda-obs-3}. The main effects of $C^{(\prime)}_{V,A}$ on these observables can be summarized as:
  \begin{itemize}
  	\item Fig.
  	\ref{Fig1hh}(e) depicts that the value of $A_{FB}^{\ell\Lambda}$ is an order of magnitude smaller than the experimentally measured $A^{\ell}_{FB}$ and
  	$A^{\Lambda}_{FB}$. After including the new $VA$ operators, we can see that its value decreases compared to its SM predictions in almost all the $s$ range. Just like $A^{\ell}_{FB}$ its zero position also shifts to the right which increases when $C_V$ becomes more negative. The value of this observable is changed throughout the $s$ region due to the change in the values of $VA$ couplings but it is insensitive to the mass of the final state $\mu$. From Table \ref{lambda-obs-2} one can find that in the low-recoil bin the numerical value of $A_{FB}^{\ell\Lambda}$ is almost free from the hadronic uncertainties.
  	\item In case of $F_{T}$ which is plotted in Fig. \ref{Fig1hh}(f), the impact of new $VA$ couplings along with the final state $\mu-$mass effects are visible only in low $s$ region. Like $F_L$, this observable has minimal dependence on the FFs especially in the bin $[15,20]$ GeV$^2$ and this can be seen clearly from the Table \ref{lambda-obs-2}.
  	As we know that $F_{L} + F_{T} \approx 1$ therefore, the behavior of $F_{T}$ and $F_L$ are expected to be opposite to each other in the presence of $VA$ couplings and it can be noticed from Fig. \ref{Fig1hh}.
  	\item The observables $\alpha_{\theta_{\ell}}$ and
  	$\alpha^{\prime}_{\theta_{\ell}}$ are plotted in Figs. \ref{Fig1hh}(g) and \ref{Fig1hh}(h), respectively. From these plots, one can see that the $\mu$-mass
  	effects are visible in low $s$ region for $\alpha_{\theta_{\ell}}$
  	but not for the $\alpha^{\prime}_{\theta_{\ell}}$. However, both
  	observables are sensitive to the $VA$ couplings and to extract
  	the imprints of NP both are significant to be measured precisely at
  	LHCb and Belle-II experiments. Furthermore, the behavior of $\alpha^{\prime}_{\theta_{\ell}}$ is similar to the $A^{\ell}_{FB}$ and it also passes
  	from the zero-position at a specific value of $s$ in the SM. Also this zero-position
  	is shifted towards the higher value of $s$ when $C_V$ is set to higher
  	negative value. In order to quantify the impact of new $VA$ couplings their numerical values along with the SM predictions are given in Table \ref{lambda-obs-2}.
  	\item For the observables $\alpha_{\theta_\Lambda},\; \alpha_{\xi},\; \alpha_L$ and $\alpha_U$ the maximum deviations from the SM predictions come only when we set
  	$C_V=-1.34$, $C_A^{\prime}=-0.4$ and $C_V^{\prime}=C_A=0$ as
  	shown by the green curves in Fig. \ref{Fig2hh}(a,b,c,d). However, in $Y_2$
  	this is the case for $C_V=-1.61$ and
  	$C_V^{\prime}=C_A=C_A^{\prime}=0$ which is drawn as an orange
  	curve in Fig. \ref{Fig2hh}(e). It can also be noticed from Table \ref{lambda-obs-2} that $Y_2$ has negligible uncertainties due to FFs. While the value of $Y_{4sc}$ is suppressed in the SM and even after adding the new $VA$ couplings still it is not in a reasonable range to be measured experimentally. The $\mu$-mass effect is also insignificant for all these observables at large-recoil.
  	\item The four-folded decay distribution defined in Eq. (\ref{eq6}) gives us a chance to single out the different physical observables
  	by studying different foldings. In semileptonic $B-$meson decays, such
  	foldings have been studied in detail, especially the penguin
  	asymmetries $(P_{i})$ where the $P_{5}^{(\prime)}$ is the most
  	important one. 
  	However, in the current study of the $\Lambda_b$ baryon, we consider only $\mathcal{P}_{3},\; \mathcal{P}_{8}$ and $\mathcal{P}_{9}$ which are the coefficients of
  	$\cos\theta_{\ell}\cos\theta_{\Lambda}$, $\cos\theta_{\ell}$ and $\cos\theta_{\Lambda}$, respectively. We can see from Eq. (\ref{observables1}) together with the expressions given in Appendix, that these observables heavily depend on the $VA$ couplings. We find that the values of $\mathcal{P}_{3}$ and $\mathcal{P}_{8}$ maximally change from their SM predictions when we set $C_V=-1.61$ and $C_V^{\prime}=C_A=C_A^{\prime}=0$ in almost all the $s$ region (orange curve in Fig. \ref{Fig2a}) and numerically it can be seen in Table \ref{lambda-obs-3}. Similar to the case of $A^{\ell}_{FB}$ and $\alpha^{\prime}_{\theta_{\ell}}$, the zero-positions of $\mathcal{P}_{3}$ and $\mathcal{P}_{8}$ move to the right from their SM zero-positions. But the effects of $VA$ couplings on $\mathcal{P}_9$ are prominent only for low values of $s$ and for this particular observable the $\mu-$mass term contribution is also quite visible in this region. In addition, from Table \ref{lambda-obs-3}, it is clear that the uncertainties in $P_3,\;P_8$ and $P_9$ are comparatively large in the large-recoil region.
  \end{itemize}
  
	
\subsection{Scalar and Pseudo-scalar Part}
In order to constraint the $SP$ couplings the golden channel is the $B_{s}\to \mu^{+}\mu^{-}$. In this decay we do not have any contributions from $C^{\prime}_{V}$ and the one
proportional to $C_{A}^{\prime}$ is helicity suppressed $(\mathcal{O}(m^2_{\ell}/m_{B}^2))$. Therefore; by
using the available experimental data of $B_{s}\to \mu^{+}\mu^{-}$
and $B \to X_{s}\mu^{+}\mu^{-}$ decay channels, the constraints on
$SP$ couplings $C_{S,P}^{(\prime)}= \big[-4.0, 4.0\big]$ are already obtained in \cite{Das:2018sms}. In the present study, we use these constraints to see the dependence of different physical observables on $SP$ couplings.
\begin{figure}
\begin{center}
\begin{tabular}{ll}
\includegraphics[height=4.8cm,width=8cm]{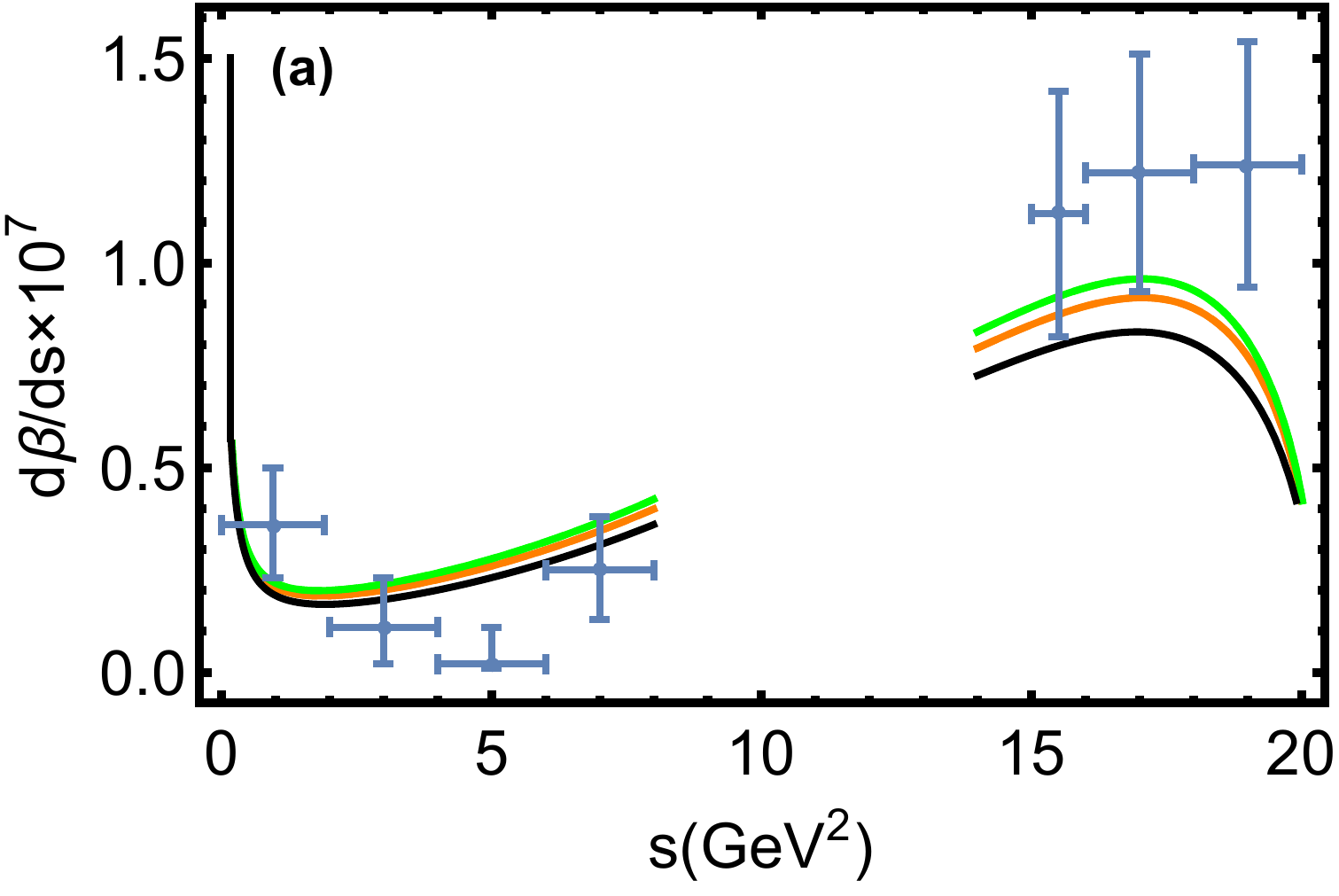}  & \ \ \includegraphics[height=4.8cm,width=8cm]{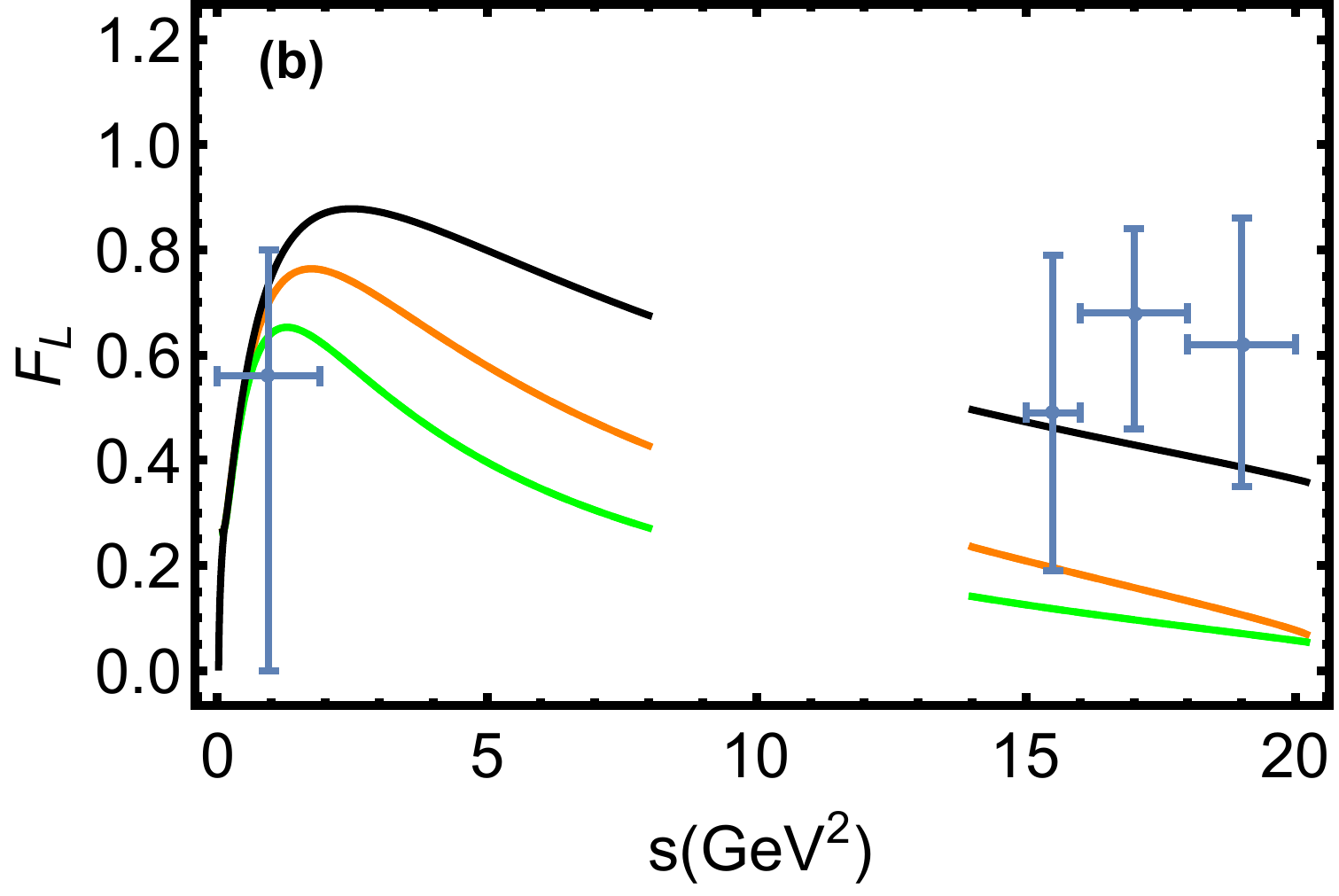} \\
\\
\includegraphics[height=4.8cm,width=8cm]{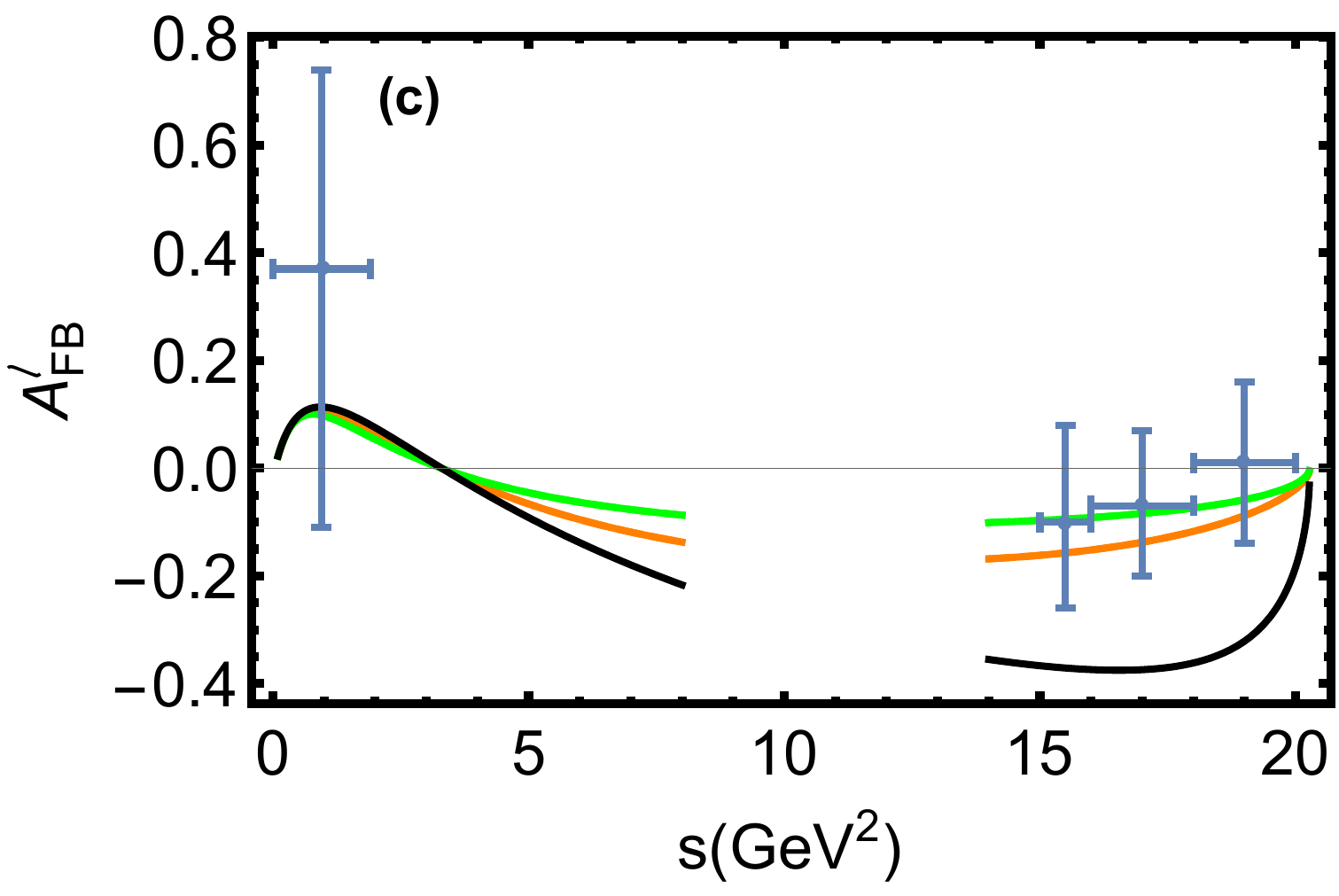}  & \ \ \includegraphics[height=4.8cm,width=8cm]{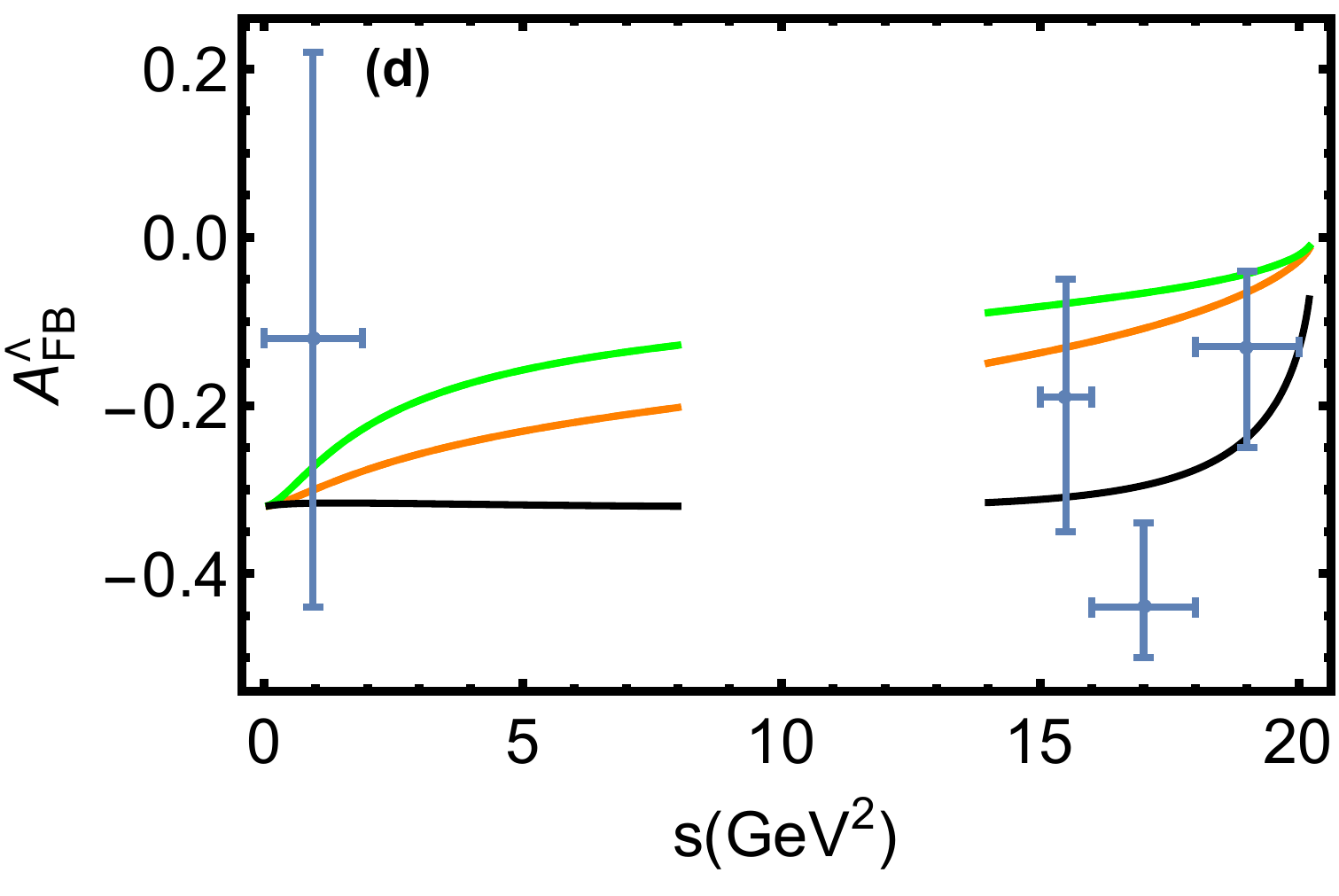} \\
\\
\includegraphics[height=4.8cm,width=8cm]{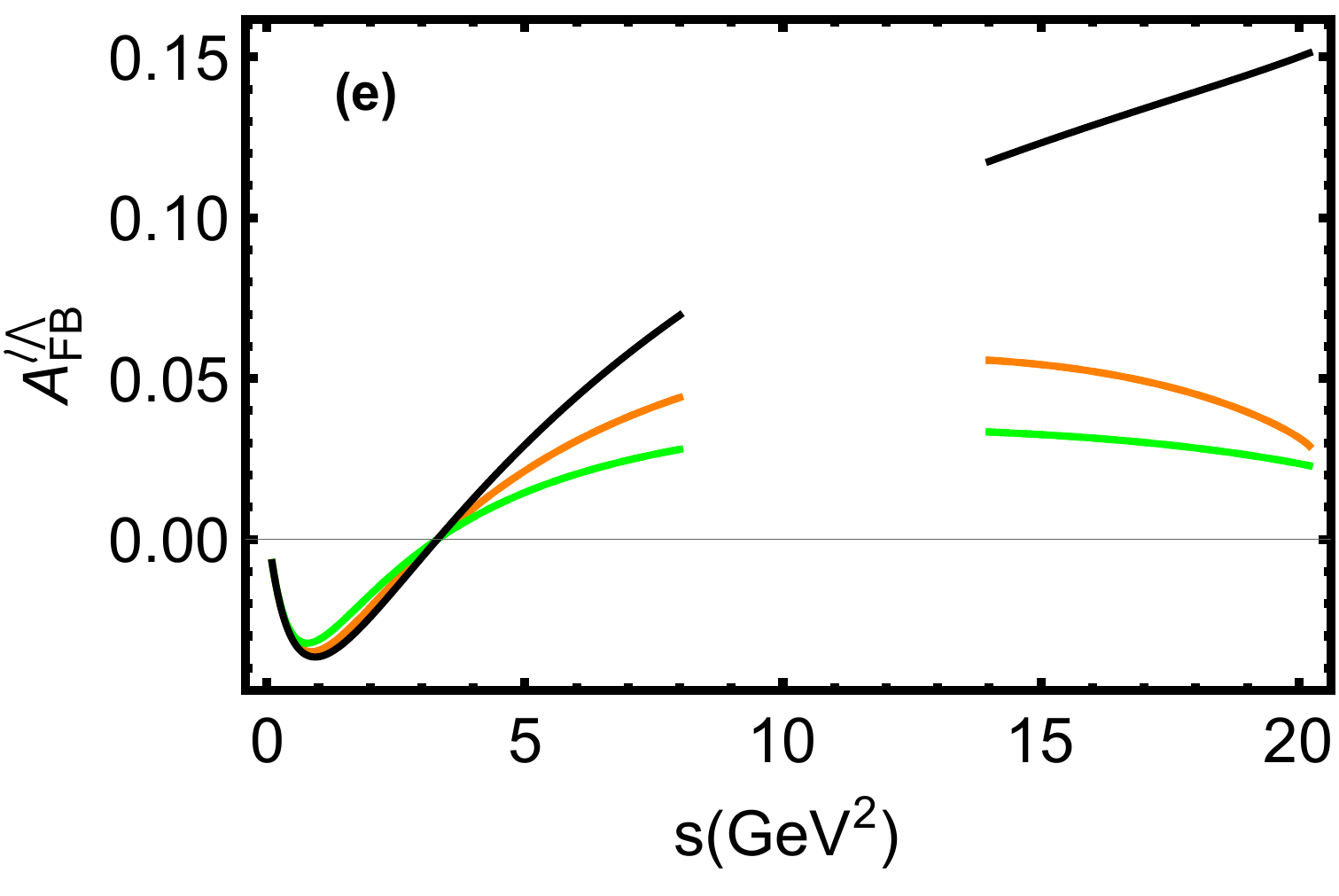} & \ \ \includegraphics[height=4.8cm,width=8cm]{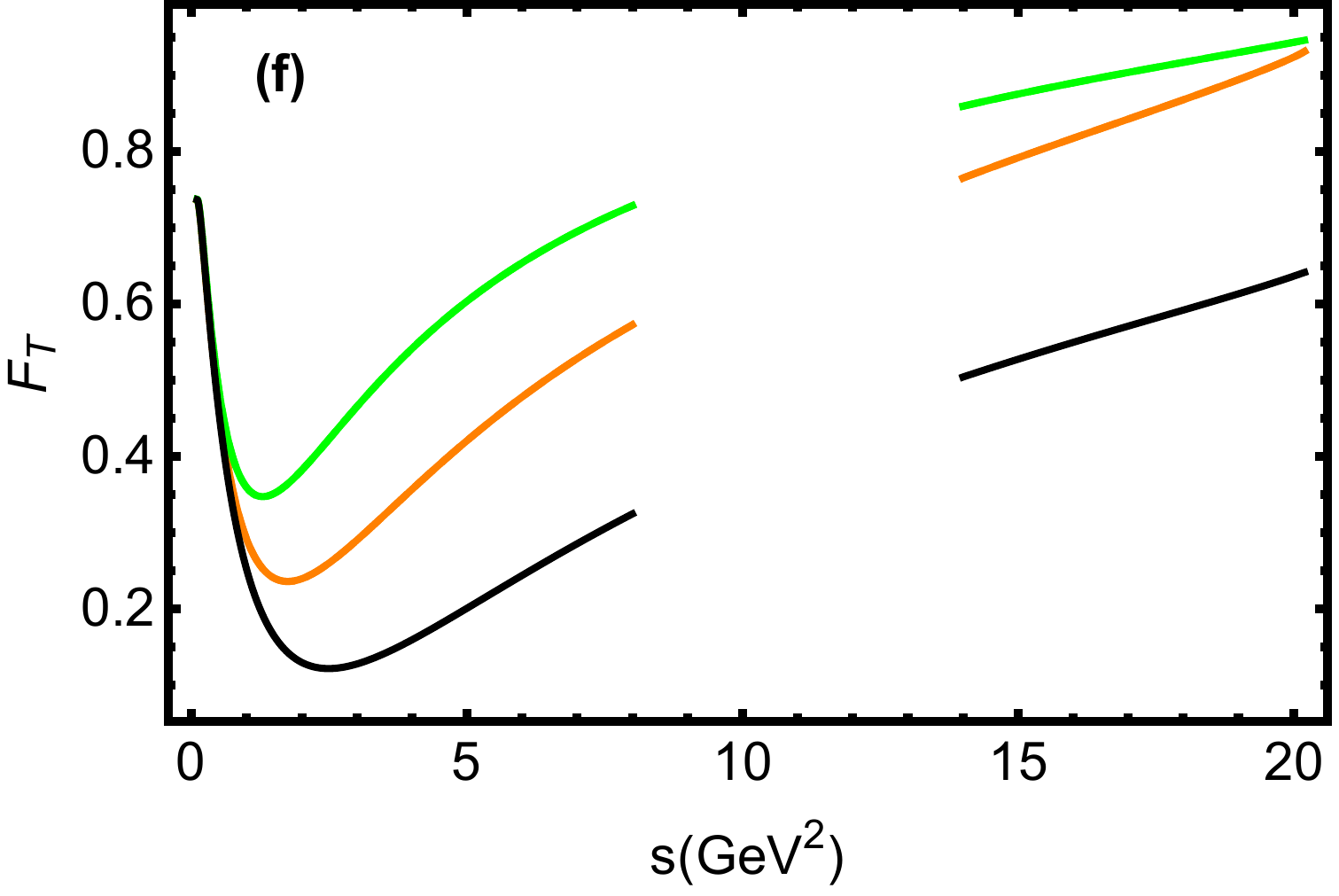} \\
\\
\includegraphics[height=4.8cm,width=8cm]{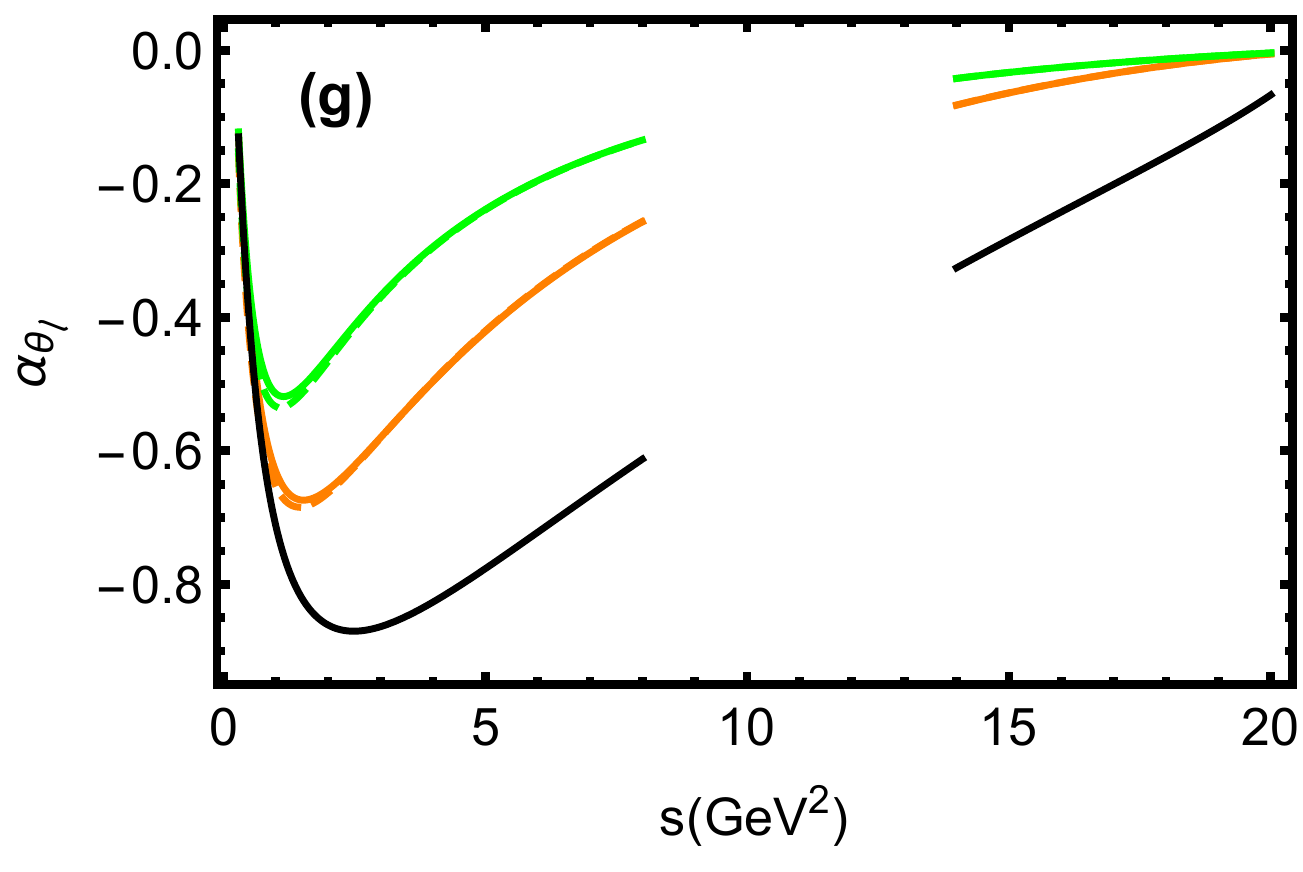} & \ \ \includegraphics[height=4.8cm,width=8cm]{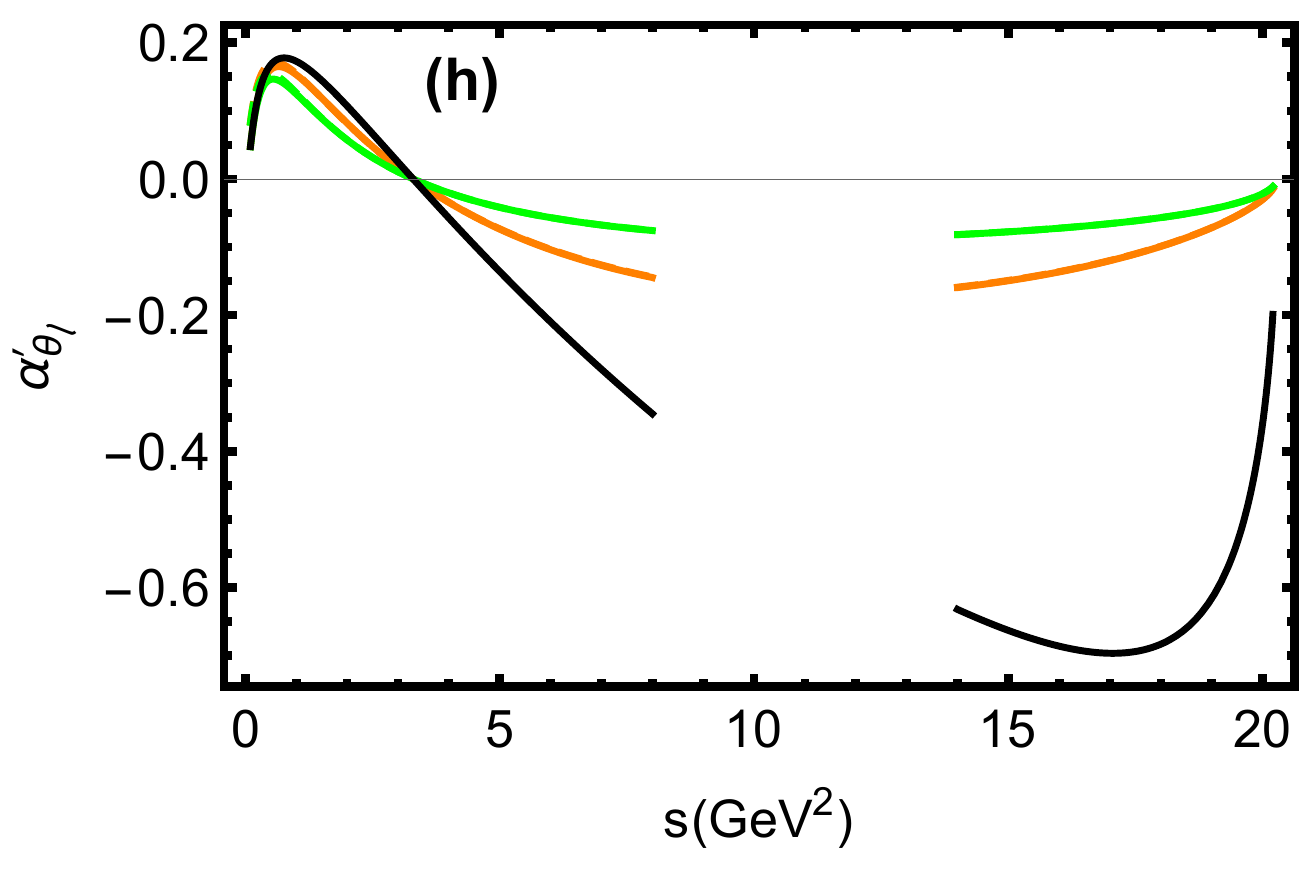} \\
\end{tabular}
\end{center}
\caption{Observables in the SM and in the presence of new SP couplings. The SM curves are denoted by black color. The orange color is obtained with $C_S=C_P=-3$ and $C_S^{\prime}=C_P^{\prime}=-3.1$ (for $d\mathcal{B}/ds$ orange color is for $C_S=C_P=-1$ and $C_S^{\prime}=C_P^{\prime}=-1.1$) and the green line is drawn when $C_S=3$ and $C_S^{\prime}=2.9$. The solid and dashed lines are for the massive and massless $\mu-$ cases respectively. }\label{Fig10}
\end{figure}
\begin{figure}
    \begin{center}
        \begin{tabular}{ll}
\includegraphics[height=4.8cm,width=8cm]{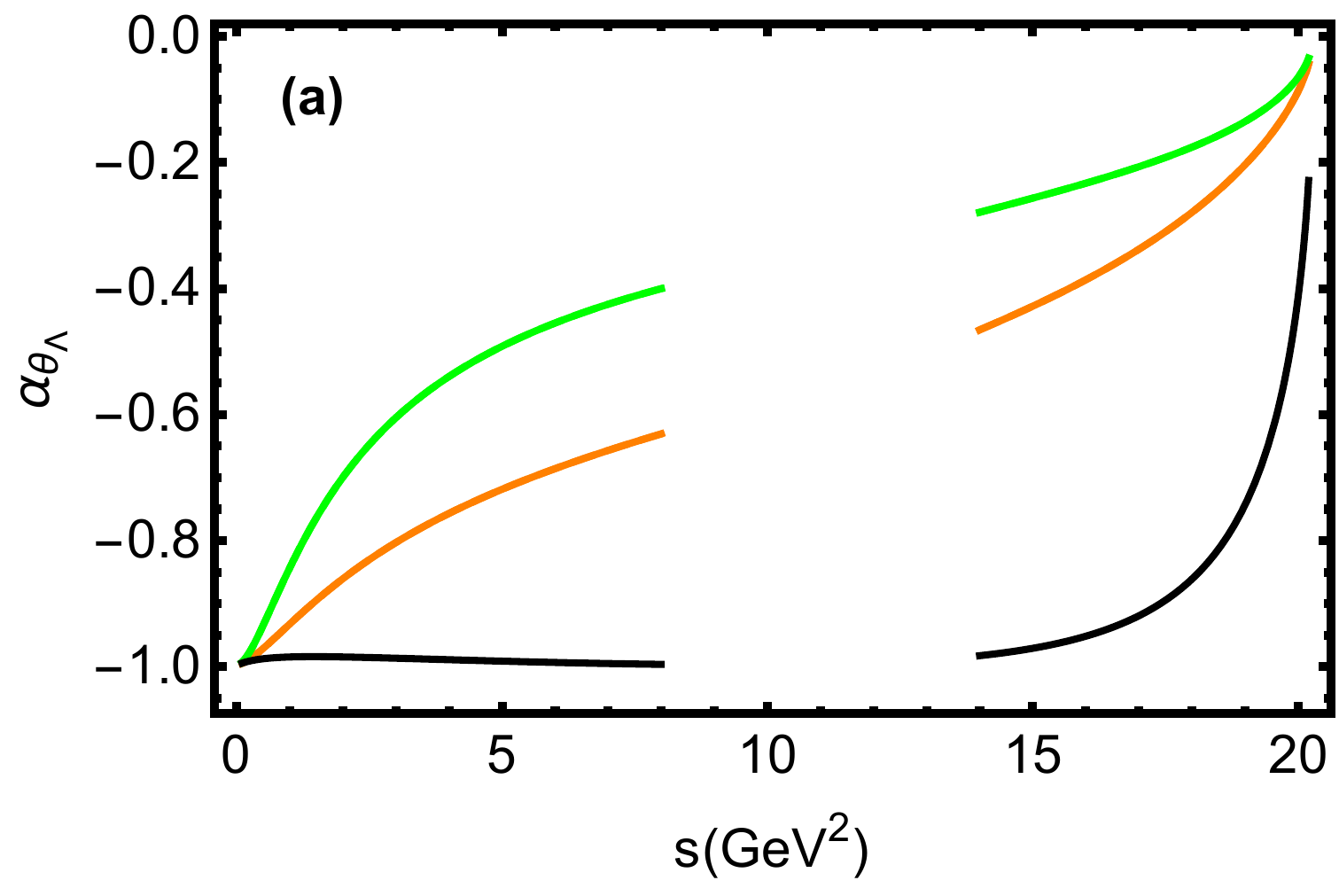}  & \ \
\includegraphics[height=4.8cm,width=8cm]{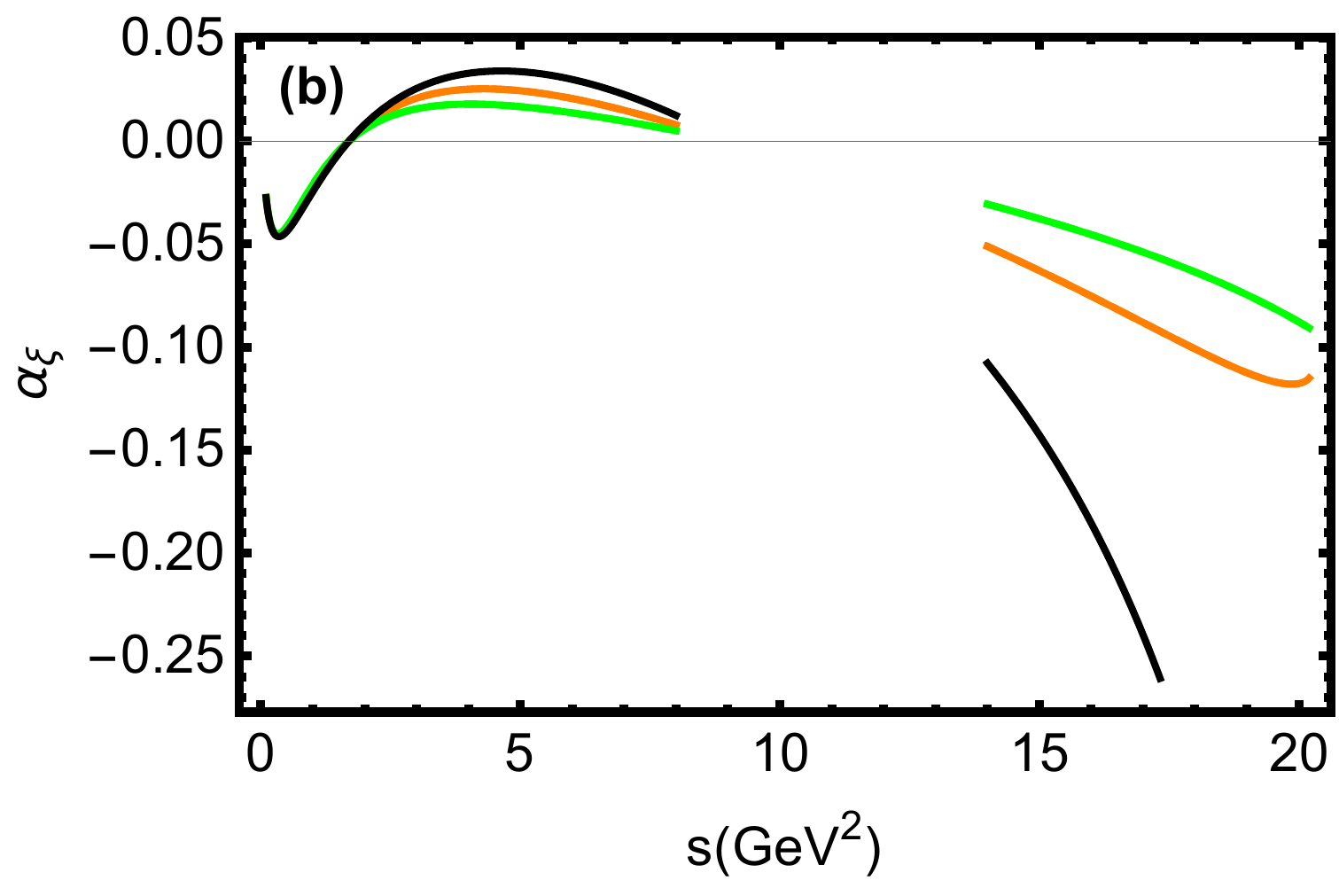} \\
           \includegraphics[height=4.8cm,width=8cm]{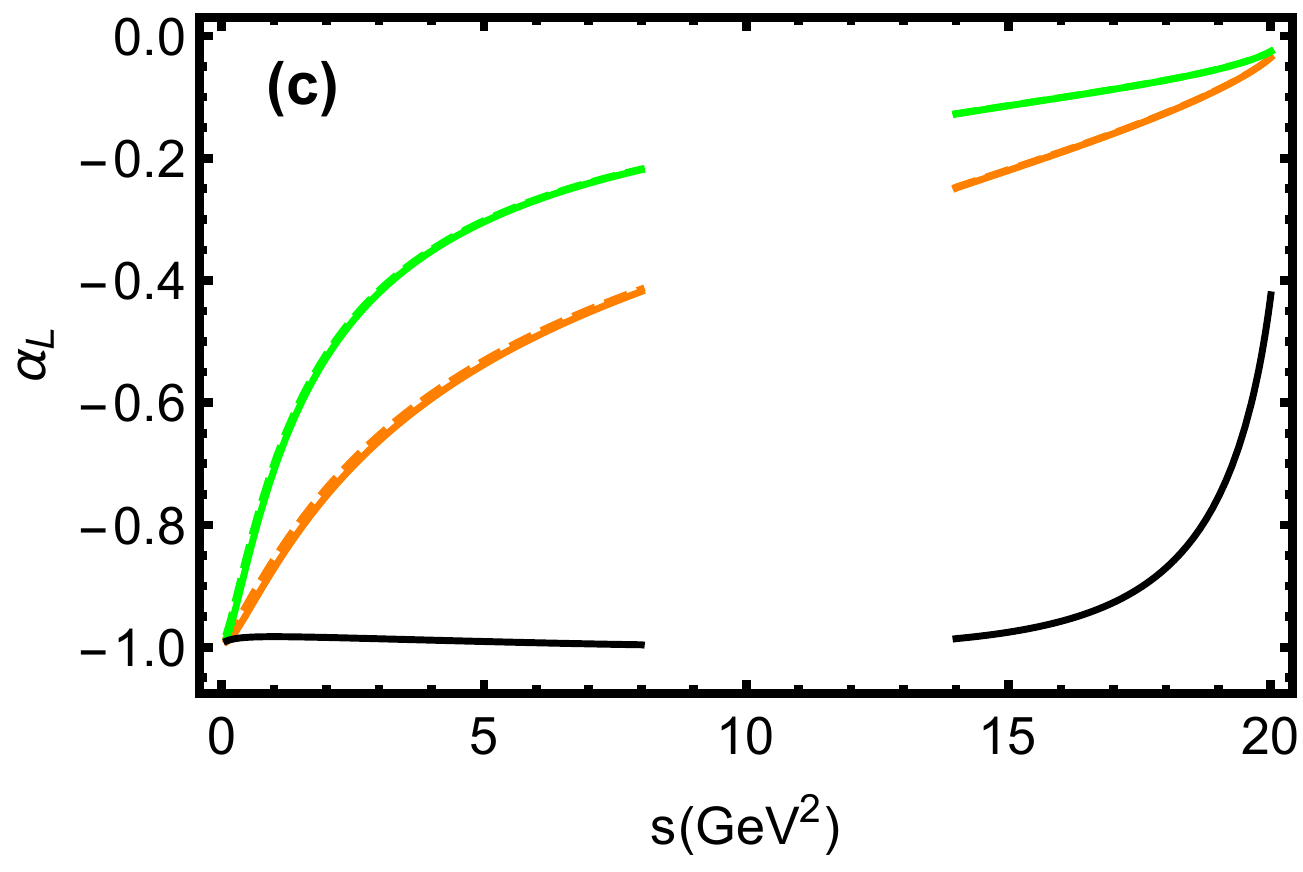} & \ \
            \includegraphics[height=4.8cm,width=8cm]{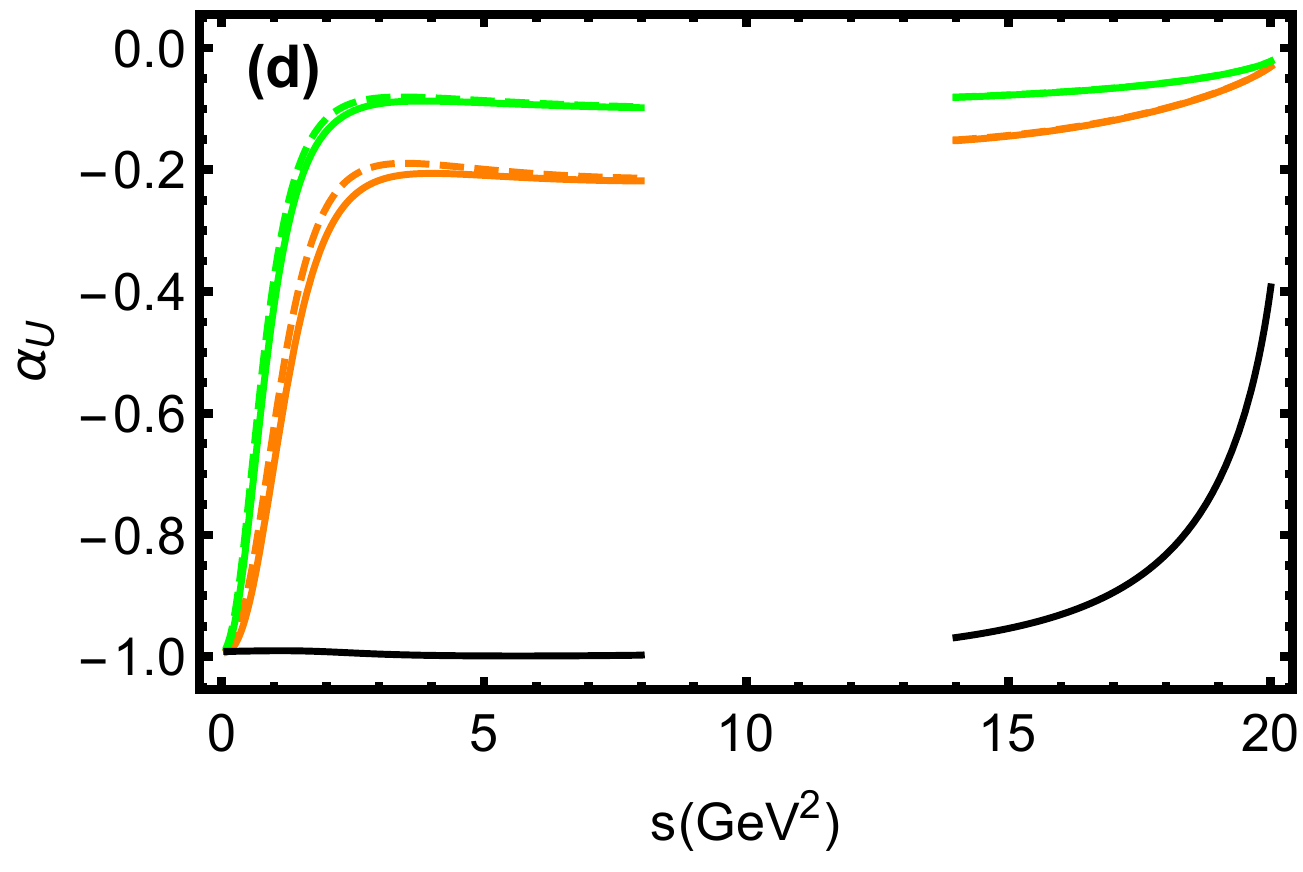}  \\
            \includegraphics[height=4.8cm,width=8cm]{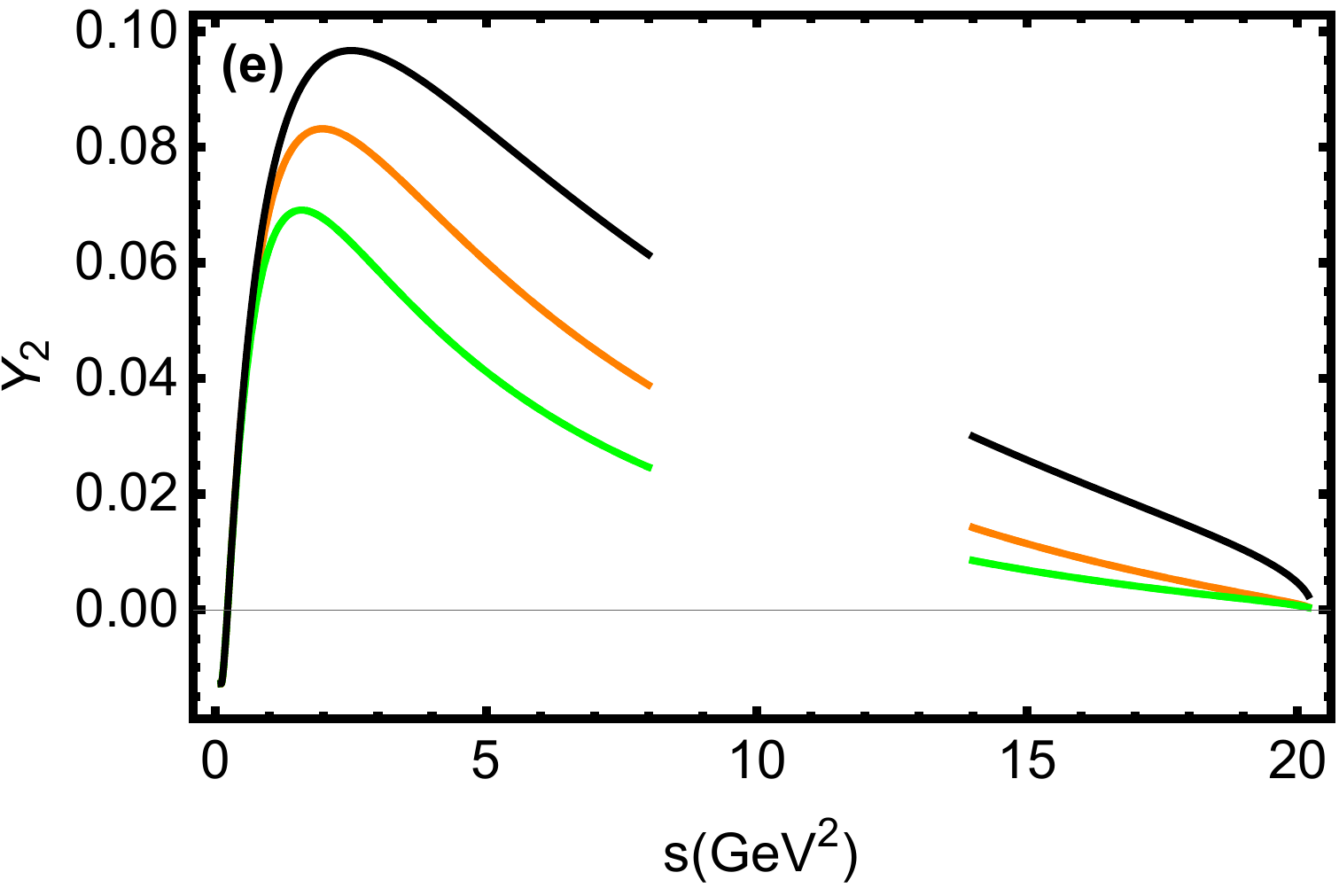} & \ \ \includegraphics[height=4.8cm,width=8cm]{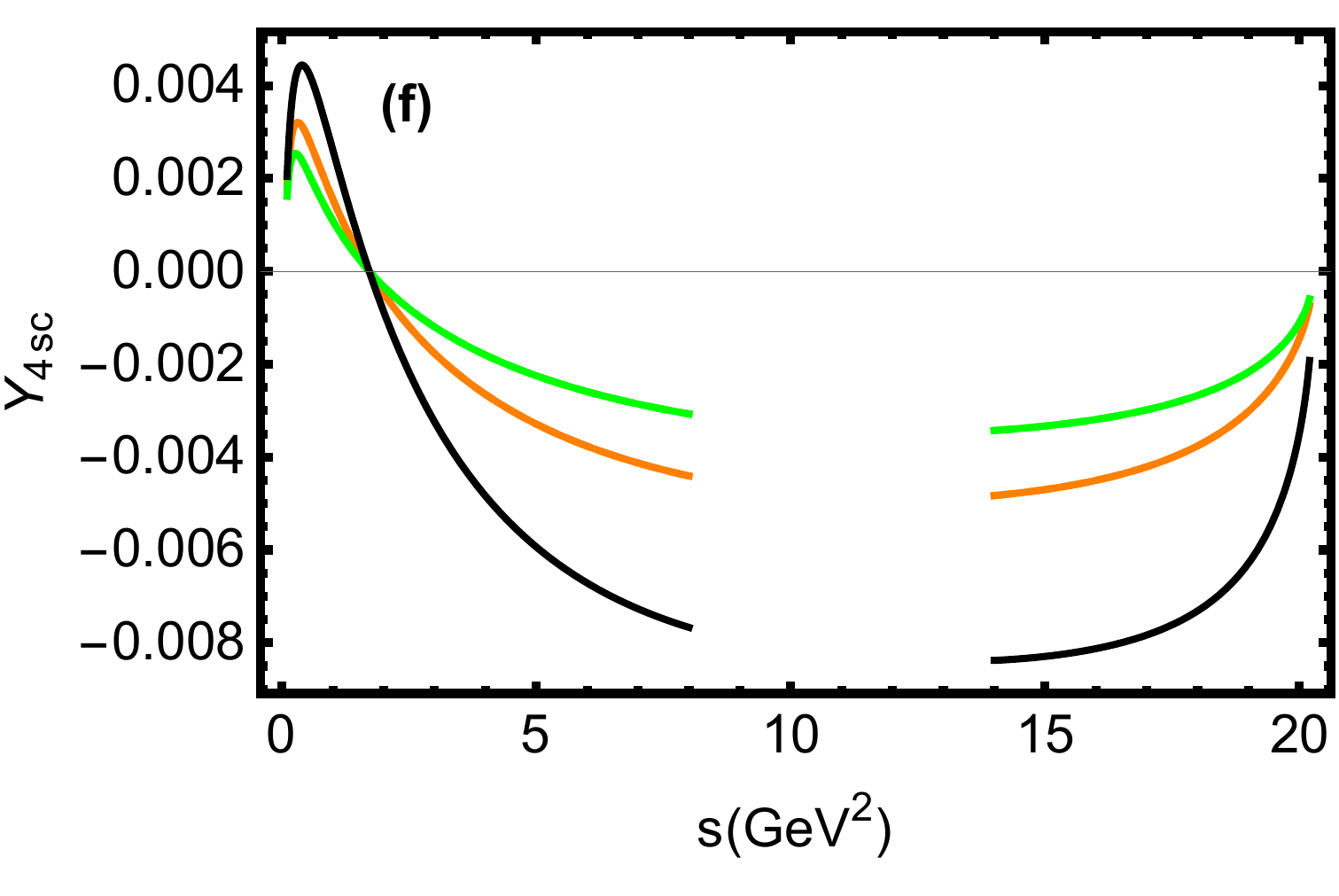}
             \end{tabular}
    \end{center}
    \caption{Observables in the SM along with SP couplings. The description of different curves is similar to the Fig. \ref{Fig10} }\label{Fig11}
\end{figure}
\begin{figure}
    \begin{center}
        \begin{tabular}{ll}
\includegraphics[height=4.8cm,width=8cm]{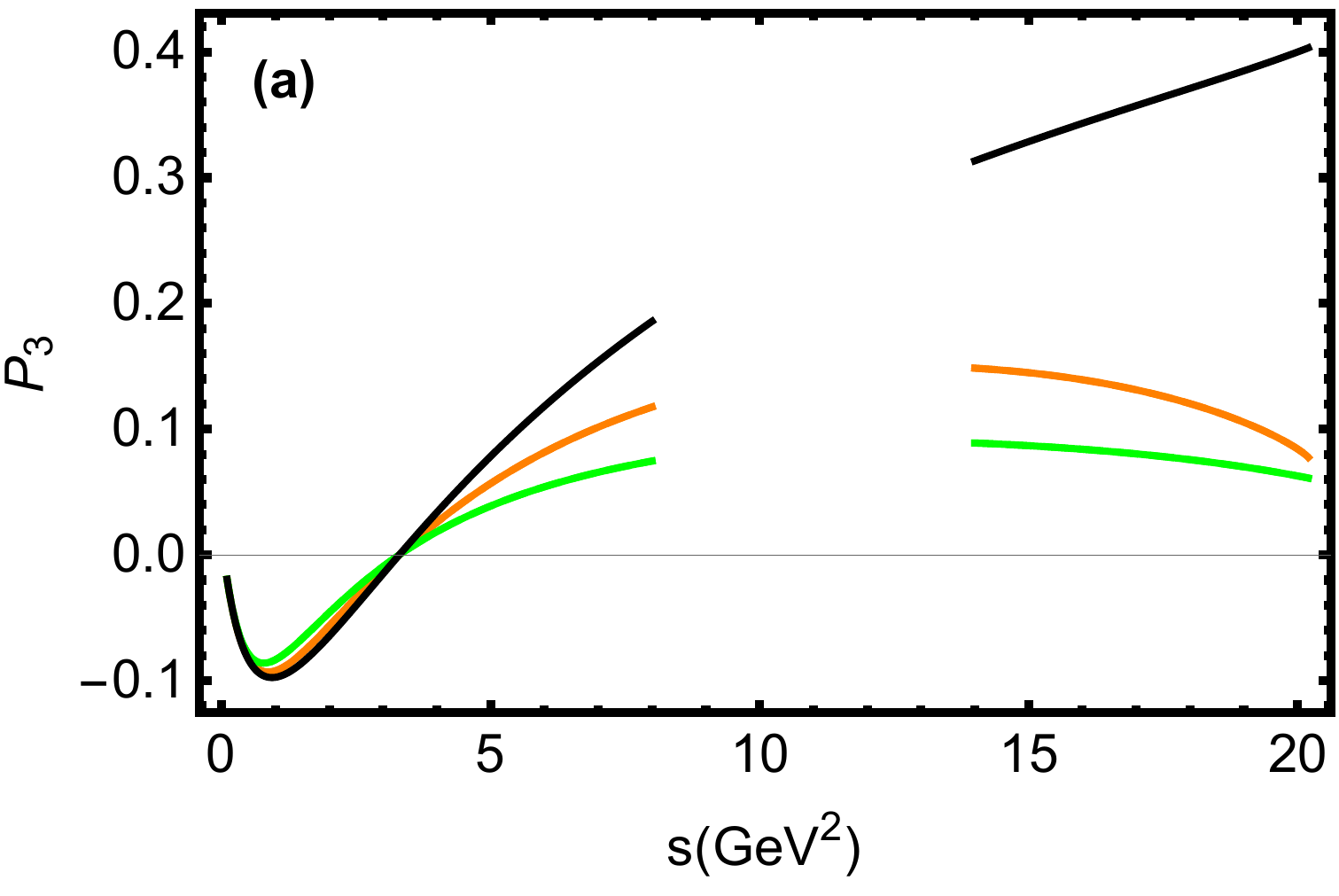} & \ \
            \includegraphics[height=4.8cm,width=8cm]{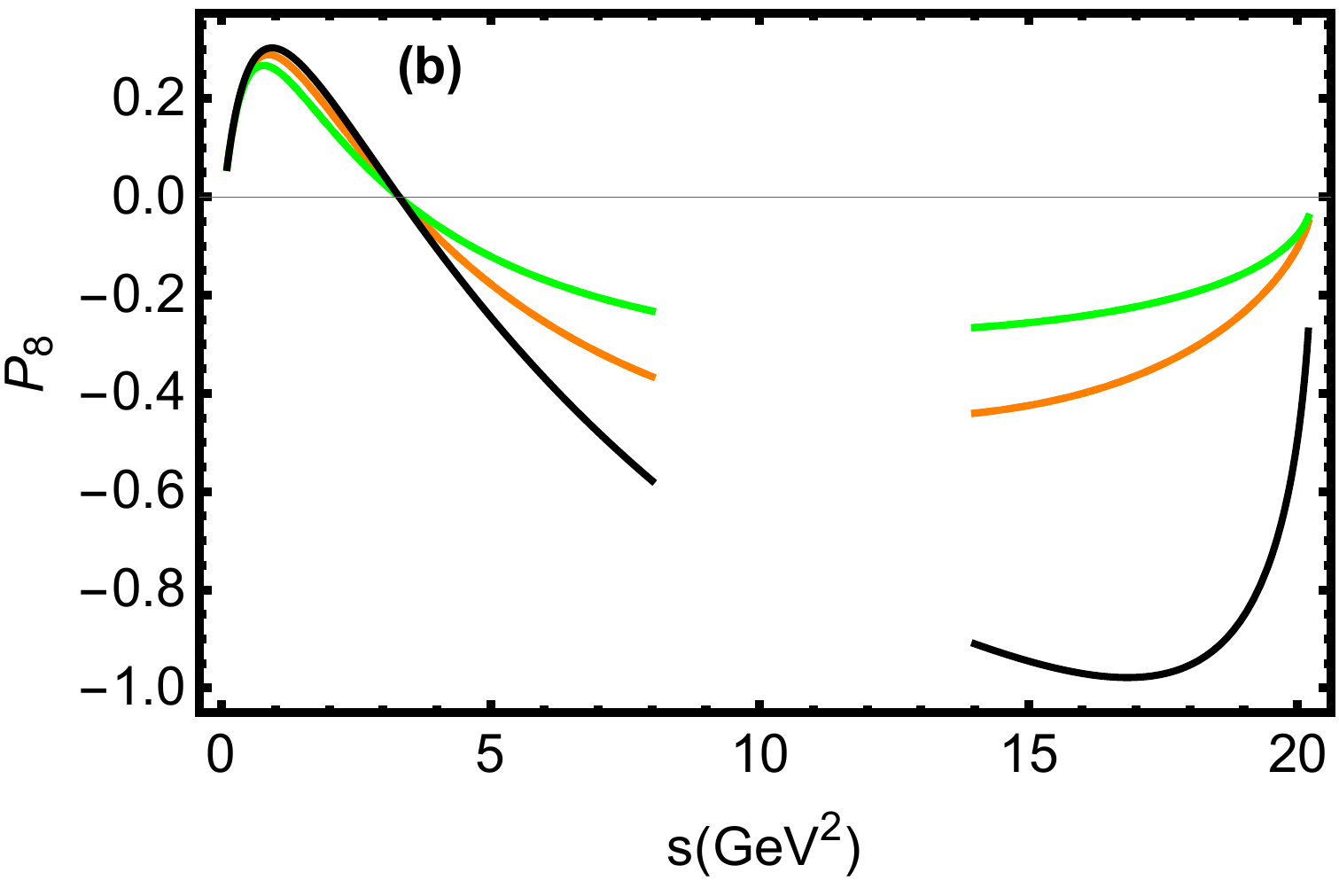} \\
            \\
            \includegraphics[height=4.8cm,width=8cm]{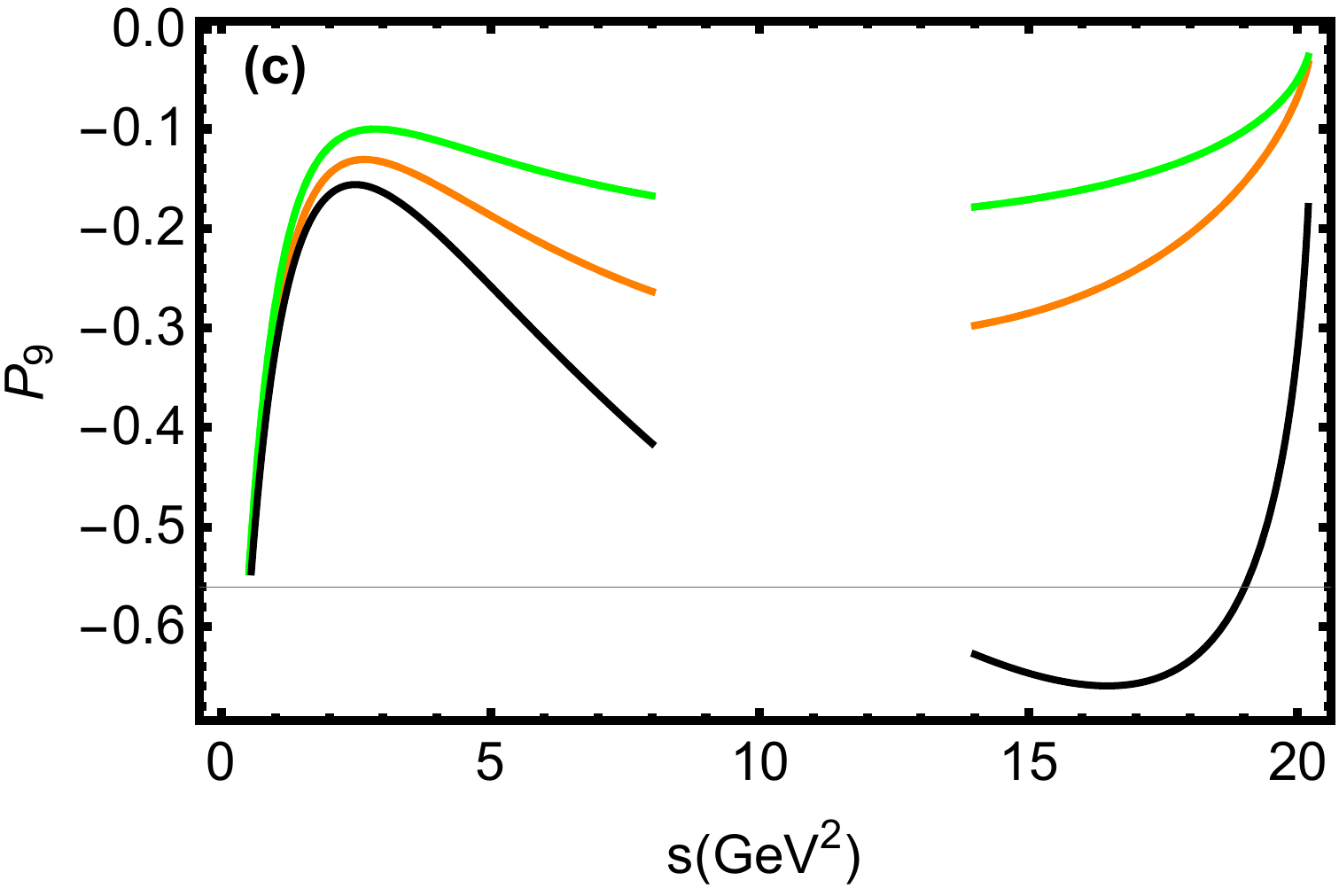} & \ \
             \\
            \end{tabular}
    \end{center}
    \caption{$\mathcal{P}_3$, $\mathcal{P}_8$ and $\mathcal{P}_9$ in the SM and in the presence of SP couplings. The description of different curves is similar to the Fig. \ref{Fig10}. }\label{Fig12}
\end{figure}

As the $SP$ couplings are absent in the SM therefore, in contrast to the new $VA$ couplings mentioned in previous section, a new angular coefficient arises which corresponds to $\cos{\theta_{\Lambda}}$. In addition to this coefficient, all the SM angular coefficients are modified except $K_{3sc}$ and $K_{4sc}$ and hence we expect that most of the physical observables show strong dependence on these $SP$ couplings. Hence, by taking $C_{S,P}^{(\prime)}=[-3.1,3]$ with the condition $|C_{S,P}-C_{S,P}^{\prime}|\lesssim 0.1$ \cite{Das:2018sms} due to having a large pull in global fits to $B-$Physics data the corresponding results are plotted in Figs. \ref{Fig10}, \ref{Fig11} and Fig. \ref{Fig12}.  We have considered two scenarios of $SP$ couplings and their results for all observables are presented in Tables \ref{lambda-obs-1}, \ref{lambda-obs-2} and \ref{lambda-obs-3} along with uncertainties. The important observations can be summarized as:
\begin{itemize}
	\item In the massless $\mu-$ limit, we can see that our results of the $d\mathcal{B}/ds$,\; $A^{\ell}_{FB}$ and $F_{L}$ for the $SP$ couplings are in agreement with the trend shown in \cite{Das:2018sms} and the values of these observables mainly change in the high $s$ region. It is clear from Fig. \ref{Fig10}(a) that the results of $d\mathcal{B}/ds$ are SM like in the low $s$ region but get closer to LHCb data for the bins $s \in [15,16]$ GeV$^2$ and $s \in [16,18]$ GeV$^2$ when $SP$ couplings are introduced. This can also be noticed from the numerical values of  $d\mathcal{B}/ds$ appended in the first column of Table \ref{lambda-obs-1}. From Fig. \ref{Fig10}(c), it can be seen that for $A_{FB}^{\ell}$ a good agreement to the data is achieved when we set $C_{S} = 3.0$ and $C_{S}^{\prime} = 2.9$ and it is displayed by the green curve. Just to mention, in contrast to the $SP$ couplings, the $VA$ couplings do not accommodate the data of $A_{FB}^{\ell}$ in high $s$ bins. However, the zero-position is not affected because the contributions from the $SP$ couplings do not contain any odd power term in $\cos\theta_{\ell}$. On the other hand, after inclusion of $SP$ couplings, $F_{L}$ agrees with the data only in $s \in [0.1,2]$ GeV$^2$ bin (c.f. Fig. \ref{Fig10}(b)) and for this particular observable the SM predictions show better trend with the data as can be read from Table \ref{lambda-obs-1}. However, the more data of these observable will reveal the future status of $SP$ couplings. For the $A_{FB}^{\Lambda}$, in contrast to the $VA$ coupling, this observable is sensitive to the $SP$ operators and it can be observed from Fig. \ref{Fig10}(d). It is important to emphasis that the changed values are still within the errors in the measurements except in one high $s$ bin; i.e., $[16,18]$ GeV$^2$. 
	
Similar to the $F_{L}$, in the high $s$ bins, the SM $A_{FB}^{\Lambda}$ curve has shown better agreement with the data than the one with $SP$ couplings. It is found that these observables are also insensitive to the mass of final state $\mu$. 
		\item Compared to the $VA$ couplings, the profile of $A_{FB}^{\ell \Lambda}$ is quite sensitive to the $SP$ couplings and it can be observed in Fig. \ref{Fig10}(e). Particularly, in the high $s$ region, we can notice from Table \ref{lambda-obs-2} that its value is approximately decreased by an order of magnitude from the corresponding SM predictions. However, similar to the $A^{\ell}_{FB}$, it zero-position is not changed because it is also proportional to the $VA$ and not to the $SP$ couplings. Also the massless or massive $\mu$ considerations do not lead to any visible change in this particular observable. 
		\item In the presence of $SP$ couplings, the behavior of $F_{T}$ is opposite to that of $F_{L}$ as it is expected due to $F_{L}+F_{T}=1$ for every value of $s$. This can be noticed in Fig. \ref{Fig10}(f) and from the second column of Tables \ref{lambda-obs-1} and \ref{lambda-obs-2}.
		\item In contrast to the $VA$ couplings, one can see from Fig. \ref{Fig10}(g) and the Fig. \ref{Fig11}(a, c, d) that $\alpha_{\theta_\ell}$, $\alpha_{\theta_\Lambda}$,  $\alpha_{L}$  and $\alpha_{U}$ are quite sensitive to the $SP$ couplings. These plots show that due to the $SP$ couplings, the values of these observables are significantly suppressed from that of the SM predictions in almost all the $s$ region. For $\alpha^\prime_{\theta_\ell}$, similar to the $A^{\ell}_{FB}$ and $A^{\ell\Lambda}_{FB}$, the zero-position depends only on $VA$ couplings and hence is not expected to be changed due to consideration of the $SP$ couplings and it can be seen in Fig. \ref{Fig10}(h). However, $\alpha^\prime_{\theta_\ell}$ is looking more sensitive to the $SP$ coupling as compared to the $VA$ couplings. Particularly, in $s\in[15,20]$ GeV$^2$ bin the value of $\alpha^\prime_{\theta_\ell}$ is almost $80\%$ suppressed from its SM predictions and it can also be read from Table \ref{lambda-obs-2}. In line with this, Fig. \ref{Fig11}(b) shows that in the high $s$ region, $\alpha_{\xi}$ is also quite sensitive to the $SP$ couplings. From Table \ref{lambda-obs-3}, one can see that for the $SP$ couplings the uncertainties in the value of  $\alpha_{\xi}$ are even larger than the actual value at the large-recoil. Just like other observables, these are also insensitive to the $\mu$ mass.
		\item For the angular observables $Y$'s only the value of $Y_{2}$ is significant to be measured at the LHCb and the future experiments, therefore, we have only plotted it in Fig. \ref{Fig11}(e). From Table \ref{lambda-obs-2}, one can observe that similar to the $\alpha$'s the value of $Y_{2}$ is significantly reduced from its numbers calculated using the SM operators. Here, we can also see that in-spite $Y_{4sc}$ is sensitive to $SP$ couplings at low-recoil its value is still too small to be measured experimentally.
		\item Similar to $\alpha$'s, $\mathcal{P}_{3,\; 8,\; 9}$ are also very sensitive to the $SP$ couplings as compared to that of the $VA$ couplings and it can be found from Fig. (\ref{Fig12}). We can see that the values of $\mathcal{P}$'s are changed from their SM predictions by a factor of $4 - 6$ (c.f. Table \ref{lambda-obs-3}). Again, the position of the zero-crossing in $\mathcal{P}_{3}$ and $\mathcal{P}_{8}$ are unchanged after inclusion of $SP$ couplings and the numerical results presented in Table \ref{lambda-obs-3} stay the same even if we take the non-zero mass for the final state $\mu$.
	\end{itemize}
		Thus, together with the $B-$meson decays, we hope that it will be interesting to look for the angular asymmetries of $\Lambda_b$ baryon decay at the LHCb which help us to get better constraints on the $SP$ couplings. Also, when experimental data of these angular observables will be available for $\Lambda_b$ baryon, we would be in better position to draw a conclusion about the future status of these couplings.
		\subsection{Tensor Part}
		Just like $SP$ couplings, the one corresponding to the tensor currents are also absent in the SM and hence they will also modify the SM angular coefficients  except $K_{3sc}$ and $K_{4sc}$. In \cite{Das:2018sms}, it has been discussed in detail that $B_s \rightarrow X_s \mu^+\mu^-$  along with $B \to X_c \ell \nu_{\ell}$ are the most important channels to find the constraints on these NP couplings and by using these channels the equation of constraints is obtained to be $C_T^2+C_{T_5}^2=0.55$ \cite{Das:2018sms}. As the constraints on these couplings are quite stringent, therefore, to see their impact on physical observables we vary the values $C_{T}$ 
	\FloatBarrier
	\begin{table}[h!]
		\caption{Observables for the decay $\Lambda_b \to \Lambda (\to p\pi)\mu^+ \mu^-$ in the SM and in different scenarios of NP couplings along with LHCb results in respective bins. Scenario $VA-1$ corresponds to $C_V=-1.61$, $C_V^{\prime}=C_A=C_A^{\prime}=0$, $VA-2$ is for $C_V=-C_A=-1$, $C_V^{\prime}=C_A^{\prime}=0$ and $VA-3$ represents the scenario when $C_V=-1.34$, $C_V^{\prime}=C_A=0$, $C_A^{\prime}=-0.4$. Similarly, in $SP-1$ case we have taken $C_S=C_P=-3$, $C_S^{\prime}=C_P^{\prime}=-3.1$ whereas $SP-2$ corresponds to $C_S=3$, $C_S^{\prime}=2.9$ and $C_P=C_P^{\prime}=0$. The tensor couplings correspond to $C_T=0.72$ and $C_{T5}=0.2$. The experimental results are taken from \cite{Aaij:2015xza}. }\label{lambda-obs-1}
		\begin{tabular}{ll||llll}
			\hline\hline
			&  & $\ \left\langle \frac{d\beta }{ds}\times 10^{-7}\right\rangle $ & $\ \
			\ \ \ \ \left\langle F_{L}\right\rangle $ & $\ \ \ \ \ \ \left\langle
			A_{FB}^{\ell}\right\rangle $ & $ \ \ \ \ \left\langle A_{FB}^{\Lambda }\right\rangle $
			\\ \hline\hline
			$\lbrack 0.1,2]$ & 
			\begin{tabular}{l}
				$SM$ \\ 
				$VA-1$ \\ 
				$VA-2$ \\ 
				$VA-3$ \\ 
				$SP-1$ \\ 
				$SP-2$ \\ 
				$T^{\prime }$ \\ 
				$LHCb$%
			\end{tabular}
			& 
			\begin{tabular}{l}
				$0.238_{-0.131}^{+0.185}$ \\ 
				$0.214_{-0.116}^{+0.163}$ \\ 
				$0.187_{-0.100}^{+0.136}$ \\ 
				$0.222_{-0.120}^{+0.168}$ \\ 
				$0.280_{-0.167}^{+0.255}$ \\ 
				$0.268_{-0.157}^{+0.236}$ \\ 
				$0.434_{-0.232}^{+0.318}$ \\ 
				$0.36_{-0.13}^{+0.14}$ \\ 
			\end{tabular}
			& 
			\begin{tabular}{l}
				$0.535_{-0.051}^{+0.025}$ \\ 
				$0.453_{-0.048}^{+0.024}$ \\ 
				$0.398_{-0.046}^{+0.025}$ \\ 
				$0.459_{-0.042}^{+0.022}$ \\ 
				$0.335_{-0.045}^{+0.084}$ \\ 
				$0.241_{-0.044}^{+0.116}$ \\ 
				$0.293_{-0.039}^{+0.020}$ \\ 
				$0.56_{-0.57}^{+0.24}$ \\ 
			\end{tabular}
			& 
			\begin{tabular}{l}
				$ \ \  0.097_{-0.003}^{+0.006}$ \\ 
				$ \ \  0.122_{-0.010}^{+0.003}$ \\ 
				$ \ \  0.101_{-0.009}^{+0.003}$ \\ 
				$ \ \  0.117_{-0.007}^{+0.002}$ \\ 
				$ \ \  0.060_{-0.010}^{+0.020}$ \\ 
				$ \ \  0.043_{-0.009}^{+0.025}$ \\ 
				$ \ \  0.053_{-0.003}^{+0.002}$ \\ 
				$ \ \  0.37_{-0.48}^{+0.37}$ \\ 
			\end{tabular}
			& 
			\begin{tabular}{l}
				$-0.310_{-0.004}^{+0.011}$ \\ 
				$-0.312_{-0.003}^{+0.008}$ \\ 
				$-0.312_{-0.003}^{+0.008}$ \\ 
				$-0.303_{-0.004}^{+0.010}$ \\ 
				$-0.194_{-0.064}^{+0.031}$ \\ 
				$-0.139_{-0.081}^{+0.029}$ \\ 
				$-0.167_{-0.002}^{+0.016}$ \\ 
				$-0.12_{-0.32}^{+0.34}$ \\ 
			\end{tabular}
			\\ \hline
			$\lbrack 15,16]$ & 
			\begin{tabular}{l}
				$SM$ \\ 
				$VA-1$ \\ 
				$VA-2$ \\ 
				$VA-3$ \\ 
				$SP-1$ \\ 
				$SP-2$ \\ 
				$T^{\prime }$ \\ 
				$LHCb$%
			\end{tabular}
			& 
			\begin{tabular}{l}
				$0.797_{-0.155}^{+0.172}$ \\ 
				$0.558_{-0.108}^{+0.120}$ \\ 
				$0.439_{-0.086}^{+0.087}$ \\ 
				$0.568_{-0.110}^{+0.168}$ \\ 
				$0.960_{-0.208}^{+0.236}$ \\ 
				$0.917_{-0.195}^{+0.220}$ \\ 
				$0.837_{-0.167}^{+0.172}$ \\ 
				$1.12_{-0.30}^{+0.30}$ \\ 
			\end{tabular}
			& 
			\begin{tabular}{l}
				$0.454_{-0.004}^{+0.005}$ \\ 
				$0.450_{-0.004}^{+0.004}$ \\ 
				$0.460_{-0.004}^{+0.005}$ \\ 
				$0.458_{-0.004}^{+0.004}$ \\ 
				$0.261_{-0.019}^{+0.025}$ \\ 
				$0.187_{-0.017}^{+0.023}$ \\ 
				$0.432_{-0.000}^{+0.007}$ \\ 
				$0.49_{-0.30}^{+0.30}$ \\ 
			\end{tabular}
			& 
			\begin{tabular}{l}
				$-0.382_{-0.001}^{+0.002}$ \\ 
				$-0.301_{-0.002}^{+0.002}$ \\ 
				$-0.379_{-0.002}^{+0.002}$ \\ 
				$-0.337_{-0.002}^{+0.002}$ \\ 
				$-0.220_{-0.018}^{+0.014}$ \\ 
				$-0.157_{-0.017}^{+0.012}$ \\ 
				$-0.362_{-0.002}^{+0.004}$ \\ 
				$-0.10_{-0.16}^{+0.18}$ \\ 
			\end{tabular}
			& 
			\begin{tabular}{l}
				$-0.307_{-0.001}^{+0.001}$ \\ 
				$-0.307_{-0.001}^{+0.001}$ \\ 
				$-0.307_{-0.001}^{+0.001}$ \\ 
				$-0.316_{-0.001}^{+0.001}$ \\ 
				$-0.177_{-0.016}^{+0.012}$ \\ 
				$-0.126_{-0.014}^{+0.011}$ \\ 
				$-0.292_{-0.001}^{+0.001}$ \\ 
				$-0.19_{-0.16}^{+0.14}$ \\ 
			\end{tabular}
			\\ \hline
			$\lbrack 16,18]$ & 
			\begin{tabular}{l}
				$SM$ \\ 
				$VA-1$ \\ 
				$VA-2$ \\ 
				$VA-3$ \\ 
				$SP-1$ \\ 
				$SP-2$ \\ 
				$T^{\prime }$ \\ 
				$LHCb$%
			\end{tabular}
			& 
			\begin{tabular}{l}
				$0.824_{-0.130}^{+0.141}$ \\ 
				$0.576_{-0.091}^{+0.098}$ \\ 
				$0.454_{-0.072}^{+0.060}$ \\ 
				$0.574_{-0.090}^{+0.098}$ \\ 
				$0.998_{-0.177}^{+0.196}$ \\ 
				$0.953_{-0.165}^{+0.182}$ \\ 
				$0.862_{-0.148}^{+0.130}$ \\ 
				$1.22_{-0.29}^{+0.29}$ \\ 
			\end{tabular}
			& 
			\begin{tabular}{l}
				$0.418_{-0.002}^{+0.003}$ \\ 
				$0.415_{-0.002}^{+0.002}$ \\ 
				$0.422_{-0.002}^{+0.003}$ \\ 
				$0.422_{-0.002}^{+0.002}$ \\ 
				$0.234_{-0.014}^{+0.018}$ \\ 
				$0.168_{-0.012}^{+0.016}$ \\ 
				$0.406_{-0.006}^{+0.005}$ \\ 
				$0.68_{-0.22}^{+0.16}$ \\ 
			\end{tabular}
			& 
			\begin{tabular}{l}
				$-0.381_{-0.001}^{+0.001}$ \\ 
				$-0.306_{-0.001}^{+0.001}$ \\ 
				$-0.381_{-0.001}^{+0.001}$ \\ 
				$-0.349_{-0.001}^{+0.001}$ \\ 
				$-0.213_{-0.014}^{+0.012}$ \\ 
				$-0.153_{-0.013}^{+0.010}$ \\ 
				$-0.361_{-0.004}^{+0.002}$ \\ 
				$-0.07_{-0.13}^{+0.14}$ \\ 
			\end{tabular}
			& 
			\begin{tabular}{l}
				$-0.289_{-0.001}^{+0.001}$ \\ 
				$-0.289_{-0.001}^{+0.001}$ \\ 
				$-0.289_{-0.001}^{+0.001}$ \\ 
				$-0.304_{-0.001}^{+0.000}$ \\ 
				$-0.162_{-0.012}^{+0.009}$ \\ 
				$-0.116_{-0.011}^{+0.008}$ \\ 
				$-0.276_{-0.001}^{+0.001}$ \\ 
				$-0.44_{-0.06}^{+0.10}$ \\ 
			\end{tabular}
			\\ \hline$\lbrack 18,20]$ & 
			\begin{tabular}{l}
				$SM$ \\ 
				$VA-1$ \\ 
				$VA-2$ \\ 
				$VA-3$ \\ 
				$SP-1$ \\ 
				$SP-2$ \\ 
				$T^{\prime }$ \\ 
				$LHCb$%
			\end{tabular}
			& 
			\begin{tabular}{l}
				$0.658_{-0.073}^{+0.078}$ \\ 
				$0.459_{-0.051}^{+0.054}$ \\ 
				$0.354_{-0.031}^{+0.009}$ \\ 
				$0.438_{-0.049}^{+0.052}$ \\ 
				$0.808_{-0.103}^{+0.110}$ \\ 
				$0.771_{-0.096}^{+0.102}$ \\ 
				$0.684_{-0.118}^{+0.029}$ \\ 
				$1.24_{-0.30}^{+0.30}$ \\ 
			\end{tabular}
			& 
			\begin{tabular}{l}
				$0.371_{-0.001}^{+0.001}$ \\ 
				$0.370_{-0.001}^{+0.001}$ \\ 
				$0.373_{-0.001}^{+0.001}$ \\ 
				$0.375_{-0.001}^{+0.001}$ \\ 
				$0.192_{-0.009}^{+0.010}$ \\ 
				$0.141_{-0.008}^{+0.009}$ \\ 
				$0.385_{-0.028}^{+0.003}$ \\ 
				$0.62_{-0.27}^{+0.24}$ \\ 
			\end{tabular}
			& 
			\begin{tabular}{l}
				$-0.317_{-0.001}^{+0.001}$ \\ 
				$-0.260_{-0.001}^{+0.001}$ \\ 
				$-0.318_{-0.001}^{+0.001}$ \\ 
				$-0.307_{-0.001}^{+0.001}$ \\ 
				$-0.164_{-0.008}^{+0.007}$ \\ 
				$-0.120_{-0.007}^{+0.006}$ \\ 
				$-0.286_{-0.019}^{+0.001}$ \\ 
				$ \ \ 0.01_{-0.15}^{+0.16}$ \\ 
			\end{tabular}
			& 
			\begin{tabular}{l}
				$-0.227_{-0.000}^{+0.000}$ \\ 
				$-0.227_{-0.000}^{+0.000}$ \\ 
				$-0.226_{-0.000}^{+0.000}$ \\ 
				$-0.248_{-0.000}^{+0.000}$ \\ 
				$-0.117_{-0.006}^{+0.005}$ \\ 
				$-0.086_{-0.005}^{+0.004}$ \\ 
				$-0.215_{-0.001}^{+0.000}$ \\ 
				$-0.13_{-0.12}^{+0.10}$ \\ 
			\end{tabular} 
			\\ \hline
			$\lbrack 15,20]$ & 
			\begin{tabular}{l}
				$SM$ \\ 
				$VA-1$ \\ 
				$VA-2$ \\ 
				$VA-3$ \\ 
				$SP-1$ \\ 
				$SP-2$ \\ 
				$T^{\prime }$ \\ 
				$LHCb$%
			\end{tabular}
			& 
			\begin{tabular}{l}
				$0.752_{-0.112}^{+0.122}$ \\ 
				$0.525_{-0.078}^{+0.085}$ \\ 
				$0.415_{-0.062}^{+0.038}$ \\ 
				$0.518_{-0.078}^{+0.084}$ \\ 
				$0.914_{-0.153}^{+0.169}$ \\ 
				$0.873_{-0.143}^{+0.158}$ \\ 
				$0.786_{-0.140}^{+0.098}$ \\ 
				$1.20_{-0.27}^{+0.27}$ \\ 
			\end{tabular}
			& 
			\begin{tabular}{l}
				$0.409_{-0.001}^{+0.001}$ \\ 
				$0.407_{-0.001}^{+0.001}$ \\ 
				$0.412_{-0.001}^{+0.001}$ \\ 
				$0.414_{-0.001}^{+0.001}$ \\ 
				$0.224_{-0.012}^{+0.015}$ \\ 
				$0.162_{-0.011}^{+0.014}$ \\ 
				$0.404_{-0.016}^{+0.004}$ \\ 
				$0.61_{-0.14}^{+0.11}$ \\ 
			\end{tabular}
			& 
			\begin{tabular}{l}
				$-0.359_{-0.002}^{+0.002}$ \\ 
				$-0.289_{-0.002}^{+0.002}$ \\ 
				$-0.359_{-0.002}^{+0.002}$ \\ 
				$-0.332_{-0.002}^{+0.002}$ \\ 
				$-0.196_{-0.011}^{+0.009}$ \\ 
				$-0.142_{-0.011}^{+0.009}$ \\ 
				$-0.337_{-0.003}^{+0.004}$ \\ 
				$-0.05_{-0.09}^{+0.09}$ \\ 
			\end{tabular}
			& 
			\begin{tabular}{l}
				$-0.271_{-0.001}^{+0.001}$ \\ 
				$-0.271_{-0.001}^{+0.001}$ \\ 
				$-0.271_{-0.001}^{+0.001}$ \\ 
				$-0.287_{-0.001}^{+0.001}$ \\ 
				$-0.149_{-0.009}^{+0.007}$ \\ 
				$-0.107_{-0.008}^{+0.007}$ \\ 
				$-0.257_{-0.001}^{+0.002}$ \\ 
				$-0.29_{-0.08}^{+0.08}$ \\ 
			\end{tabular}%
		\end{tabular}
	\end{table}
	\FloatBarrier
	\FloatBarrier
	\begin{table}[t!]
		\caption{Observables in the SM and in the presence of new $VA$ and $SP$ couplings for which experimental results are not available. The description of different scenarios is same as in Table \ref{lambda-obs-1}.} \label{lambda-obs-2}
		\begin{eqnarray*}
			&&%
			\begin{tabular}{ll||llll}
				\hline\hline
				&  & $\ \left\langle A_{FB}^{\ell\Lambda }\right\rangle $ & $\ \ \ \ \ \
				\left\langle F_{T}\right\rangle $ & $\left\langle Y_{3sc}\times
				10^{-3}\right\rangle $ & \ \ \ $\left\langle Y_{4sc}\right\rangle $ \\ \hline\hline
				$\lbrack 1.1,6]$ & 
				\begin{tabular}{l}
					$SM$ \\ 
					$VA-1$ \\ 
					$VA-2$ \\ 
					$VA-3$ \\ 
					$SP-1$ \\ 
					$SP-2$%
				\end{tabular}
				& 
				\begin{tabular}{l}
					$-0.002_{-0.004}^{+0.003}$ \\ 
					$-0.034_{-0.003}^{+0.002}$ \\ 
					$-0.022_{-0.004}^{+0.003}$ \\ 
					$-0.026_{-0.003}^{+0.002}$ \\ 
					$-0.001_{-0.003}^{+0.001}$ \\ 
					$-0.001_{-0.001}^{+0.001}$ \\ 
				\end{tabular}
				& 
				\begin{tabular}{l}
					$0.182_{-0.001}^{+0.002}$ \\ 
					$0.266_{-0.011}^{+0.502}$ \\ 
					$0.236_{-0.019}^{+0.517}$ \\ 
					$0.240_{-0.003}^{+0.008}$ \\ 
					$0.581_{-0.165}^{+0.073}$ \\ 
					$0.722_{-0.158}^{+0.057}$ \\ 
				\end{tabular}
				& 
				\begin{tabular}{l}
					$0.326_{-0.053}^{+0.101}$ \\ 
					$0.238_{-0.037}^{+0.071}$ \\ 
					$0.524_{-0.063}^{+0.118}$ \\ 
					$0.250_{-0.038}^{+0.071}$ \\ 
					$0.167_{-0.051}^{+0.138}$ \\ 
					$0.111_{-0.037}^{+0.117}$ \\ 
				\end{tabular}
				& 
				\begin{tabular}{l}
					$ \ \ 0.004_{-0.004}^{+0.008}$ \\ 
					$ \ \ 0.003_{-0.004}^{+0.007}$ \\ 
					$ \ \ 0.006_{-0.005}^{+0.009}$ \\ 
					$ \ \ 0.004_{-0.004}^{+0.007}$ \\ 
					$ \ \ 0.002_{-0.002}^{+0.007}$ \\ 
					$ \ \ 0.001_{-0.001}^{+0.005}$ \\ 
				\end{tabular}
				\\ \hline
				$\lbrack 15,20]$ & 
				\begin{tabular}{l}
					$SM$ \\ 
					$VA-1$ \\ 
					$VA-2$ \\ 
					$VA-3$ \\ 
					$SP-1$ \\ 
					$SP-2$%
				\end{tabular}
				& 
				\begin{tabular}{l}
					$ \ \ 0.143_{-0.000}^{+0.000}$ \\ 
					$ \ \ 0.116_{-0.000}^{+0.000}$ \\ 
					$ \ \ 0.144_{-0.000}^{+0.000}$ \\ 
					$ \ \ 0.126_{-0.000}^{+0.000}$ \\ 
					$ \ \ 0.078_{-0.004}^{+0.005}$ \\ 
					$ \ \ 0.057_{-0.004}^{+0.004}$ \\ 
				\end{tabular}
				& 
				\begin{tabular}{l}
					$0.591_{-0.001}^{+0.001}$ \\ 
					$0.593_{-0.001}^{+0.001}$ \\ 
					$0.587_{-0.001}^{+0.001}$ \\ 
					$0.586_{-0.001}^{+0.001}$ \\ 
					$0.776_{-0.015}^{+0.012}$ \\ 
					$0.838_{-0.014}^{+0.011}$ \\ 
				\end{tabular}
				& 
				\begin{tabular}{l}
					$0.018_{-0.001}^{+0.001}$ \\ 
					$0.020_{-0.001}^{+0.001}$ \\ 
					$0.027_{-0.002}^{+0.002}$ \\ 
					$0.021_{-0.002}^{+0.002}$ \\ 
					$0.010_{-0.000}^{+0.000}$ \\ 
					$0.007_{-0.000}^{+0.000}$ \\ 
				\end{tabular}
				& 
				\begin{tabular}{l}
					$-0.010_{-0.000}^{+0.000}$ \\ 
					$-0.010_{-0.000}^{+0.000}$ \\ 
					$-0.010_{-0.000}^{+0.000}$ \\ 
					$-0.010_{-0.000}^{+0.000}$ \\ 
					$-0.005_{-0.000}^{+0.000}$ \\ 
					$-0.004_{-0.000}^{+0.000}$ \\ 
				\end{tabular}%
			\end{tabular}
			\\
			&&%
			\begin{tabular}{ll||llll}
				\hline\hline
				&  & $\ \ \ \ \ \ \left\langle Y_{2}\right\rangle $ & $\ \ \ \ \ \
				\left\langle \alpha _{\theta _{\Lambda }}\right\rangle $ & $\ \ \ \ \ \
				\left\langle \alpha _{\theta _{l}}\right\rangle $ & $\ \ \ \ \ \
				\left\langle \alpha _{\theta _{l}}^{\prime }\right\rangle $ \\ \hline\hline
				$\lbrack 1.1,6]$ & 
				\begin{tabular}{l}
					$SM$ \\ 
					$VA-1$ \\ 
					$VA-2$ \\ 
					$VA-3$ \\ 
					$SP-1$ \\ 
					$SP-2$%
				\end{tabular}
				& 
				\begin{tabular}{l}
					$ \ \ 0.084_{-0.003}^{+0.001}$ \\ 
					$ \ \ 0.071_{-0.004}^{+0.002}$ \\ 
					$ \ \ 0.076_{-0.005}^{+0.002}$ \\ 
					$ \ \ 0.070_{-0.004}^{+0.002}$ \\ 
					$ \ \ 0.043_{-0.007}^{+0.015}$ \\ 
					$ \ \ 0.028_{-0.006}^{+0.015}$ \\ 
				\end{tabular}
				& 
				\begin{tabular}{l}
					$-0.961_{-0.015}^{+0.039}$ \\ 
					$-0.966_{-0.013}^{+0.034}$ \\ 
					$-0.961_{-0.015}^{+0.039}$ \\ 
					$-0.920_{-0.020}^{+0.044}$ \\ 
					$-0.492_{-0.168}^{+0.078}$ \\ 
					$-0.327_{-0.167}^{+0.062}$ \\ 
				\end{tabular}
				& 
				\begin{tabular}{l}
					$-0.800_{-0.001}^{+0.003}$ \\ 
					$-0.704_{-0.004}^{+0.011}$ \\ 
					$-0.750_{-0.006}^{+0.018}$ \\ 
					$-0.727_{-0.004}^{+0.011}$ \\ 
					$-0.259_{-0.167}^{+0.058}$ \\ 
					$-0.152_{-0.122}^{+0.036}$ \\ 
				\end{tabular}
				& 
				\begin{tabular}{l}
					$ \ \ 0.014_{-0.017}^{+0.031}$ \\ 
					$ \ \ 0.163_{-0.015}^{+0.024}$ \\ 
					$ \ \ 0.107_{-0.017}^{+0.030}$ \\ 
					$ \ \ 0.131_{-0.016}^{+0.027}$ \\ 
					$ \ \ 0.004_{-0.005}^{+0.019}$ \\ 
					$ \ \ 0.003_{-0.003}^{+0.013}$ \\ 
				\end{tabular}
				\\ \hline
				$\lbrack 15,20]$ & 
				\begin{tabular}{l}
					$SM$ \\ 
					$VA-1$ \\ 
					$VA-2$ \\ 
					$VA-3$ \\ 
					$SP-1$ \\ 
					$SP-2$%
				\end{tabular}
				& 
				\begin{tabular}{l}
					$ \ \ 0.016_{-0.000}^{+0.000}$ \\ 
					$ \ \ 0.015_{-0.000}^{+0.000}$ \\ 
					$ \ \ 0.016_{-0.000}^{+0.000}$ \\ 
					$ \ \ 0.016_{-0.000}^{+0.000}$ \\ 
					$ \ \ 0.009_{-0.001}^{+0.001}$ \\ 
					$ \ \ 0.006_{-0.000}^{+0.001}$ \\ 
				\end{tabular}
				& 
				\begin{tabular}{l}
					$-0.844_{-0.002}^{+0.003}$ \\ 
					$-0.844_{-0.002}^{+0.003}$ \\ 
					$-0.844_{-0.002}^{+0.003}$ \\ 
					$-0.895_{-0.002}^{+0.002}$ \\ 
					$-0.463_{-0.028}^{+0.023}$ \\ 
					$-0.335_{-0.026}^{+0.021}$ \\ 
				\end{tabular}
				& 
				\begin{tabular}{l}
					$-0.161_{-0.002}^{+0.002}$ \\ 
					$-0.159_{-0.002}^{+0.002}$ \\ 
					$-0.168_{-0.002}^{+0.002}$ \\ 
					$-0.171_{-0.002}^{+0.002}$ \\ 
					$-0.048_{-0.006}^{+0.004}$ \\ 
					$-0.030_{-0.003}^{+0.003}$ \\ 
				\end{tabular}
				& 
				\begin{tabular}{l}
					$-0.679_{-0.004}^{+0.004}$ \\ 
					$-0.548_{-0.003}^{+0.004}$ \\ 
					$-0.677_{-0.004}^{+0.004}$ \\ 
					$-0.626_{-0.004}^{+0.004}$ \\ 
					$-0.212_{-0.022}^{+0.017}$ \\ 
					$-0.126_{-0.014}^{+0.010}$ \\ 
				\end{tabular}%
			\end{tabular}%
		\end{eqnarray*}%
	\end{table}
	\FloatBarrier
and $C_{T_5}$ such that the above equation of constraints is satisfied. Doing this we find that the maximum impact on the different observables is achieved when we select $C_T=0.72$ and $C_{T5}=0.2$ \cite{Das:2018sms}.
\begin{table}[h]
		\caption{Values of some observables in the SM and in the presence of new VA and SP couplings. Description of different scenarios is same as in Table \ref{lambda-obs-1}.} \label{lambda-obs-3}
		
		\begin{tabular}{ll||lllllll}
			\hline\hline
			&  & $\ \left\langle \alpha _{\xi }\right\rangle $ & $\ \ \
			\left\langle \alpha _{\xi }^{\prime }\times 10^{-2}\right\rangle $ & $\ \ \
			\ \ \ \left\langle \alpha _{U}\right\rangle $ & $\left\langle \alpha
			_{L}\right\rangle $ & $ \ \left\langle P_{3}\right\rangle $ & $ \ \ \ \ \ \ \left\langle
			P_{8}\right\rangle $ & $ \ \ \ \ \ \ \left\langle P_{9}\right\rangle $ \\ \hline\hline
			$\lbrack 1.1,6]$ & 
			\begin{tabular}{l}
				$SM$ \\ 
				$VA-1$ \\ 
				$VA-2$ \\ 
				$VA-3$ \\ 
				$SP-1$ \\ 
				$SP-2$%
			\end{tabular}
			& 
			\begin{tabular}{l}
				$ \ \ 0.087_{-0.030}^{+0.050}$ \\ 
				$ \ \ 0.042_{-0.019}^{+0.029}$ \\ 
				$ \ \ 0.069_{-0.026}^{+0.042}$ \\ 
				$ \ \ 0.053_{-0.018}^{+0.028}$ \\ 
				$ \ \ 0.044_{-0.021}^{+0.054}$ \\ 
				$ \ \ 0.029_{-0.014}^{+0.044}$ \\ 
			\end{tabular}
			& 
			\begin{tabular}{l}
				$-0.202_{-0.110}^{+0.057}$ \\ 
				$-0.260_{-0.138}^{+0.072}$ \\ 
				$-0.262_{-0.138}^{+0.073}$ \\ 
				$-0.242_{-0.125}^{+0.066}$ \\ 
				$-0.103_{-0.120}^{+0.042}$ \\ 
				$-0.069_{-0.098}^{+0.029}$ \\ 
			\end{tabular}
			& 
			\begin{tabular}{l}
				$-0.973_{-0.018}^{+0.063}$ \\ 
				$-0.983_{-0.011}^{+0.038}$ \\ 
				$-0.979_{-0.014}^{+0.046}$ \\ 
				$-0.944_{-0.021}^{+0.053}$ \\ 
				$-0.084_{-0.087}^{+0.022}$ \\ 
				$-0.043_{-0.044}^{+0.011}$ \\ 
			\end{tabular}
			& 
			\begin{tabular}{l}
				$-0.960_{-0.015}^{+0.037}$ \\ 
				$-0.963_{-0.013}^{+0.034}$ \\ 
				$-0.959_{-0.015}^{+0.038}$ \\ 
				$-0.917_{-0.020}^{+0.043}$ \\ 
				$-0.310_{-0.182}^{+0.066}$ \\ 
				$-0.182_{-0.135}^{+0.041}$ \\ 
			\end{tabular}
			& 
			\begin{tabular}{l}
				$-0.005_{-0.011}^{+0.008}$ \\ 
				$-0.089_{-0.008}^{+0.007}$ \\ 
				$-0.059_{-0.010}^{+0.008}$ \\ 
				$-0.070_{-0.007}^{+0.006}$ \\ 
				$-0.003_{-0.000}^{+0.004}$ \\ 
				$-0.002_{-0.007}^{+0.002}$ \\ 
			\end{tabular}
			& 
			\begin{tabular}{l}
				$ \ \ 0.025_{-0.030}^{+0.055}$ \\ 
				$ \ \ 0.284_{-0.025}^{+0.041}$ \\ 
				$ \ \ 0.190_{-0.030}^{+0.051}$ \\ 
				$ \ \ 0.231_{-0.027}^{+0.046}$ \\ 
				$ \ \ 0.013_{-0.015}^{+0.045}$ \\ 
				$ \ \ 0.008_{-0.010}^{+0.034}$ \\ 
			\end{tabular}
			& 
			\begin{tabular}{l}
				$-0.226_{-0.005}^{+0.011}$ \\ 
				$-0.325_{-0.001}^{+0.002}$ \\ 
				$-0.279_{-0.004}^{+0.002}$ \\ 
				$-0.291_{-0.003}^{+0.007}$ \\ 
				$-0.116_{-0.038}^{+0.018}$ \\ 
				$-0.077_{-0.038}^{+0.014}$ \\ 
			\end{tabular}
			\\ \hline
			$\lbrack 15,20]$ & 
			\begin{tabular}{l}
				$SM$ \\ 
				$VA-1$ \\ 
				$VA-2$ \\ 
				$VA-3$ \\ 
				$SP-1$ \\ 
				$SP-2$%
			\end{tabular}
			& 
			\begin{tabular}{l}
				$-0.304_{-0.002}^{+0.002}$ \\ 
				$-0.251_{-0.001}^{+0.001}$ \\ 
				$-0.308_{-0.002}^{+0.002}$ \\ 
				$-0.228_{-0.002}^{+0.001}$ \\ 
				$-0.167_{-0.012}^{+0.001}$ \\ 
				$-0.120_{-0.011}^{+0.008}$ \\ 
			\end{tabular}
			& 
			\begin{tabular}{l}
				$-0.021_{-0.005}^{+0.004}$ \\ 
				$-0.030_{-0.007}^{+0.006}$ \\ 
				$-0.028_{-0.007}^{+0.006}$ \\ 
				$-0.030_{-0.006}^{+0.006}$ \\ 
				$-0.012_{-0.004}^{+0.003}$ \\ 
				$-0.008_{-0.003}^{+0.002}$ \\ 
			\end{tabular}
			& 
			\begin{tabular}{l}
				$-0.805_{-0.008}^{+0.003}$ \\ 
				$-0.807_{-0.002}^{+0.003}$ \\ 
				$-0.803_{-0.002}^{+0.002}$ \\ 
				$-0.864_{-0.002}^{+0.002}$ \\ 
				$-0.211_{-0.022}^{+0.017}$ \\ 
				$-0.130_{-0.015}^{+0.011}$ \\ 
			\end{tabular}
			& 
			\begin{tabular}{l}
				$-0.860_{-0.002}^{+0.003}$ \\ 
				$-0.860_{-0.002}^{+0.003}$ \\ 
				$-0.861_{-0.002}^{+0.003}$ \\ 
				$-0.908_{-0.002}^{+0.002}$ \\ 
				$-0.256_{-0.025}^{+0.020}$ \\ 
				$-0.161_{-0.018}^{+0.013}$ \\ 
			\end{tabular}%
			& 
			\begin{tabular}{l}
				$ \ \ 0.382_{-0.001}^{+0.001}$ \\ 
				$ \ \ 0.310_{-0.001}^{+0.001}$ \\ 
				$ \ \ 0.383_{-0.001}^{+0.001}$ \\ 
				$ \ \ 0.336_{-0.001}^{+0.001}$ \\ 
				$ \ \ 0.198_{-0.197}^{+0.011}$ \\ 
				$ \ \ 0.151_{-0.010}^{+0.012}$ \\ 
			\end{tabular}
			& 
			\begin{tabular}{l}
				$-0.956_{-0.004}^{+0.005}$ \\ 
				$-0.770_{-0.004}^{+0.005}$ \\ 
				$-0.956_{-0.004}^{+0.005}$ \\ 
				$-0.885_{-0.004}^{-0.005}$ \\ 
				$-0.523_{-0.030}^{+0.025}$ \\ 
				$-0.375_{-0.029}^{+0.023}$ \\ 
			\end{tabular}
			& 
			\begin{tabular}{l}
				$-0.611_{-0.003}^{+0.003}$ \\ 
				$-0.614_{-0.003}^{+0.003}$ \\ 
				$-0.606_{-0.003}^{+0.003}$ \\ 
				$-0.650_{-0.002}^{+0.003}$ \\ 
				$-0.335_{-0.019}^{+0.016}$ \\ 
				$-0.242_{-0.018}^{+0.015}$ \\ 
			\end{tabular}
		\end{tabular}
	\end{table}

\begin{figure}
	\begin{center}
		\begin{tabular}{ll}
			\includegraphics[height=4.8cm,width=8cm]{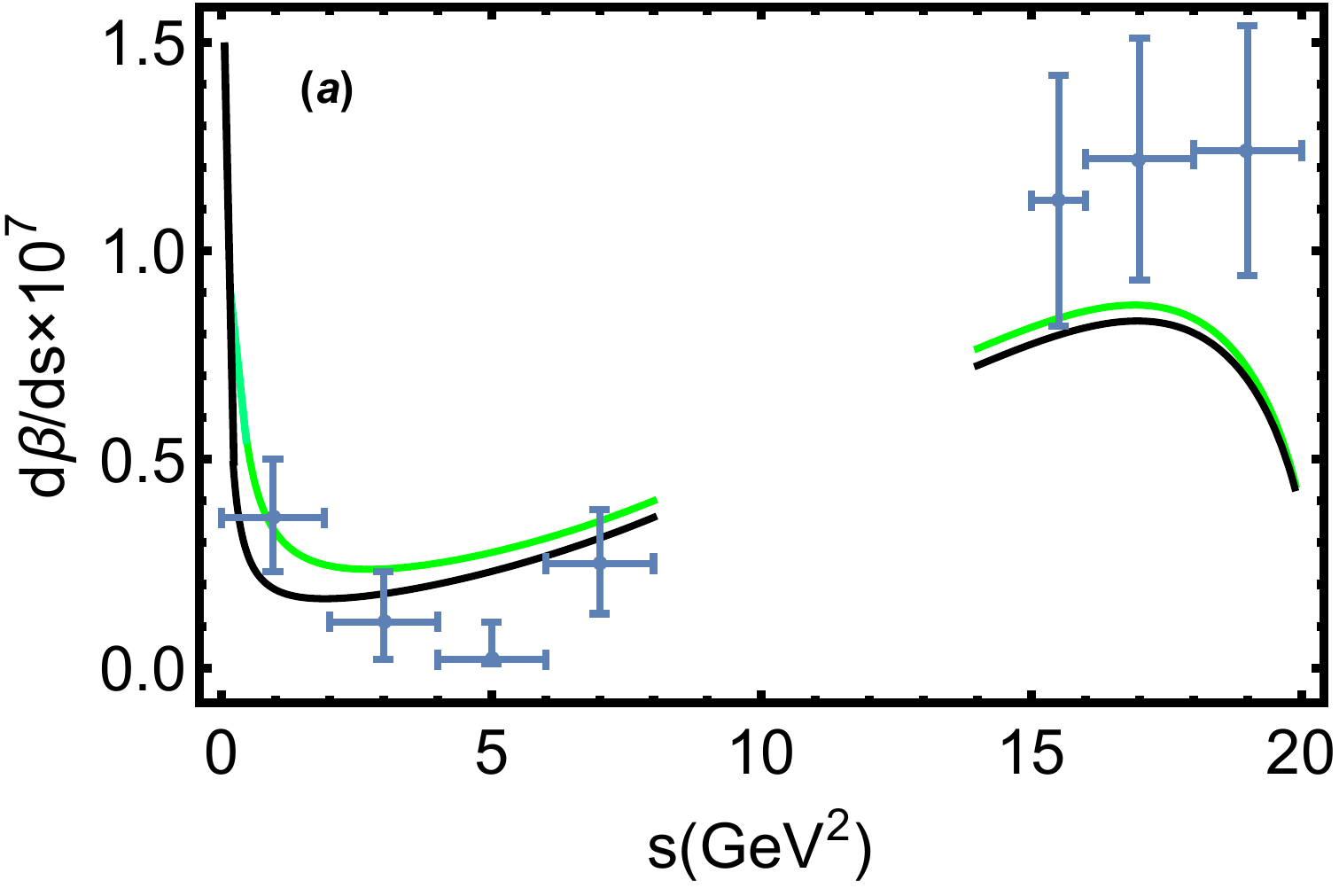} & { \ \ \ \ \ \ } \includegraphics[height=4.8cm,width=8cm]{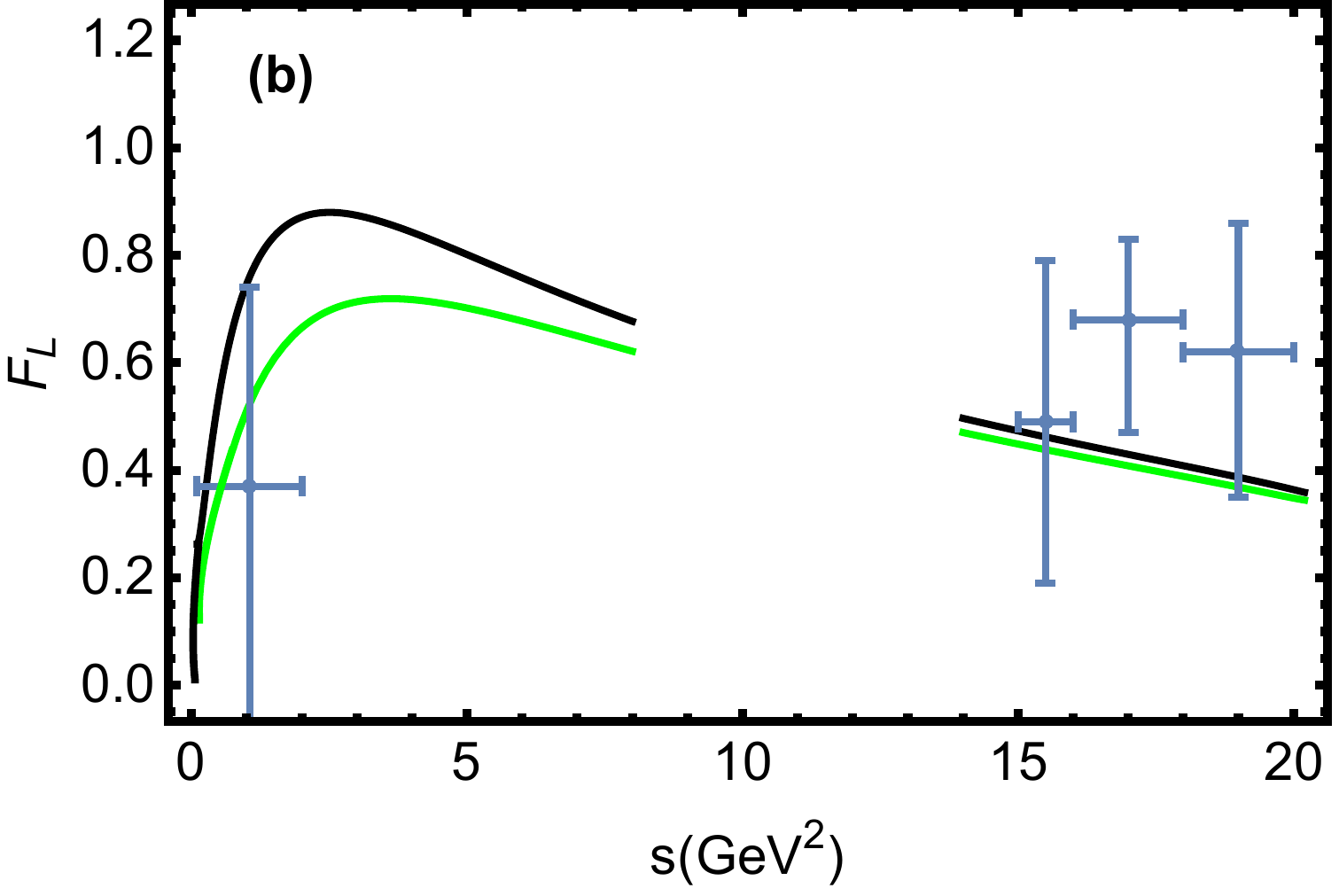} \\
			\\
			\includegraphics[height=4.8cm,width=8cm]{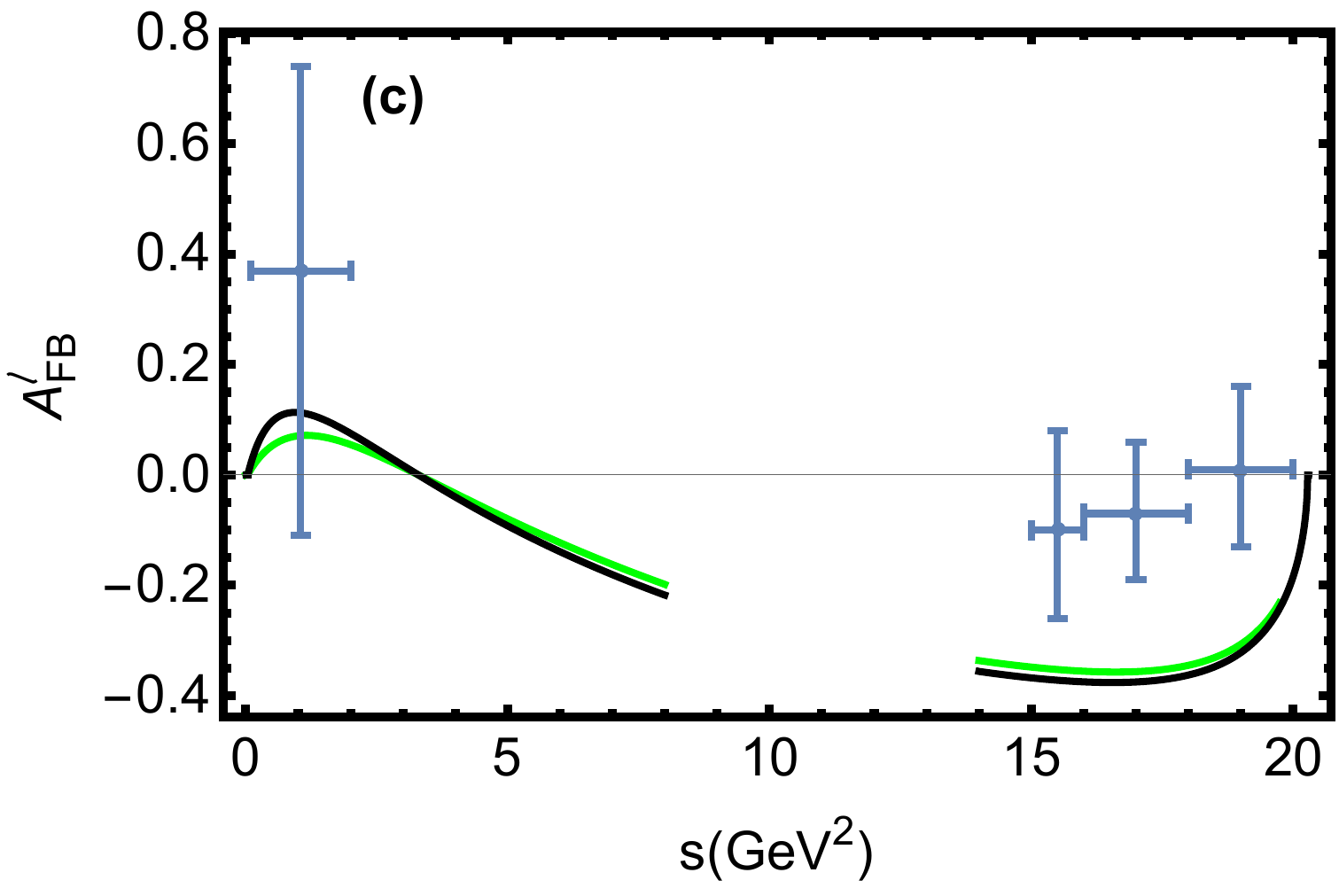} & { \ \ \ \ \ \ } \includegraphics[height=4.8cm,width=8cm]{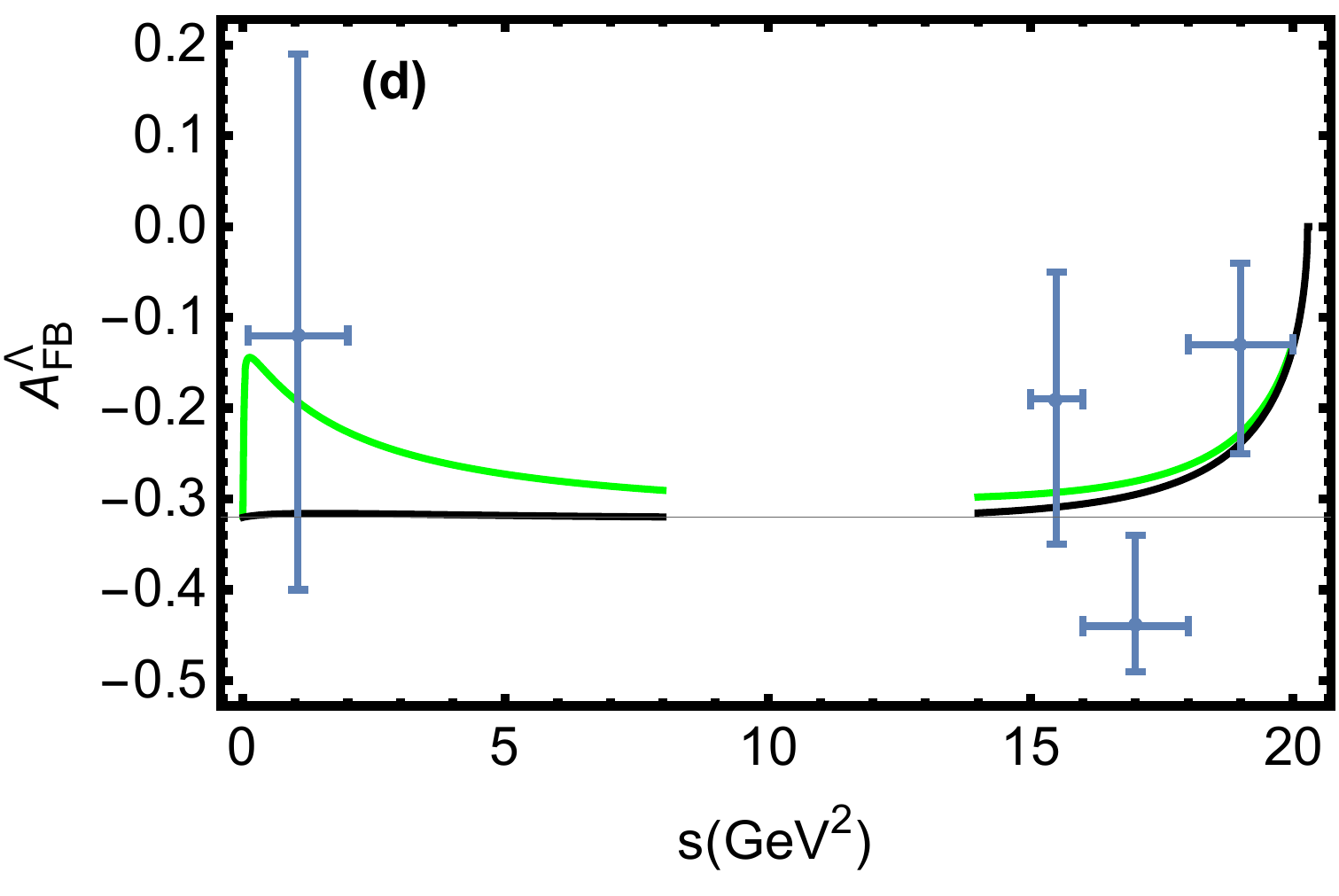}
		\end{tabular}
	\end{center}
	\caption{These plots are constructed by taking $C_T=0.72$ and $C_{T5}=0.2$ and the black color indicates SM.}\label{Fig19}
\end{figure}

We have also explored that in contrast to the $VA$ and $SP$ couplings, very few observables are affected by the tensor couplings only in low $s$ region. The values of most of the observables do not show any dependence on the tensor couplings and remain close to their SM predictions. For example, in case $d\mathcal{B}/ds$, $F_L$, $A^{\ell}_{FB}$ and $A^\Lambda_{FB}$ on which the experimental data is available, the imprints of tensor couplings are shown in Fig. \ref{Fig19} and the numerical results in different experimental bins are provided in Table \ref{lambda-obs-1}. Here, we can observe that our analysis coincides with \cite{Das:2018sms} for $d\mathcal{B}/ds$ and $F_L$. One can also see that these four observables are sensitive to tensor couplings only in low $s$ region. However, the effects on $d\mathcal{B}/ds$ and $A^{\ell}_{FB}$ are mild as compared to $F_L$ and $A^\Lambda_{FB}$, particularly in $s \in [0.1,3]$ GeV$^2$ bin where the effects in the $A_{FB}^{\Lambda}$ are very prominent. In this region, after inclusion of tensor coupling, the value of $A_{FB}^{\Lambda}$ looks slightly better in agreement with the experimental observations as compared to that of the SM predictions alone. In short, our analysis shows that the effects of the tensor couplings are not very prominent for the angular observables except $F_L$ and $A^\Lambda_{FB}$ in low $s$ region.

\subsection{Combined effects of $VA$-$SP$ couplings on angular observables}
	As the uncertainties in the experimental data of $d\mathcal{B}/ds$, $A_{FB}^{\ell}$, $F_{L}$ and $A_{FB}^{\Lambda}$ are significantly large and based on the analysis performed above, we can say that any individual set of new couplings can not accommodate all the available data. This situation is somewhat more problematic in high $s$ bins. In this case, from Figs. \ref{Fig1hh}, \ref{Fig10} and \ref{Fig19} one can quantify the situation for these observables in Table \ref{DataAccom} that lead to the following findings:
	\begin{itemize}
\item The $VA$ couplings accommodate $d\mathcal{B}/ds$ data only in three bins of $s$ at large-recoil. The data of $F_L$ and $A^\Lambda_{FB}$ can be accommodated in low and high $s$ bins except $s\in[16,18]$ GeV$^2$ region. The data of $A^{\ell}_{FB}$ can be taken care of only in $s\in[0.1,2]$ GeV$^2$ bin. It means that the $VA$ couplings only satisfy LHCb data in those bins where the SM can also accommodate the data to the same extent. Hence, the addition of new $VA$ couplings to the SM is not sufficient alone.
		\item Just like the SM, the $SP$ couplings satisfy the $d\mathcal{B}/ds$ data in three low $s$ bins. In addition, these $SP$ couplings accommodate the data in $s \in [15,16]$ GeV$^2$ where the SM predictions do not match with experimental measurements. The data of $F_L$ can be taken care of only in low $s$ bin $s \in [0.1,2]$ GeV$^2$. The LHCb data of $A^{\ell}_{FB}$ could be fully accommodated but in case of $A^\Lambda_{FB}$ it is also possible in all bins except for $s\in[16,18]$ GeV$^2$ region.
		\item Similar to the results of the SM, the $T$ coupling only accommodate $d\mathcal{B}/ds$ data in three low $s$ bins. The data of $A^{\ell}_{FB}$ is satisfied in $s\in[0.1,2]$ GeV$^2$ bin only however, for that of $F_L$ and $A^{\Lambda}_{FB}$ it can be accommodated in all bins except $s\in[16,18]$ GeV$^2$.
	\end{itemize}
\FloatBarrier	
\begin{table}[h!]
	\caption{Data accommodated by new couplings in different bins.}\label{DataAccom}
	\begin{tabular}{|c|c|c|c|c|c|c|c|c|c|c|c|c|c|c|c|c|}
		\hline
		\multirow{2}{*}{} $\mathcal{O}$ & \multicolumn{4}{c|}{$d\mathcal{B}/ds$} & %
		\multicolumn{4}{c|}{$F_{L}$} & \multicolumn{4}{c|}{$A_{FB}^{\ell}$} & \multicolumn{4}{c|}{$A_{FB}^{\Lambda}$} \\
		\cline{1-17}
		bins (GeV$^2$) & SM & $VA$ & $SP$ & $T$ & SM & $VA$ & $SP$ & $T$ & SM & $VA$ & $SP$ & $T$ & SM & $VA$ & $SP$ & $T$\\
		\hline
		$[0.1-2]$ & \XSolidBrush & \ding{52} & \XSolidBrush & \XSolidBrush &  \ding{52} & \ding{52} & \ding{52} & \ding{52} & \ding{52} & \ding{52} & \ding{52} & \ding{52} & \ding{52} & \ding{52} & \ding{52} & \ding{52} \\
		\hline
		$[2-4]$ & \XSolidBrush & \XSolidBrush & \XSolidBrush & \XSolidBrush & \textbf{--} & \textbf{--} & \textbf{--} & \textbf{--} & \textbf{--} & \textbf{--} & \textbf{--} & \textbf{--} & \textbf{--} & \textbf{--} & \textbf{--} & \textbf{--} \\
		\hline
		$[4-6]$ & \XSolidBrush & \XSolidBrush & \XSolidBrush & \XSolidBrush & \textbf{--} & \textbf{--} & \textbf{--} & \textbf{--} & \textbf{--} & \textbf{--} & \textbf{--} & \textbf{--} & \textbf{--} & \textbf{--} & \textbf{--} & \textbf{--} \\
		\hline
		$[6-8]$ & \XSolidBrush & \ding{52} & \XSolidBrush & \XSolidBrush & \textbf{--} & \textbf{--} & \textbf{--} & \textbf{--} & \textbf{--} & \textbf{--} & \textbf{--} & \textbf{--} & \textbf{--} & \textbf{--} & \textbf{--} & \textbf{--} \\
		\hline
		$[15-16]$& \XSolidBrush & \XSolidBrush & \XSolidBrush & \XSolidBrush & \XSolidBrush & \XSolidBrush & \ding{52} & \XSolidBrush & \XSolidBrush & \XSolidBrush & \ding{52} & \XSolidBrush & \ding{52} & \ding{52} & \ding{52} & \ding{52} \\
		\hline
		$[16-18]$& \XSolidBrush & \XSolidBrush & \XSolidBrush & \XSolidBrush & \XSolidBrush & \XSolidBrush & \XSolidBrush & \XSolidBrush & \XSolidBrush & \XSolidBrush & \ding{52} & \XSolidBrush & \XSolidBrush & \XSolidBrush & \XSolidBrush & \XSolidBrush \\
		\hline
		$[18-20]$& \XSolidBrush & \XSolidBrush & \XSolidBrush & \XSolidBrush & \ding{52} & \ding{52} & \XSolidBrush & \ding{52} & \XSolidBrush & \XSolidBrush & \ding{52} & \XSolidBrush & \ding{52} & \ding{52} & \ding{52} & \ding{52} \\
		\hline
	\end{tabular}
\end{table}
\FloatBarrier
Based on these observations, we can see that taking new couplings separately is not a favorable option in the presence of available data of the different physical observables in $\Lambda_b \to \Lambda (\to p\pi^-)\mu^{+}\mu^{-}$ decay. Therefore, it is useful to see if two new couplings are turned on together, does this situation improves or not? In order to do so, the constraints on new WCs corresponding to vector, axial-vector, scalar and pseudo-scalar operators are once again chosen from the one adopted by \cite{Das:2018sms} and by using the global fit presented in \cite{Altmannshofer:2017fio}; i.e., 
\begin{eqnarray}
C_V = [-1.61,-1] \hspace{1cm} C_S^{(\prime)} = [-4,4] \hspace{1cm}  C_P^{(\prime)} = [-4,4] \label{range}
\end{eqnarray}
with $|C_{S,P}-C_{S,P}^{\prime}|\leq 0.1$. We have not included $C_V^{\prime},\; C_A,\; C_A^{\prime},\; C_T$ and $C_{T5}$ in the forthcoming numerical analysis as the severe constraints from $B-$physics on these WCs do not allow us to vary them significantly. Thus, from Eq. (\ref{range}) the following ten combinations are possible:
\begin{eqnarray}
& \text{(i)}: (C_S,C_V),\quad \text{(ii)}: (C_{S}^{\prime},C_V),\quad \text{(iii)}: (C_P,C_V),\quad \text{(iv)}: (C_{P}^{\prime},C_V),\quad \text{(v)}: (C_S,C_{S'})&\label{comb} \\
& \text{(vi)}: (C_{S},C_{P}^{\prime}),\quad \text{(vii)}: (C_{S}^{\prime},C_P),\quad \text{(viii)}: (C_S,C_P),\quad \text{(ix)}: (C_{S'},C_{P'}),\quad \text{(x)}: (C_{P},C_{P}^{\prime}).& \notag
\end{eqnarray}
Among these combinations, we are interested in looking for the combination(s) which maximally accommodate the current available data of all the four observables mentioned above. With this condition, by exploring the various combinations given in Eq. (\ref{comb}), it is found that there is not a single choice which could explain the full data of all four observables simultaneously. However, we found from Eq. (\ref{combleft}) that there are six combinations of new WCs which could accommodate the data of the four observables;, i.e., $d\mathcal{B}/ds$, $F_L$, $A^{\ell}_{FB}$ and $A^\Lambda_{FB}$ simultaneously in the bins $s\in[0.1,2]$ GeV$^2$ and $s\in[15,16]$ GeV$^2$ and three observables (excluding $d\mathcal{B}/ds$) in $s\in[18,20]$ GeV$^2$ bin. In this case, for the bin $s\in[16,18]$ GeV$^2$ only $F_L$ and $A^\Lambda_{FB}$ can be taken care of. On the other hand, if we would like to accommodate the data of $d\mathcal{B}/ds$ as well, we have to choose one from the other three observables. Therefore, based on these observations the six possible combinations of new couplings which accommodate almost all the data of above mentioned three or four observables simultaneously are
	\begin{eqnarray}
	&\text{(i)}: (C_S,C_V),\quad \text{(ii)}: (C_{S}^{\prime},C_V),\quad \text{(iii)}: (C_{P}^{\prime},C_V),\quad&\notag\\
	& \text{(iv)}: (C_S,C_{S}^\prime),\quad   \text{(v)}: (C_{S},C_{P}^{\prime}),\quad \text{(vi)}: (C_{S}^{\prime},C_P)&.\label{combleft}
	\end{eqnarray}
	The impact of these combinations of new couplings on the experimentally measured and other physical observables in low (high-recoil region) and high (low-recoil region) $s$ bins will be discussed from here onwards. The whole analysis is performed by taking the central values of the FFs and that of the experimental data of $d\mathcal{B}/ds$, $F_L$, $A^{\ell}_{FB}$ and $A^\Lambda_{FB}$.
	
\subsubsection{High recoil region}
	In this region, we focus only on the bin $s \in [0.1,2]$ GeV$^2$ because the LHCb data in this particular region is available for all the four observables mentioned above. First, we have examined all the six combinations given in Eq. (\ref{combleft}) by tweaking them in their current allowed ranges (c.f. Eq. (\ref{range})), to see if they could simultaneously accommodate the available data of these observables or not. At the next step, we have calculated the values of these observables for these combinations accordingly and their results are presented in Fig. \ref{Figpoint1to2}. From these plots, we have made the following observations:
	\begin{figure}[h!]
		\begin{center}
			\begin{tabular}{ll}
				\includegraphics[height=5cm,width=7cm]{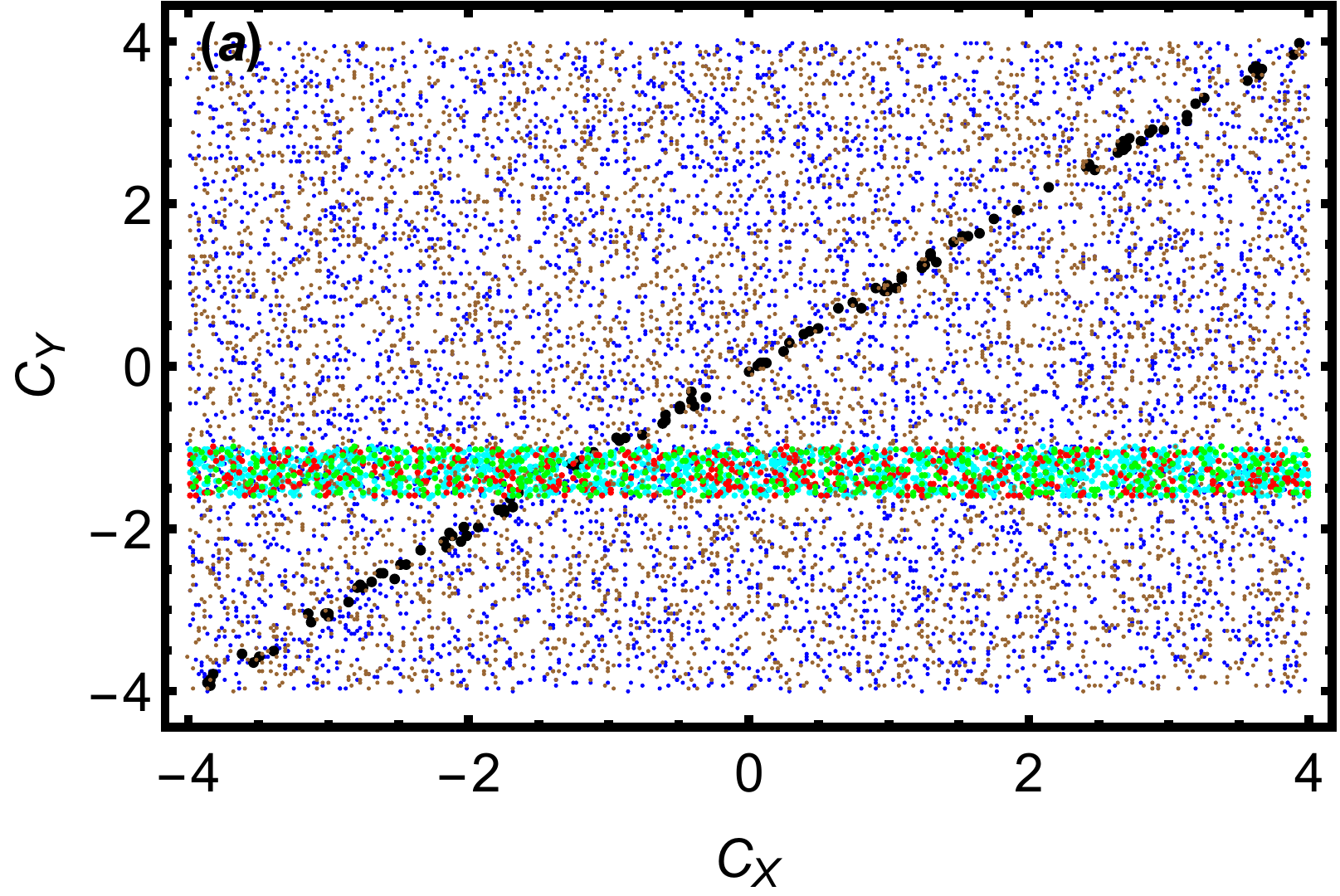}  & \ \ \includegraphics[height=5cm,width=7cm]{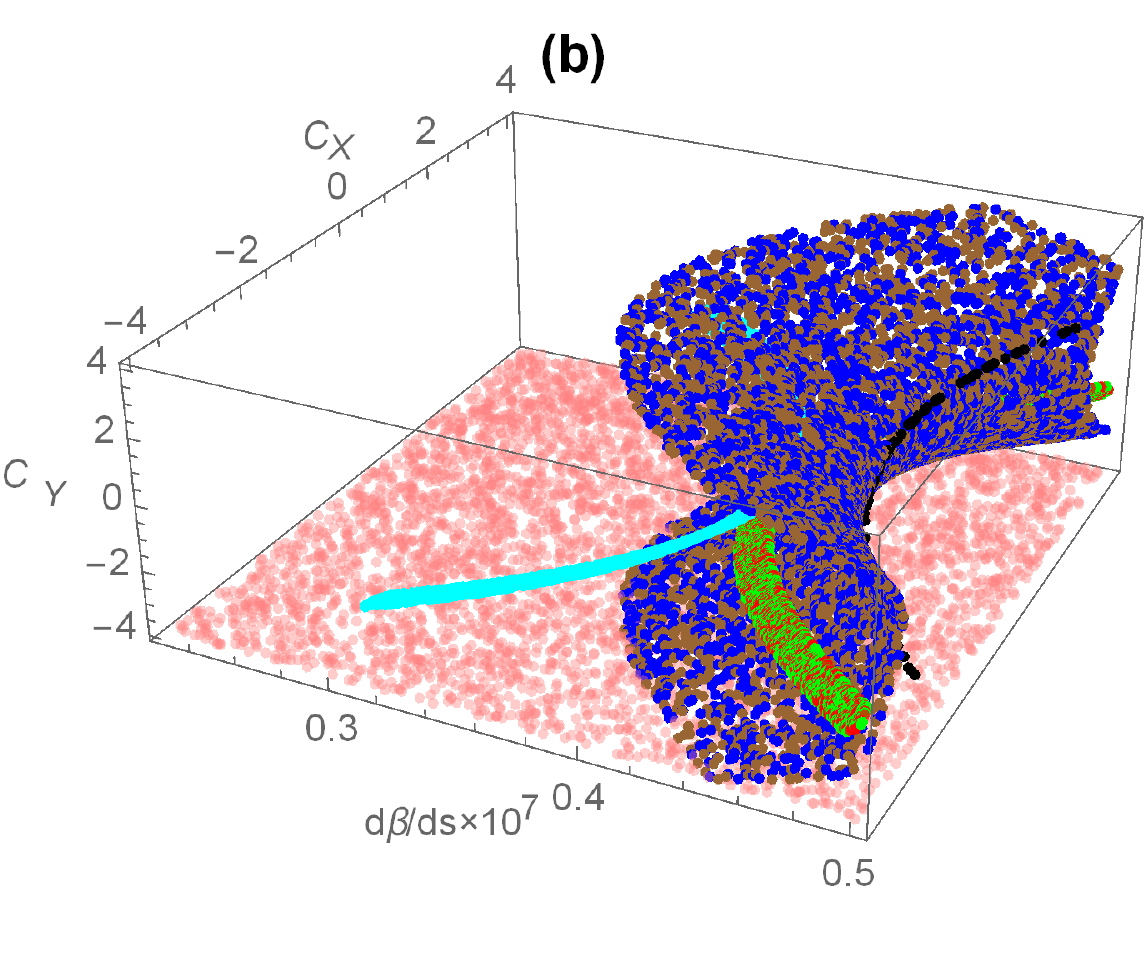} \\
				\\
				\includegraphics[height=5cm,width=7cm]{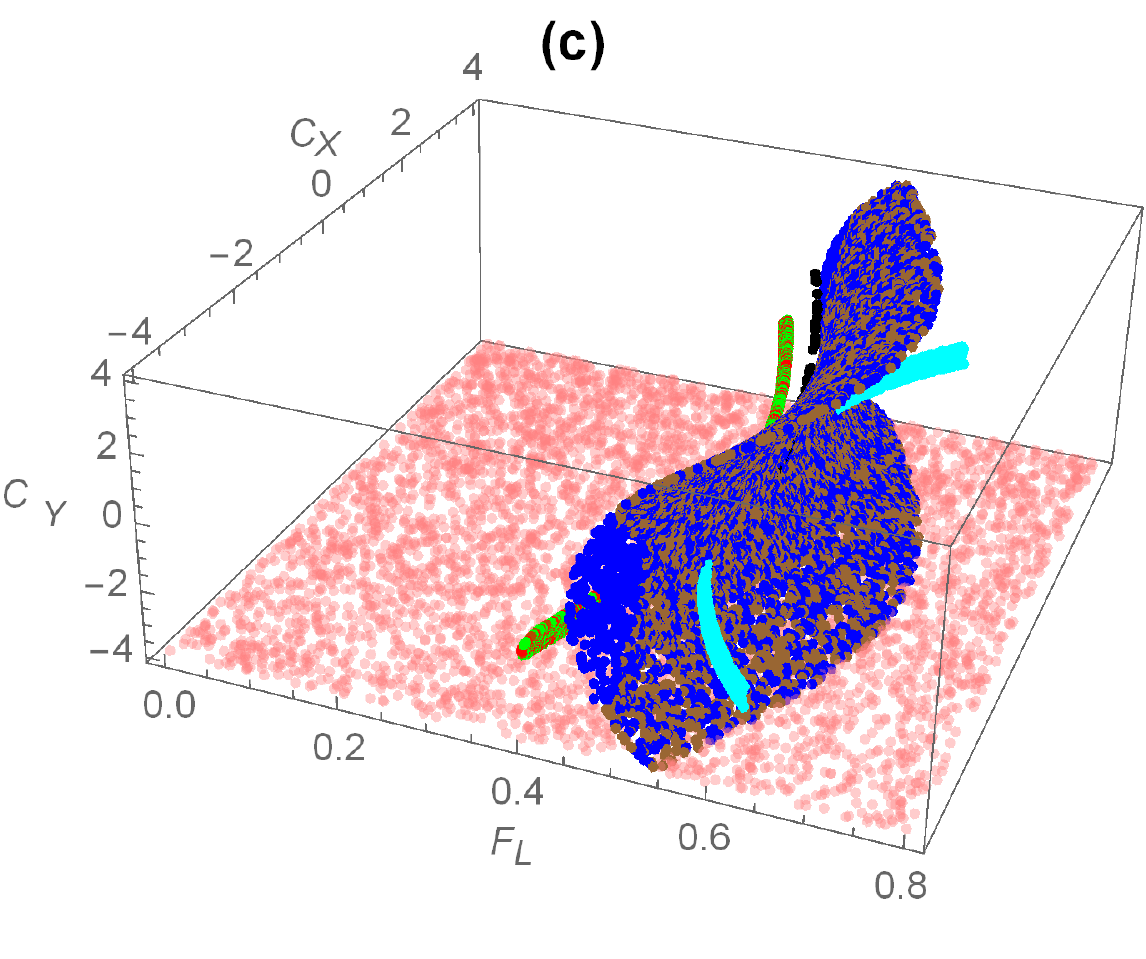}	 & \ \ \includegraphics[height=5cm,width=7cm]{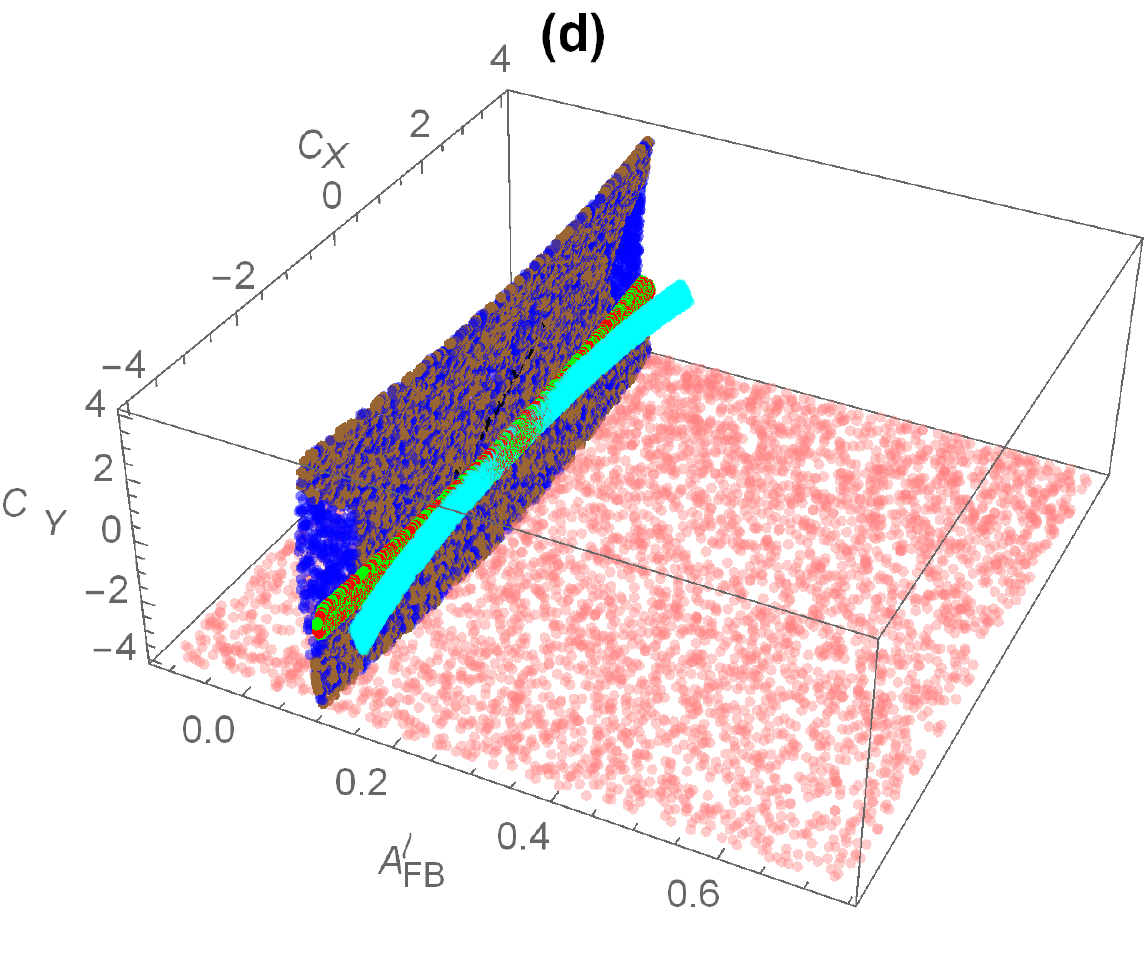} \\
				\\
				\includegraphics[height=5cm,width=7cm]{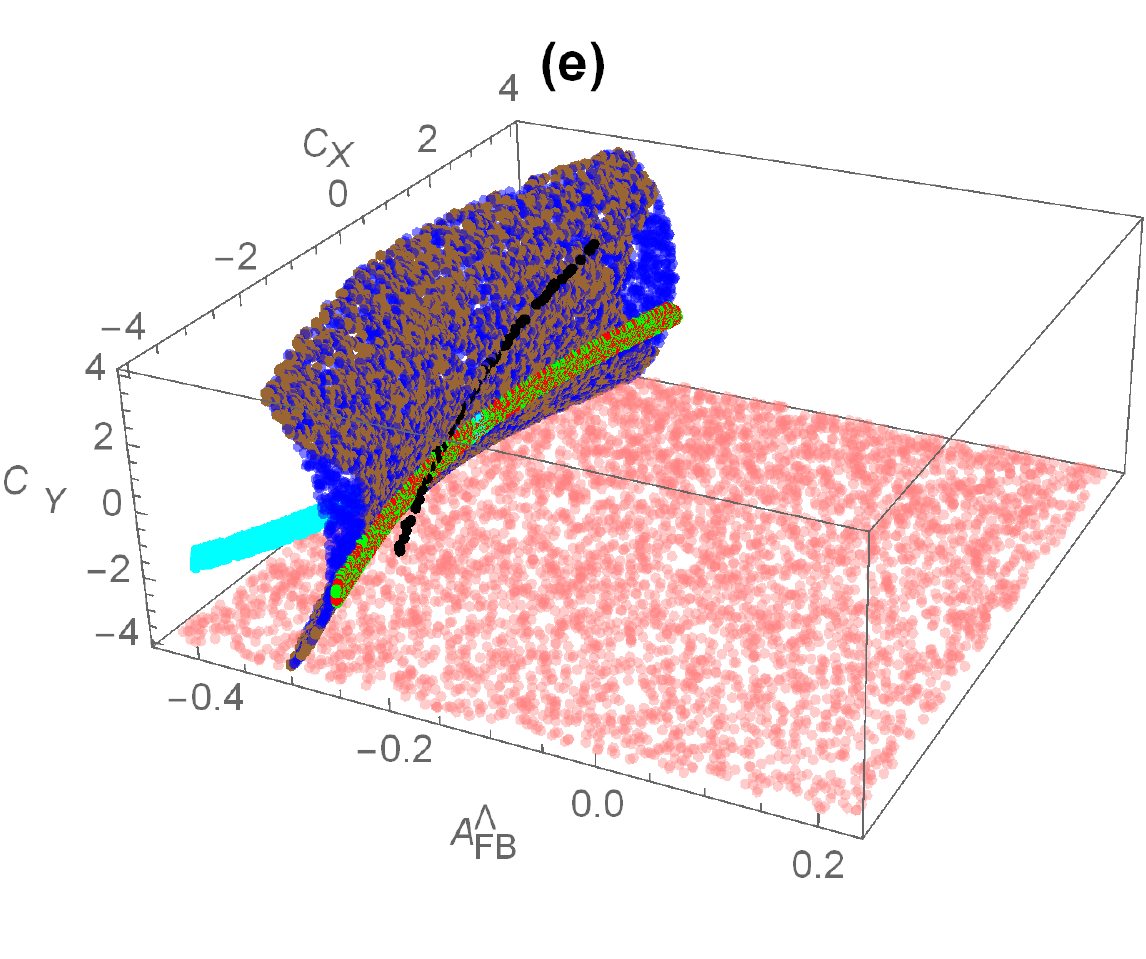}
			\end{tabular}
		\end{center}
		\caption{(a) The parametric space of $(C_X,C_Y)$ allowed from $B-$physics constraints on new WCs and that satisfy the data of $d\mathcal{B}/ds$, $F_L$, $A_{FB}^{\ell}$ and $A_{FB}^{\Lambda}$ simultaneously in the bin $s\in[0.1,2]$ GeV$^2$. Different colors in the plots represent different combinations of new WCs: Red, blue, green, cyan, brown and black dots represent the  $(C_X,\; C_Y)$ $=$ $(C_S,\; C_V)$, $(C_S,\; C_{P}^{\prime})$, $(C_{S}^{\prime},\; C_V)$, $(C_{P}^{\prime},\;C_V)$, $(C_{S}^{\prime},\;C_P)$ and $(C_S,\; C_{S}^{\prime})$, respectively.  The plots (b)-(e) present the predictions of $d\mathcal{B}/ds$, $F_L$, $A_{FB}^{\ell}$ and $A_{FB}^{\Lambda}$ in the bin $s\in[0.1,2]$ GeV$^2$ against the WCs collected in (a) where the pink flat curves reflect the measured values of $d\mathcal{B}/ds$, $F_L$, $A_{FB}^{\Lambda}$ and $A_{FB}^{\ell}$, along with the uncertainties at the LHCb.}\label{Figpoint1to2}
	\end{figure}
	
	\begin{itemize}
		\item Fig. \ref{Figpoint1to2}(a) reflects the complete range of each combination of new WCs given in Eq. (\ref{combleft}) which is allowed by $B-$physics data (c.f. Eq. (\ref{range})) with the condition $|C_{S}-C_{S}^{\prime}|\leq 0.1$. The range of $C_X$, $C_Y$ and the corresponding color schemes are given in the caption of the figure. We found that the full allowed ranges of WCs from $B-$meson decays simultaneously satisfy the the data of $d\mathcal{B}/ds$, $F_{L}$, $A_{FB}^{\ell}$ and $A_{FB}^{\Lambda}$ in $s \in [0.1,2]$ GeV$^2$ bin.
		\item By using these allowed values for each combination of new WCs, we predict the values of $d\mathcal{B}/ds$, $F_{L}$, $A_{FB}^{\ell}$ and $A_{FB}^{\Lambda}$ by plotting them against the $C_X$ and $C_Y$ in Fig. \ref{Figpoint1to2}(b)-(e). The pink plane in each plot corresponds to the measured experimental range of the observable. The central SM values of $d\mathcal{B}/ds$, $F_{L}$, $A_{FB}^{\ell}$ and $A_{FB}^{\Lambda}$ are $0.24 \times 10^{-7}$, $0.54$, $0.10$ and $-0.31$, respectively, in $s \in [0.1,2]$ GeV$^2$ as shown in Table \ref{lambda-obs-1}.
		\item Fig. \ref{Figpoint1to2}(b) shows that the value of $d\mathcal{B}/ds$ varies from $0.27 \times 10^{-7}$ to $0.50 \times 10^{-7}$ inside the experimentally allowed region when $C_X$ and $C_Y$ are varied in their range constrained by the analysis of different $B-$ meson decays. Hence it can be inferred that the experimentally allowed region $0.22 \times 10^{-7} <d\mathcal{B}/ds < 0.27 \times 10^{-7}$ is excluded by the present analysis.
		\item Fig. \ref{Figpoint1to2}(c) represents the variation in the values of $F_L$ against each combination of $C_X$ and $C_Y$. It can be noticed that the value of $F_L$ approximately varies from  $0.39 - 0.77$ when we vary the values of $C_X$ and $C_Y$ in their allowed ranges. It means current constraints on the new WCs suggest that the value of $F_L$ is $0.39 < F_L < 0.77$ therefore, it excludes the experimental measured ranges that are above and below this range of $F_L$.
		\item  Fig. \ref{Figpoint1to2}(d) depicts that the value of $A^\ell_{FB}$ is not very sensitive to the combinations of NP couplings and its value remains close to its SM prediction which is $0.097$ (central value). Therefore, the larger experimental values of this observable can not be accommodated in light of the current constraints on new WCs.
		\item In case of $A^{\Lambda}_{FB}$, the combinations $(C_P^{\prime},\; C_V)$ and $(C_S,\; C_S^{\prime})$ (cyan and black dots) change the value of this observable from its SM predictions to some extent while for other combinations of new WCs this value remains close to the SM predictions and it can be seen in Fig. \ref{Figpoint1to2}(e). In this high-recoil bin, the maximum and the minimum values of $A^{\Lambda}_{FB}$ are found to be $-0.25$ and $-0.4$, respectively. Therefore, the positive value of this observable and the value greater than $-0.25$ seems to be excluded by the current constraints on these new WCs.
		\end{itemize}
	In short, the observables $d\mathcal{B}/ds$, $F_{L}$, $A_{FB}^{\ell}$ and $A_{FB}^{\Lambda}$ in high recoil region are very interesting to tell us more about the possible values of the new $VA$ and $SP$ couplings. Particularly, in this bin the study of the observables of $\Lambda_b$ decay do not put additional constraints on the range of WCs obtained from the analysis of $B-$meson decays.
	
\subsubsection{Low recoil region}

We have already mentioned in Sect. \ref{sec1} that the QCD uncertainties arising from the non-factorizable part of the amplitude has been neglected in our analysis. In case of the very well studied $B \to K^{\ast}\ell^{+}\ell^{-}$ decay, these effects are quite challenging theoretically and it is difficult to control them \cite{Bobeth:2017vxj} and this is even more daunting task for the $\Lambda_b \to \Lambda(\to p \pi)\ell^{+}\ell^{-}$ decay. Eventually, these effects may question the NP analysis in $q^2 < m^{2}_{J/\psi}$ region.  This is the reason why the most recent phenomenological study carried out on $\Lambda_b \to \Lambda \ell^{+}\ell^{-}$ in connection to $B-$physics anomalies restricts any consideration to the low-recoil region only \cite{Blake:2019guk}. Due to this reason, we have also attempted to see if we can accommodate the data of $d\mathcal{B}/ds$, $F_{L}$, $A_{FB}^{\ell}$ and $A_{FB}^{\Lambda}$ by using the above mentioned constraints on the new WCs.
 
\begin{figure}[h!]
 	\begin{center}
 		\begin{tabular}{ll}
 			\includegraphics[height=5cm,width=7cm]{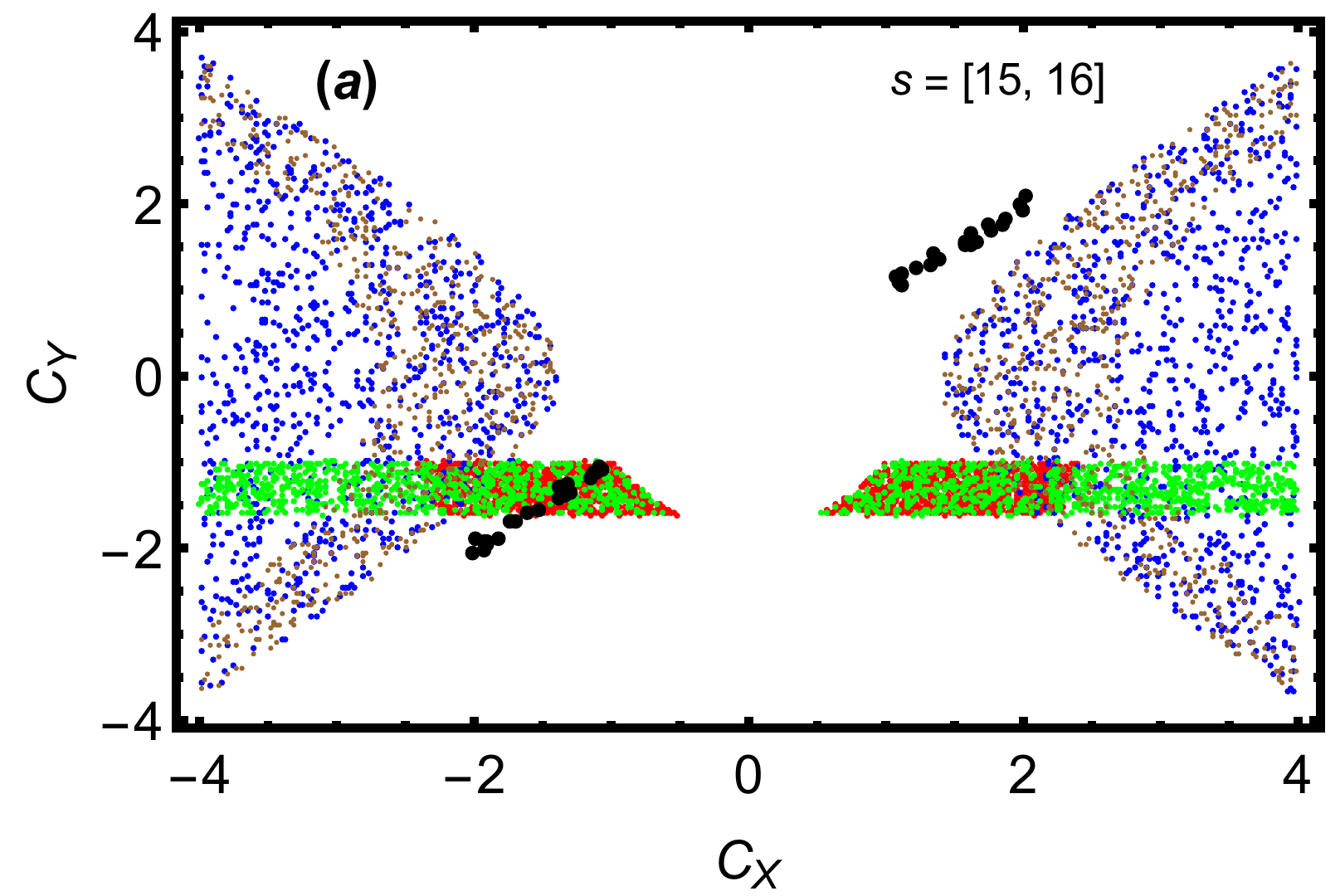}  & \ \ \includegraphics[height=5cm,width=7cm]{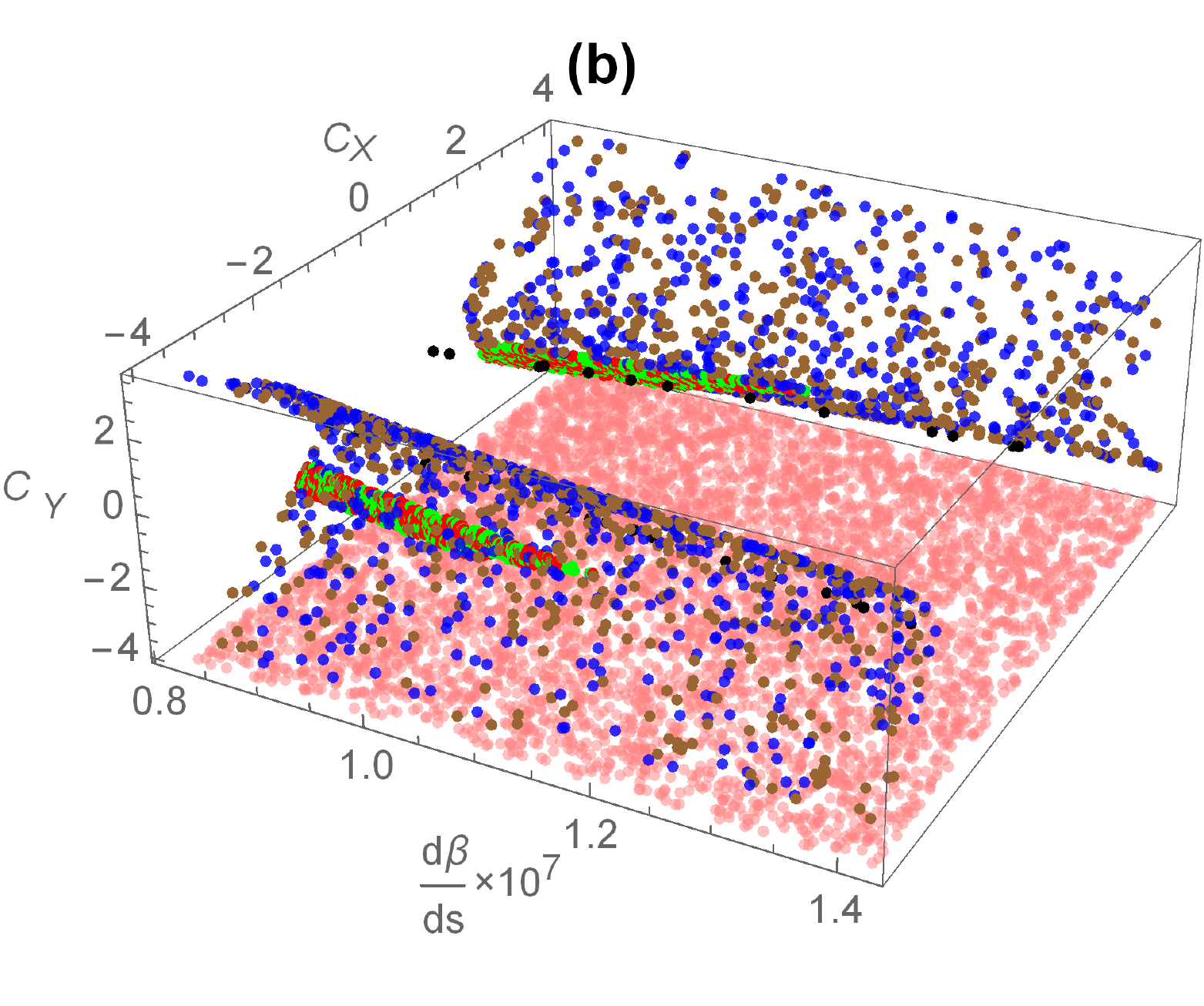} \\
 			\\
 			\includegraphics[height=5cm,width=7cm]{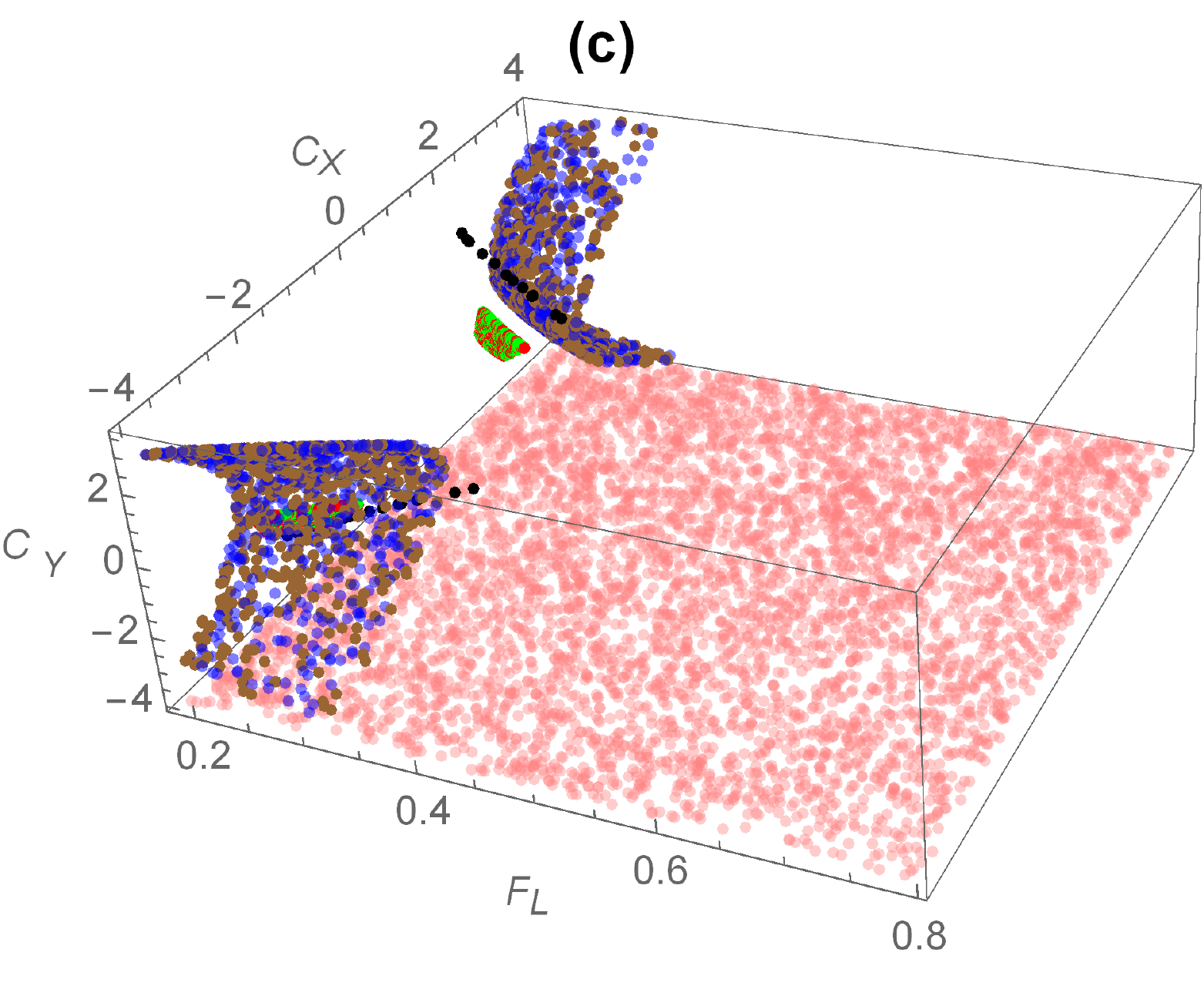} & \ \ \includegraphics[height=5cm,width=7cm]{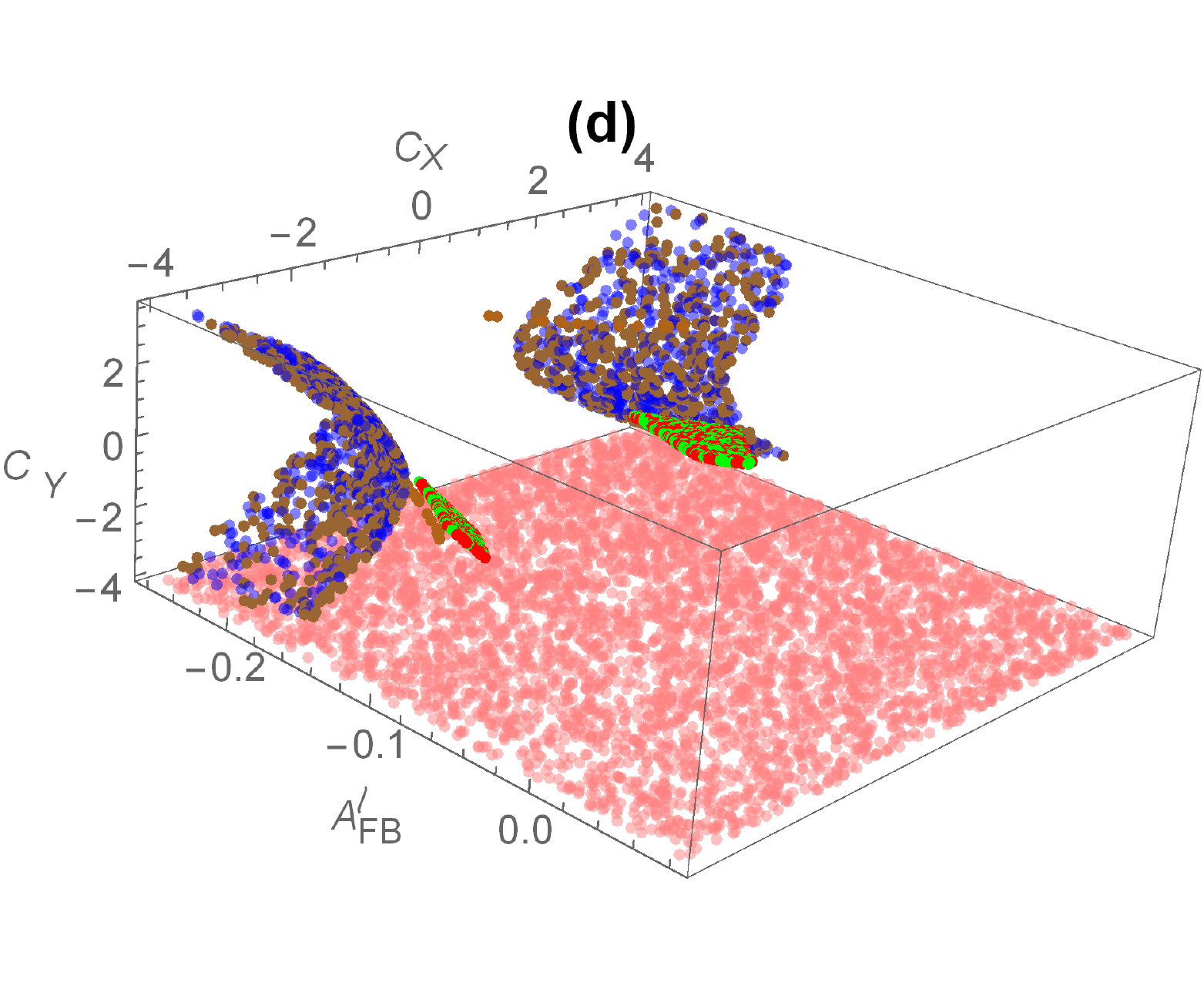} \\
 			\\
 			\includegraphics[height=5cm,width=7cm]{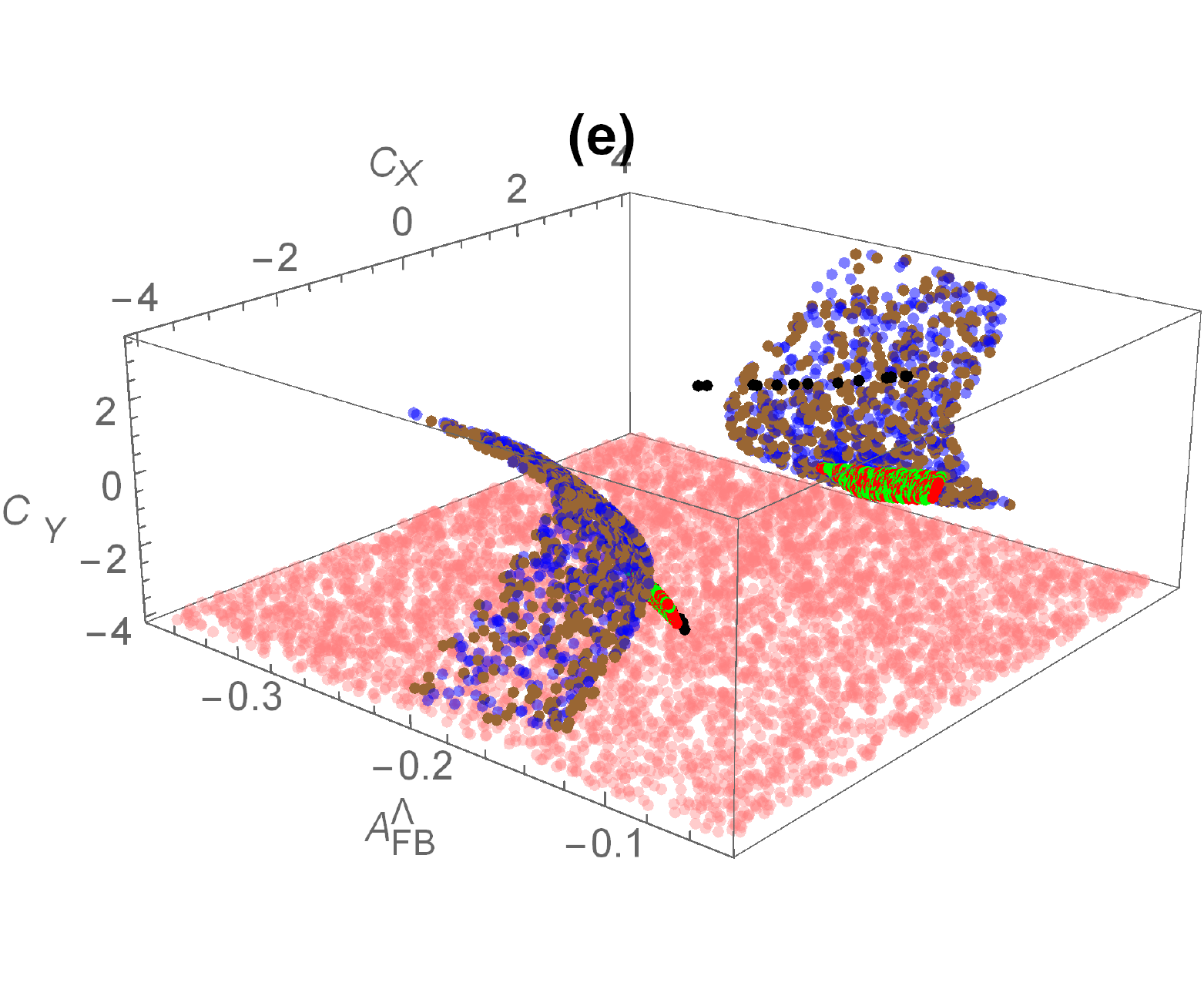}
 		\end{tabular}
 	\end{center}
 	\caption{Plot (a) shows the parametric space of $(C_X,C_Y)$ allowed from $B-$physics constraints on new WCs and which also satisfy the data of $d\mathcal{B}/ds$, $F_L$, $A_{FB}^{\ell}$ and $A_{FB}^{\Lambda}$, simultaneously, in the bin $s\in[15,16]$ GeV$^2$. (b)-(e) are the predictions of $d\mathcal{B}/ds$, $F_L$, $A_{FB}^{\ell}$ and $A_{FB}^{\Lambda}$ in the bin $s\in[15,16]$ GeV$^2$ against the WCs collected in (a). The legends are same as Fig. \ref{Figpoint1to2}. }\label{Fig15to16}
 \end{figure}

 For the low-recoil \textbf{bin $s \in [15,16]$ GeV$^2$}, one can make the following observations from Fig. \ref{Fig15to16}
	\begin{itemize}
		\item In this bin, the available data of all four observables could be accommodated by the combinations of $VA$ and $SP$ couplings given in Eq. (\ref{combleft}) with the exception of $(C_{P}^{\prime},C_V)$. However, when we try to accommodate the available data of the observables by these combinations the range of new WCs allowed by $B-$physics is further reduced and it can be seen from Fig. \ref{Fig15to16}(a).
		\item It can be noticed from the blue and brown dots in Fig. \ref{Fig15to16}(a) that for the combinations $(C_{S},C_{P}^{\prime})$ and $(C_{S}^{\prime},C_{P})$, the parametric space of $C_{S}$ is reduced to $[\pm4,\pm2.6]$ and when $C_S$ is close to its maximum value, i.e., $\pm4$, then full range of $C_{P}^{\prime}\in[+4,-4]$ is allowed. On the other hand when $C_{S}^{\prime}$ is close to $\pm4$, the $C_{P}^{\prime}$ is allowed to be varied between $[\pm3,\pm4]$ (see brown dots). It can be further seen that if $C_{S}^{(\prime)}$ reaches to $\pm2.6$ then $C_{P}^{\prime}$ goes to zero.
		\item  In the combinations of $(C_{S}^{(\prime)},C_V)$, the parametric space of $C_V$ is unchanged which can be seen from the red and green dots in Fig. \ref{Fig15to16}(a) while the parametric ranges of $C_{S}$, $C_{S}^{\prime}$ are reduced to $\pm4<C_S<\pm0.5$ and $\pm2<C_{S}^{\prime}<\pm0.5$ which can be observed from the red and green dots, respectively.
		\item The constraint on the combination $(C_S,C_{S}^{\prime})$ is already severe due to the condition $|C_{S}-C_{S}^{\prime}|\leq 0.1$ and it further narrow down when we try to explore the data of above mentioned observables. The new allowed range of this combination is between $\pm2$ to $\pm1$ with $|C_{S}-C_{S}^{\prime}|\leq 0.1$ condition. This can be seen from the black dots in Fig. \ref{Fig15to16}(a).

		\item By using these new allowed ranges of model independent WCs, we have predicted the values of all four observables in the bin $s \in [15,16]$ GeV$^2$ and plotted them in Fig. \ref{Fig15to16}(b-e). In this bin the SM value of $d\mathcal{B}/ds$ is $0.80\times10^{-7}$ and from Fig. \ref{Fig15to16}(b) we can see that this value varies between $(0.82-1.42)\times10^{-7}$ by varying the values of the combinations $(C_{S},C_{P}^{\prime})$, $(C_{S}^{\prime},C_{P})$ and $(C_{S},C_{S}^{\prime})$ in their allowed parametric space shown in Fig. \ref{Fig15to16}(a). It can also be noticed that the combinations $(C_{S},C_{P}^{\prime})$, $(C_{S}^{\prime},C_{P})$ and $(C_{S},C_{S}^{\prime})$ allow full experimental range of $d\mathcal{B}/ds$ which can be seen by blue, brown and black dots. In contrast to this, the combinations $(C_{S}^{(\prime)},C_V)$ allow the region $(0.82-1.15)\times10^{-7}$ of experimental measurements that is displayed by red and green dots in the same plot.
		\item Just like $d\mathcal{B}/ds$, the values of $F_L$ are also predicted in $s \in [15,16]$ GeV$^2$ bin and plotted in Fig. \ref{Fig15to16}(c). The SM value of $F_L$ is $0.45$ and it varies in the range $0.19 - 0.32$ for the combinations $(C_{S},C_{P}^{\prime})$, $(C_{S}^{\prime},C_{P})$ and $(C_S,C_S^{\prime})$ (see blue, brown and black dots). On the other hand choosing the combinations $(C_{S}^{(\prime)},C_V)$ the value of $F_L$ does not vary too much and predicted to be about $0.19-0.24$. This is displayed by the red and green color dots in the plot.
		\item Similarly, the values of $A^\ell_{FB}$ are predicted and plotted in Fig. \ref{Fig15to16}(d). The SM value of this observable in this bin is $-0.38$ and by using the values of combinations $(C_{S},C_{P}^{\prime})$ and $(C_{S}^{\prime},C_{P})$ it varies between $-0.15$ to $-0.26$ which is shown by blue and brown dots. For the combinations $(C_{S}^{(\prime)},C_V)$, the predicted range of the value of $A^\ell_{FB}$ is $-0.19$ to $-0.12$ (red and green dots).
		\item Fig. \ref{Fig15to16}(e) represents the predicted values of $A^\Lambda_{FB}$ by using the combinations of new WCs. The SM value of this observable in this bin is $-0.31$ and by using the allowed values of $(C_{S},C_{P}^{\prime})$ and $(C_{S}^{\prime},C_{P})$ combinations, it changes from $- 0.22$ to $-0.13$ (blue and brown dots) and by $(C_{S}^{(\prime)},C_V)$ combinations the range of the value of $A^\Lambda_{FB}$ is found to be $-0.17$ to $-0.12$ (red and green dots).
	\end{itemize}

	It is important to mention here that the values of the observables do not depend on the signs of the new WCs and it can be observed from Fig. \ref{Fig15to16}(b - e). However, when more precise data will be available from the Run 3 of the LHC, we expect that the values of the observables in this bin can be used to further constraining the new WCs particularly, the scalar type couplings. 
	
	\textbf{In $s \in [16,18]$ GeV$^2$ bin:}
	\begin{itemize}
		\item The SM values of the $d\mathcal{B}/ds$, $F_L$, $A_{FB}^{\ell}$ and $A_{FB}^{\Lambda}$ in this bin are $0.82\times10^{-7}$, $0.42$, $-0.38$ and $-0.29$, respectively. As we have mentioned earlier that we are interested only in those bins where all four and if not at least three observables could be accommodated simultaneously by using the parametric space of new WCs which is allowed by the $B-$physics data. In this particular bin we have found that only the data of two observables, $F_L$ and $A^\ell_{FB}$ favor our choice. Therefore, based on this fact we can say that this region is not good to predict the values of different angular observables. However, in future when more accurate data will be available in this range of $s$ it will be look for the possible NP effects due to these new WCs in the decay under consideration.
	\end{itemize}

	\begin{figure}[h]
		\begin{center}
			\begin{tabular}{ll}
				\includegraphics[height=5cm,width=7cm]{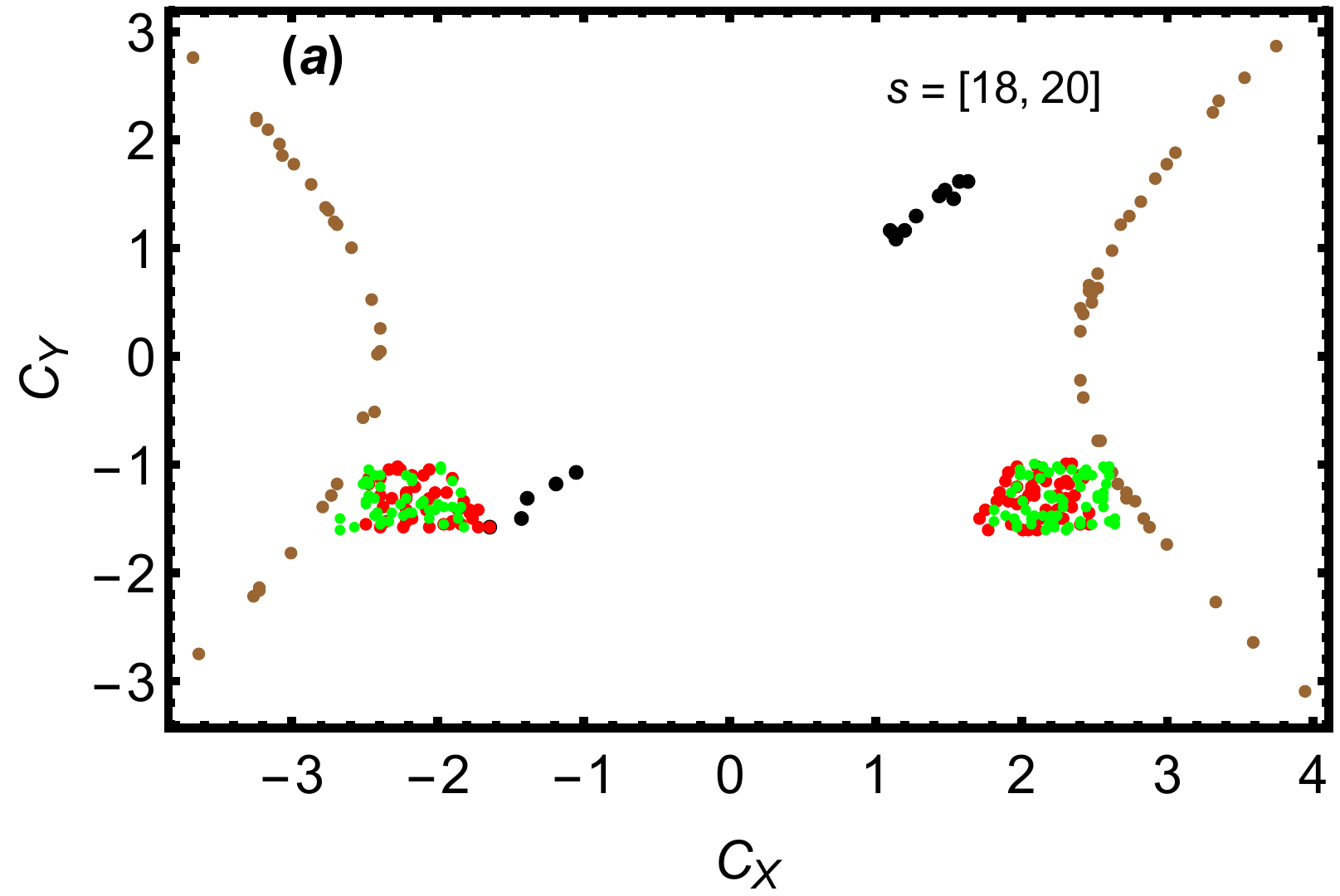}  & \ \ \includegraphics[height=5cm,width=7cm]{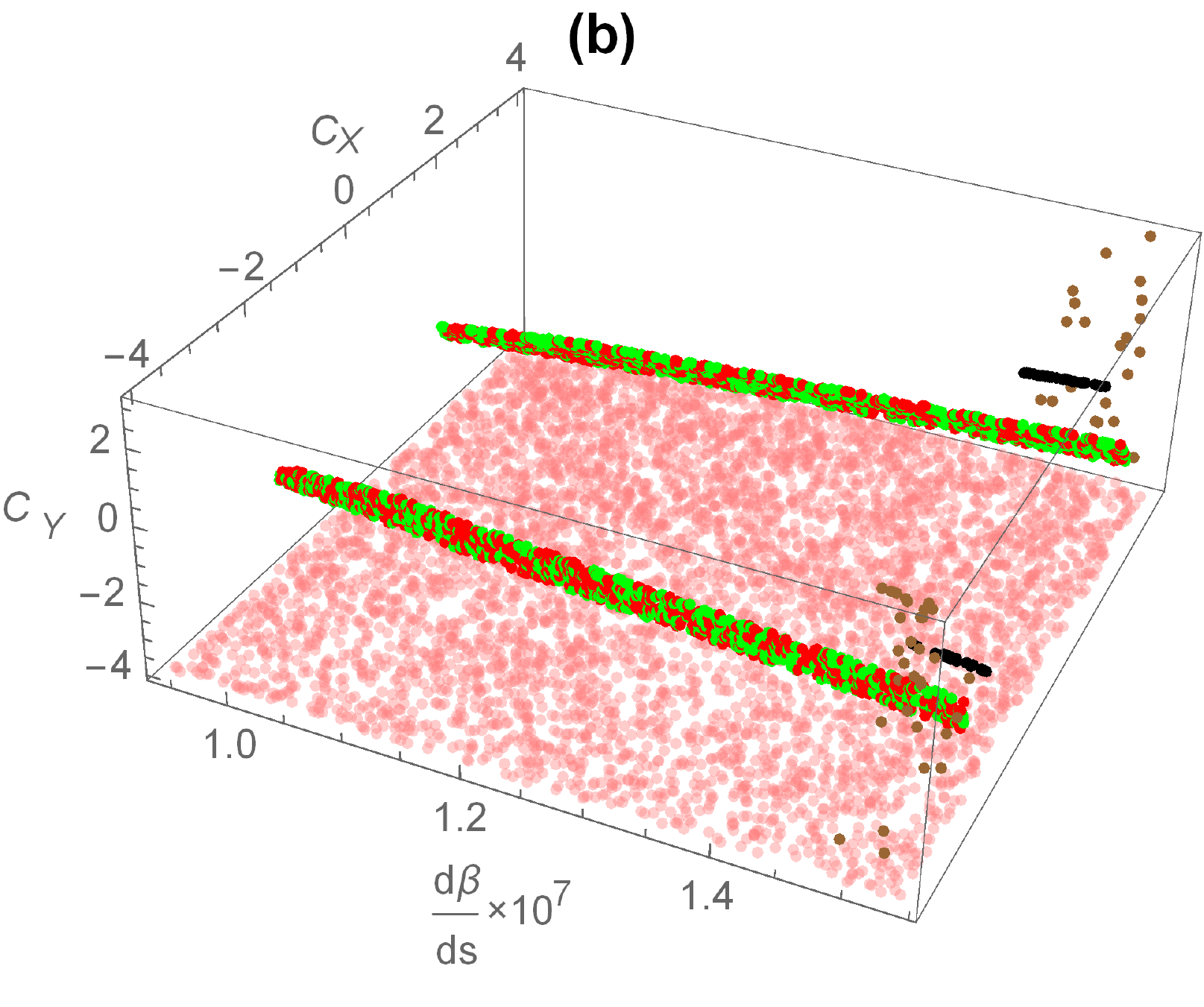} \\
				\\
				\includegraphics[height=5cm,width=7cm]{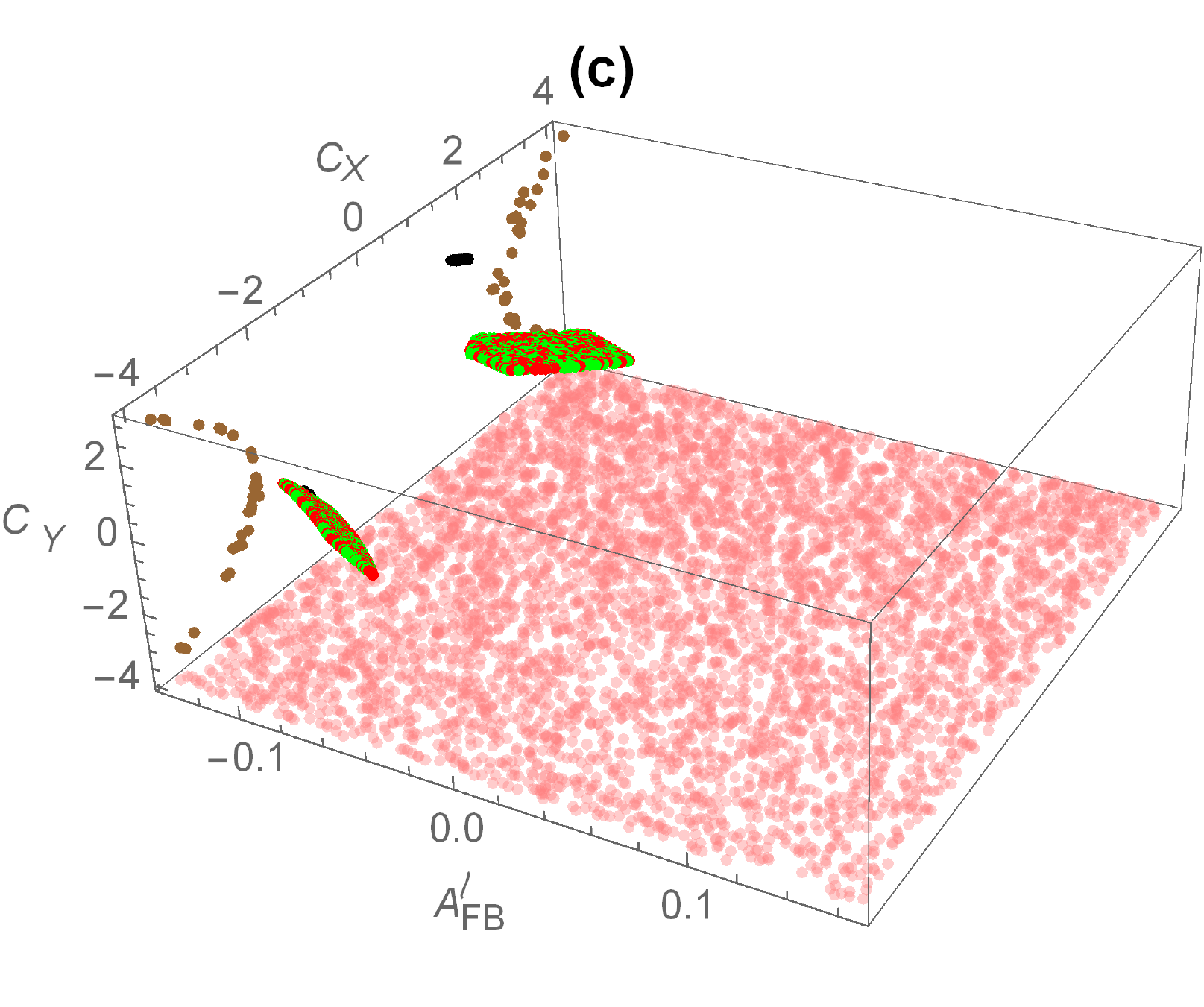} & \ \ \includegraphics[height=5cm,width=7cm]{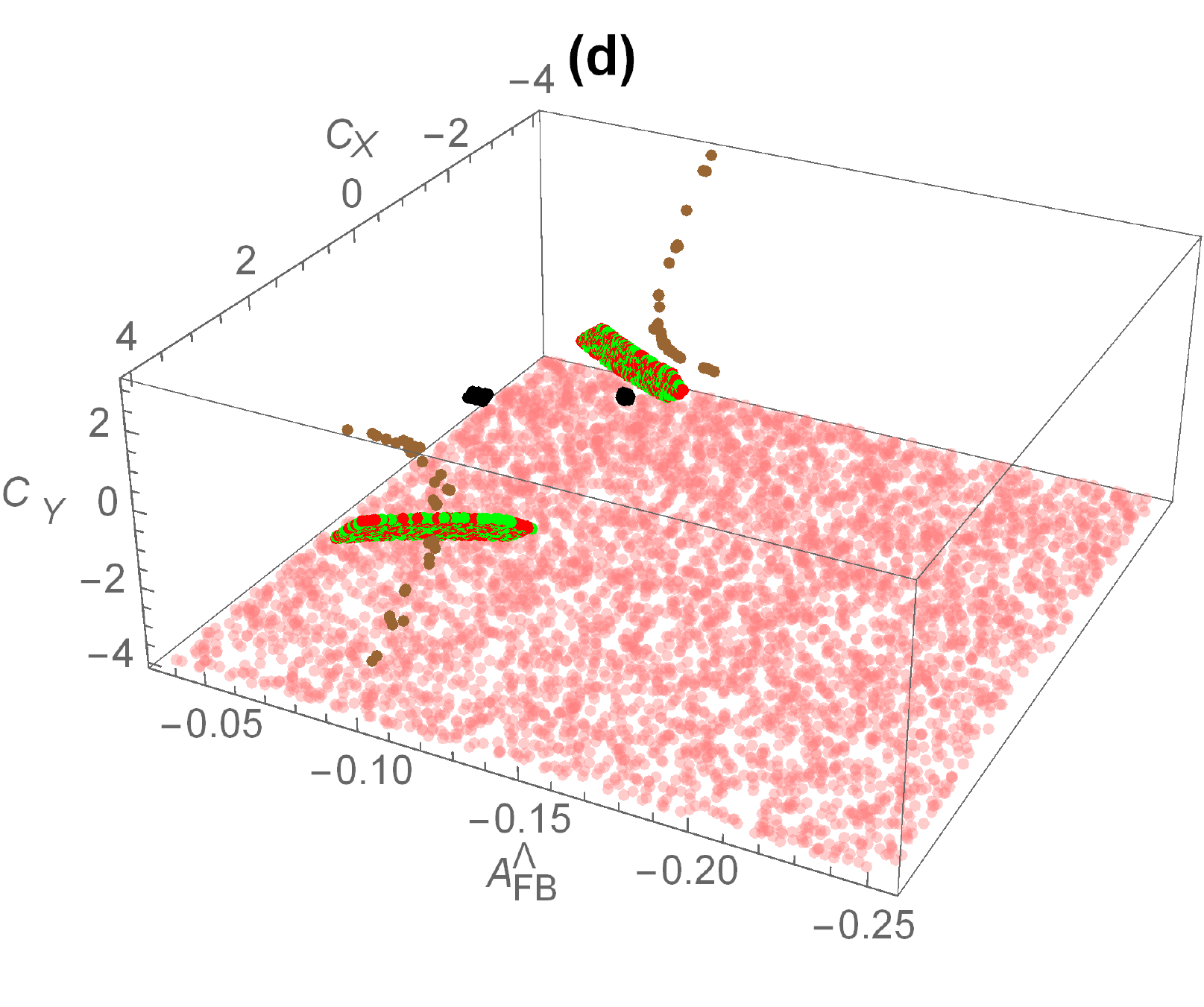}
			\end{tabular}
		\end{center}
		\caption{Plot (a) shows the parametric space of $(C_X,C_Y)$ allowed from $B$-physics constraints on the new WCs which also satisfy the data of $F_L$, $A_{FB}^{\ell}$ and $A_{FB}^{\Lambda}$ simultaneously in the bin $s\in[18-20]$ GeV$^2$. (b)-(d) are the predictions of $F_L$, $A_{FB}^{\ell}$ and $A_{FB}^{\Lambda}$ in the bin $s\in[18-20]$ GeV$^2$ against the WCs collected in (a). The legends are same as Fig. \ref{Figpoint1to2}. }\label{Fig18to20}
	\end{figure}
	\textbf{In $s \in [18-20]$ GeV$^2$ bin:}
	
	In this bin, by excluding the data of $F_L$ the available data of remaining three observables; i.e., $d\mathcal{B}/ds$, $A_{FB}^{\ell}$ and $A_{FB}^{\Lambda}$ could be accommodated simultaneously for the combinations of new WCs given in Eq. (\ref{combleft}). However, except $(C_S,C_{P}^{\prime})$ and $(C_{P}^{\prime},C_V)$ the other four combinations of new WCs can take care of LHCb data of these observables. We have also explored the case by including the data of $F_L$ but in this situation it is found that only one more observable can be accommodated at a time with it. Furthermore, as a result of satisfying the data, this bin provide more severe constraints on the new WCs that are still allowed by $B-$physics and it can be seen from Fig. \ref{Fig18to20}(a). The important observations in this case are the following:
	\begin{itemize}
		\item It can be noticed from the brown dots in Fig. \ref{Fig18to20}(a) that for the combination $(C_{S}^{\prime},C_{P})$ the parametric space of $C_{S}^{\prime}$ is reduced to $[\pm4,\pm2.4]$ and that of $C_{P}$ is $[+3,-3]$ with the severe parabolic condition $5.057(C_{S}^{\prime}-2.384)\simeq C_{P}^2$. On the other hand, for the other combinations the allowed region of $C_{S}^{(\prime)}$ is further narrow down (black and brown dots) while the region of $C_V$ still remains the same as restricted by $B-$physics data (red and green dots). Therefore, similar to the $s \in [15,16]$ GeV$^2$ bin, the $s \in [18-20]$ GeV$^2$ is also important for the scalar type new WCs.
		\item  By using the allowed ranges of new WCs shown in Fig. \ref{Fig18to20}(a) and discussed above, the predictions of the observables $d\mathcal{B}/ds$, $A_{FB}^{\ell}$ and $A_{FB}^{\Lambda}$ are plotted in Fig. \ref{Fig18to20}(b-d) in $s \in [18-20]$ GeV$^2$ bin. In this bin the SM value of $d\mathcal{B}/ds$ is $0.66\times10^{-7}$ and it can be observed from Fig. \ref{Fig18to20}(b) that by using the allowed values of $(C_{S}^{\prime}\; ,C_{P})$ and $(C_{S},\; C_{S}^{\prime})$ combinations, the value of $d\mathcal{B}/ds$ varies in a very small region of experimental range, i.e., roughly $(1.47 - 1.54)\times10^{-7}$ (brown and black dots, respectively). In contrast, the combinations $(C_{S}^{(\prime)},\; C_V)$ allow the full region of experimental values and in this case the range of the value is found to be $(0.94 - 1.54)\times10^{-7} $ that is displayed by the red and green dots in Fig. \ref{Fig18to20}(b).
		\item Similarly, the values of $A^\ell_{FB}$ are predicted and plotted in Fig. \ref{Fig18to20}(c). The SM value of $A^\ell_{FB}$ in this bin is $-0.32$ and by using the allowed values of the combinations $(C_{S}^{\prime},C_{P})$ and $(C_{S},C_{S}^{\prime})$ the value is found to be $\simeq-0.13$ which is shown by brown and black dots. For the combinations $(C_{S}^{(\prime)},C_V)$ the predicted range of $A^\ell_{FB}$ is $-0.07$ to  $-0.14$ which is plotted by red and green dots. One can further notice that the allowed range of combinations predict only the negative value of $A^\ell_{FB}$ which satisfies a very small experimental region of this observable, particularly, the positive experimental value of this observable is not possible to accommodate if we use the above six combinations.
		\item Fig. \ref{Fig18to20}(d) represents the predicted values of $A^\Lambda_{FB}$ by using the combinations of new WCs. The SM value of $A^\Lambda_{FB}$ in this particular bin is $-0.23$ and by using the allowed values of $(C_{S}^{\prime},C_{P})$ and $(C_{S},C_{S}^{\prime})$ combinations, this value is roughly to be $-0.10$ (see the brown and black dots) and by $(C_{S}^{(\prime)},C_V)$ combinations its values change from $-0.07$  to $-0.12$ which can be seen by the red and green dots. The higher experimental values of this observable is also not possible to be reproduced by using the current constraints on any possible combination of new WCs.
	\end{itemize}
	To summarize, in this particular bin, the analysis of above mentioned observables helps us to put the additional constraints on the new scalar type couplings. On the other hand, the parametric space of the vector type couplings does not change and remains to be the same as constrained by the $B-$physics data. Moreover, similar to the case of $s \in [15,16]$ GeV$^2$ bin, the numerical values of the observables in this case are also independent of the sign of new WCs.
	
\subsection{Lepton mass effects}
It has already been mentioned that we have calculated the expressions of different physical observables by taking the mass of final state leptons to be non-zero which is not the case in \cite{Das:2018sms} and hence our study can be easily extended to the semileptonic $\Lambda_b \to \Lambda \tau^{+}\tau^{-}$ case. Based on our analysis of $\Lambda_b \to \Lambda \mu^{+}\mu^{-}$ we find that $\mu-$mass effects in the angular observables of the four body decay $\Lambda_b \to \Lambda (\to p\pi^-)\mu^{+}\mu^{-}$ are not prominent consequently in the case of muons, the lepton mass terms can be safely ignored like in \cite{Das:2018sms}. For the sake of completeness, we have also calculated the values of angular observables in low-recoil region for the case of $\Lambda_b \to \Lambda (\to p\pi^-)\tau^{+}\tau^{-}$ decay in SM and also by considering the new WCs corresponding to the model independent approach.
\FloatBarrier
\begin{table}[h!]
	\caption{Observables with and without lepton mass for the decay $\Lambda_b \to \Lambda (\to p\pi)\tau^+ \tau^-$ in the SM and in different scenarios of NP couplings where the case $m_{\ell}\neq 0$ corresponds to $m_{\ell} =m_{\tau}$ in $s\in [15,20]$ GeV$^2$ bin. Scenario $VA-1$ corresponds to $C_V=-1.61$, $C_V^{\prime}=C_A=C_A^{\prime}=0$, $VA-2$ is the case when $C_V=-C_A=-1$, $C_V^{\prime}=C_A^{\prime}=0$ and $VA-3$ represent $C_V=-1.34$, $C_V^{\prime}=C_A=0$, $C_A^{\prime}=-0.4$. Similarly, in $SP-1$ case we have taken $C_S=C_P=-3$, $C_S^{\prime}=C_P^{\prime}=-3.1$ whereas $SP-2$ contain $C_S=3$, $C_S^{\prime}=2.9$ and $C_P=C_P^{\prime}=0$. Tensor couplings correspond to $C_T=0.72$ and $C_{T5}=0.2$. }\label{tau-obs-1}
	\begin{tabular}{l||l|l|l|l|l|l}
		\hline
		& $\frac{d\mathcal{B}}{ds}\times 10^{-7}$ & $ \ \ \ \ \ \  F_{L} \ \ \ \  $ & $ \ \ \ \ A_{FB}^{\ell }$ \ \ \ & $ \ \ \ \ \ A_{FB}^{\Lambda } \ \ $ & \  $%
		\ \ \ A_{FB}^{\ell \Lambda }$ \ \ \ & $ \ \ \ \ \ F_{T}$ \ \  \\ \hline\hline
		$SM_{m_{\ell =0}}$ & $ \ \ \ \  0.75$ & $ \ \ \ \ 0.41$ & $ \ -0.35$ & $  \ -0.26$ &  \ \ \ $0.14$ & $ \ \ \ 0.60$
		\\
		$SM_{m_{\ell \neq 0}}$ & \ \ \ \ $0.53$ & \ \ \ \  $0.35$ & \ $-0.13$ & \ $-0.26$ & \ \ \ $0.06$ & \ \ \ $0.65
		$ \\ \hline
		$VA-1_{m_{\ell =0}}$ & \ \ \ \ $0.52$ & \ \ \ \ $0.40$ & \ $-0.28$ & \ $-0.26$ & \ \ \  $0.12$ & \ \ \ $0.60$
		\\
		$VA-1_{m_{\ell \neq 0}}$ & \ \ \ \  $0.37$ & \ \ \ \ $0.35$ & \ $-0.11$ & \ $-0.26$ & \ \ \ $0.04$ & $%
		\ \ \	0.65$ \\ \hline
		$VA-2_{m_{\ell =0}}$ &  \ \ \ \ $0.42$ & \ \ \ \ $0.41$ & \  $-0.35$ &  \ $-0.26$ & \ \ \ $0.14$ & \ \ \  $0.59$
		\\
		$VA-2_{m_{\ell \neq 0}}$ & \ \ \ \ $0.29$ & \ \ \ \  $0.35$ & \ $-0.13$ & \ $-0.26$ & \ \ \ $0.06$ & $%
		\ \ \	0.65$ \\ \hline
		$VA-3_{m_{\ell =0}}$ & \ \ \ \  $0.52$ & \ \ \ \ $0.41$ & \ $-0.33$ & \ $-0.28$ & \ \ \ $0.13$ & \ \ \ $0.59$
		\\
		$VA-3_{m_{\ell \neq 0}}$ & \ \ \ \ $0.36$ & \ \ \ \  $0.35$ & \ $-0.12$ & \ $-0.28$ & \ \ \  $0.05$ & $%
		\ \ \	0.65$ \\ \hline
		$SP-1_{m_{\ell =0}}$ & \ \ \ \ $0.91$ & \ \ \ \  $0.22$ & \ $-0.20$ & \ $-0.15$ & \ \ \  $0.14$ & \ \ \ $0.59$
		\\
		$SP-1_{m_{\ell \neq 0}}$ & \ \ \ \ $0.68$ & \ \ \ \  $0.15$ &  \  $-0.05$ & \   $-0.12$ & \ \ \ $0.03$ & $%
		\ \ \ 0.85$ \\ \hline
		$SP-2_{m_{\ell =0}}$ & \ \ \ \ $0.87$ & \ \ \ \ $0.16$ & \ $-0.14$ & \ $-0.11$ & \ \ \ $0.06$ & \ \ \ $0.84$
		\\
		$SP-2_{m_{\ell \neq 0}}$ & \ \ \ \ $0.90$ & \ \ \ \ $0.21$ & \ $-0.08$ & \ $-0.16$ & \ \ \ $0.03$ & $%
		\ \ \ 0.79$ \\ \hline
		$T^\prime _{m_{\ell =0}}$ & \ \ \ \ $0.79$ & \ \ \ \  $0.39$ & \ $-0.33$ & \ $-0.25$ &  & \ \ \
		\\
		$T^\prime _{m_{\ell \neq 0}}$ & \ \ \ \  $0.54$ & \ \ \ \ $0.34$ & \ $-0.13$ & \ $-0.26$ &  &
	\end{tabular}
\end{table}
\FloatBarrier
\FloatBarrier
\begin{table}
	\caption{Observables by taking the massive and massless $\tau$ in $\Lambda_b \to \Lambda (\to p\pi)\tau^+ \tau^-$ decay in the SM and also in different NP scenarios. Description of couplings is similar to Table \ref{tau-obs-1}. }\label{tau-obs-2}
	\begin{tabular}{l||l|l|l|l|l|l}
		\hline
		& $Y_{3sc}\times 10^{-3}$ & $Y_{4sc}\times 10^{-2}$ & \ \ \ \ \ $Y_{2}$ \ \ \ \  & \ \ \ \ $\alpha _{\theta
			_{\Lambda }}$ \ \ \  & \ \ \ \ \ $\alpha _{\theta _{\ell }}$ \ \ \ & \ \ \ \ \ $\alpha _{\theta _{\ell
			}}^{\prime }$ \ \ \ \\ \hline\hline
			$SM_{m_{\ell =0}}$ & \ \ \ \  $0.02$ & \ \ \ $-0.96$   & \ \ \ $0.02$ & \  $-0.82$ & \ $-0.15$ & \ \ $-0.67$
			\\
			$SM_{m_{\ell \neq 0}}$ & \ \ \ \ $0.00$ & \ \ \ $-0.19$ & \ \ \ $0.00$ & \  $-0.82$ & \ $-0.03$ & $%
			\ 	-0.26$ \\ \hline
			$VA-1_{m_{\ell =0}}$ & \ \ \ \ $0.02$ & \ \ \  $-0.94$ & \ \ \  $0.02$ & \ $-0.82$ & \ $-0.15$ & \ $-0.54
			$ \\
			$VA-1_{m_{\ell \neq 0}}$ & \ \ \ \ $0.00$ & \ \ \  $-0.18$ & \ \ \ $0.00$ & \ $-0.82$ & \ $-0.03$ & $%
			\ 	-0.21$ \\ \hline
			$VA-2_{m_{\ell =0}}$ & \ \ \ \ $0.03$ & \ \ \  $-1.00$ & \ \ \ $0.02$ & \ $-0.82$  & \ $-0.16$ & \ $-0.67
			$ \\
			$VA-2_{m_{\ell \neq 0}}$ & \ \ \ \ $0.00$ & \ \ \ $-0.19$ & \ \ \ $0.00$ & \ $-0.82$ & \ $-0.03$ & $%
			\ 	-0.26$ \\ \hline
			$VA-3_{m_{\ell =0}}$ & \ \ \ \ $0.02$ & \ \ \ $-1.00$ & \ \ \ $0.02$ & \ $-0.87$ & \ $-0.16$ & \ $-0.62
			$ \\
			$VA-3_{m_{\ell \neq 0}}$ & \ \ \ \ $0.00$ & \ \ \ $-0.19$ & \ \ \ $0.00$ & \ $-0.87$ & \ $-0.03$ & $%
			\	-0.24$ \\ \hline
			$SP-1_{m_{\ell =0}}$ & \ \ \ \ $0.01$ & \ \ \ $-0.10$ & \ \ \ $0.01$ & \ $-0.46$ & \ $-0.05$ & \ $-0.21
			$ \\
			$SP-1_{m_{\ell \neq 0}}$ & \ \ \ \ $0.00$ & \ \ \ $-0.07$ & \ \ \ $0.00$ & \  $-0.32$ & \  $-0.01$ & $%
			\ -0.05$ \\ \hline
			$SP-2_{m_{\ell =0}}$ & \ \ \ \ $0.01$ & \ \ \ $-0.50$ & \ \ \ $0.01$ & \ $-0.34$ & \ $-0.03$ & \ $-0.13
			$ \\
			$SP-2_{m_{\ell \neq 0}}$ & \ \ \ \ $0.00$ & \ \ \ $-0.11$ & \ \ \ $0.00$ & \ $-0.51$ & \ $-0.01$ & \ $%
			-0.09$%
		\end{tabular}%
	\end{table}%
	\FloatBarrier
	\FloatBarrier
	\begin{table}[h!]
		\caption{Observables by taking the massive and massless $\tau$ in $\Lambda_b \to \Lambda (\to p\pi)\tau^+ \tau^-$ decay in the SM and also in different NP scenarios. Description of couplings is similar to Table \ref{tau-obs-1}. }\label{tau-obs-3}
		\begin{tabular}{ll|l|l|l|l|l|l}
			\hline
			& \multicolumn{1}{||l|}{ $  \ \ \ \ \ \alpha _{\xi } \ \ \ \ $} & $\alpha _{\xi }^{\prime
			}\times 10^{-3}$ & $ \ \ \ \ \alpha _{U} \ \ \ $ & $ \ \ \ \alpha _{L} \ \ \ \ $ & $ \ \ \ \ \mathcal{P}_{3} \ \ \ $ & $ \ \ \ \ \mathcal{P}_{8} \ \ \ \ \ $ & $ \ \ \ \ \mathcal{P}_{9} \ \ \ \ $
			\\ \hline\hline
			\multicolumn{1}{l||}{$SM_{m_{\ell =0}}$} & \ \ $-0.32$ & $ \ \ -0.20$ &  \ $-0.79$ & $%
			\ 	-0.84$ &\ \  $0.38$ & \ $-0.94$ & \ $-0.60$ \\
			\multicolumn{1}{l||}{$SM_{m_{\ell \neq 0}}$} & \ \  $-0.13$ & \ $-0.07$  & \ $-0.82$ &
			\ $-0.83$ & \ \ $0.14$ &  \ $-0.35$ & \ $-0.69$ \\ \hline
			\multicolumn{1}{l||}{$VA-1_{m_{\ell =0}}$} & \ \  $-0.26$ &  \ $-0.29$ & \ $-0.79$ & $%
			\ -0.84$ & \ \ $0.31$ & \ $-0.76$ & \ $-0.60$ \\
			\multicolumn{1}{l||}{$VA-1_{m_{\ell \neq 0}}$} & \ \ $-0.11$ & \ $-0.11$ & \ $-0.82$
			& \ $-0.83$ & \ \ $0.12$ & \ $-0.29$ & \ $-0.69$ \\ \hline
			\multicolumn{1}{l||}{$VA-2_{m_{\ell =0}}$} & \ \ $-0.32$ & \ $-0.27$ & \ $-0.79$ & $%
			\ -0.84$ & \ \  $0.37$ & \ $-0.94$ & $ \ -0.59$ \\
			\multicolumn{1}{l||}{$VA-2_{m_{\ell \neq 0}}$} & \ \ $-0.13$ & \ $-0.01$ & \ $-0.82$
			& \ $-0.83$ & \ \ $0.15$ & \ $-0.36$ & \ $-0.58$ \\ \hline
			\multicolumn{1}{l||}{$VA-3_{m_{\ell =0}}$} & \ \ $-0.25$ & \  $-4.69$ & \ $-0.85$ & $%
			\ 	-0.89$ & \ \ $0.34$ & \ $-0.87$ & \ $-0.64$ \\
			\multicolumn{1}{l||}{$VA-3_{m_{\ell \neq 0}}$} & \ \ $-0.10$ & \  $-1.78$ & \ $-0.87$
			& \ $-0.88$ & \ \ $0.13$ & \  $-0.33$ & \ $-0.73$ \\ \hline
			\multicolumn{1}{l||}{$SP-1_{m_{\ell =0}}$} & \ \ $-0.16$ & \ $-0.12$ & \ $-0.21$ & $%
			\ 	-0.26$ & \ \ $0.20$ & \ $-0.52$ & \ $-0.34$ \\
			\multicolumn{1}{l||}{$SP-1_{m_{\ell \neq 0}}$} & \ \  $-0.05$ & \  $-0.03$ &  \   $-0.14$ & $%
			\  -0.14$ &  \ \ $0.06$ & \  $-0.14$ & $ \  -0.27$ \\ \hline
			\multicolumn{1}{l||}{$SP-2_{m_{\ell =0}}$} & \ \   $-0.12$ & \ $-0.08$ & \ $-0.13$ & $%
			\ 	-0.16$ & \ \ $0.15$ & \ $-0.37$ & \ $-0.24$ \\
			\multicolumn{1}{l||}{$SP-2_{m_{\ell \neq 0}}$} & $ \ \ -0.07$ & $ \ -0.05$ & $ \ -0.28$
			& $ \ -0.30$ & $ \ \ 0.09$ & $ \  -0.21$ & $ \ -0.44$%
		\end{tabular}%
	\end{table}
	\FloatBarrier
	By using the central values of the FFs the calculated values of different physical observables are listed in Tables \ref{tau-obs-1} - \ref{tau-obs-3}. From the first row in each of the Tables \ref{tau-obs-1} - \ref{tau-obs-3}, one can notice that the magnitude of the SM values of observables $A_{FB}^{\ell}$, $F_T$, $Y_{4sc}$, $\alpha_{\theta_{\ell}}$, $\alpha^{\prime}_{\theta_{\ell}}$, $\alpha_{\xi}$, $\alpha_{\xi}^{\prime}$ and $P_8$ are increased due to the non-zero $\tau$'s mass whereas the values of $A_{FB}^{\Lambda}$ and $\alpha_{\theta_\Lambda}$ do not receive tauon mass effect. Similar effects can also be noticed in these tables when we include the $VA$ (rows 2 - 4) and $T$ (row 7) couplings along with the SM couplings. It is noticed that in the case of $SP-1$, the values of $F_L, \; A_{FB}^{\ell},\; A_{FB}^{\Lambda},\; F_T,\; Y_{4sc},\; \alpha_{\theta_{\Lambda}},\; \alpha_{\theta_{\ell}},\; \alpha_{\theta_{\ell}}^{\prime}, \alpha_{\xi},\; \alpha_{\xi}^{\prime},\; \alpha_L,\; P_8$ and $P_9$ increase when tau mass effects are included whereas $d\mathcal{B}/ds,\; A_{FB}^{\ell \Lambda},\; \alpha_U$ and $P_3$ values decrease.
	
	For the second possibility of scalar couplings $(SP-2)$, the values of the observables $A_{FB}^{\ell},$, $Y_{4sc}$, $\alpha_{\theta_\ell}$, $\alpha_{\theta_{\ell}}^{\prime}$, $\alpha_\xi$, $\alpha^\prime_\xi$, $\alpha_U$ and $P_8$ are increased due to the $\tau$ mass while the values of the observables $d\mathcal{B}/ds$, $F_L$, $A^\Lambda_{FB}$, $A^{\ell\Lambda}_{FB}$, $F_T$, $\alpha_{\theta_\Lambda}$, $\alpha_L$, $P_3$ and $P_9$ are decreased. However, there are no effects of non-zero tauon mass observed in the calculated values of $Y_2$ and $Y_{3sc}$ in both scenarios of SP couplings (c.f. sixth row of Table \ref{tau-obs-3}). In short, we have found that the effects of $\tau$ mass are significantly large in the decay $\Lambda_b \to \Lambda (\to p\pi)\tau^+ \tau^-$ therefore, in contrast to the case of muons, to pursue the NP effects in the angular observables of $\Lambda_b \to \Lambda (\to p\pi)\tau^+ \tau^-$ decay it is indispensable to include the lepton mass terms in the expressions of different physical observables. Consequently, it is worthy to derive the expressions by taking the lepton mass to be non zero in the semileptonic decay $\Lambda_b \to \Lambda (\to p\pi)\ell^+ \ell^-$ where $\ell=e,\mu,\tau$.
	
	\subsection{Most favorable pair of Wilson coefficients}	
\begin{figure}[h!]
		\begin{center}
			\begin{tabular}{ll}
				\includegraphics[height=5cm,width=7cm]{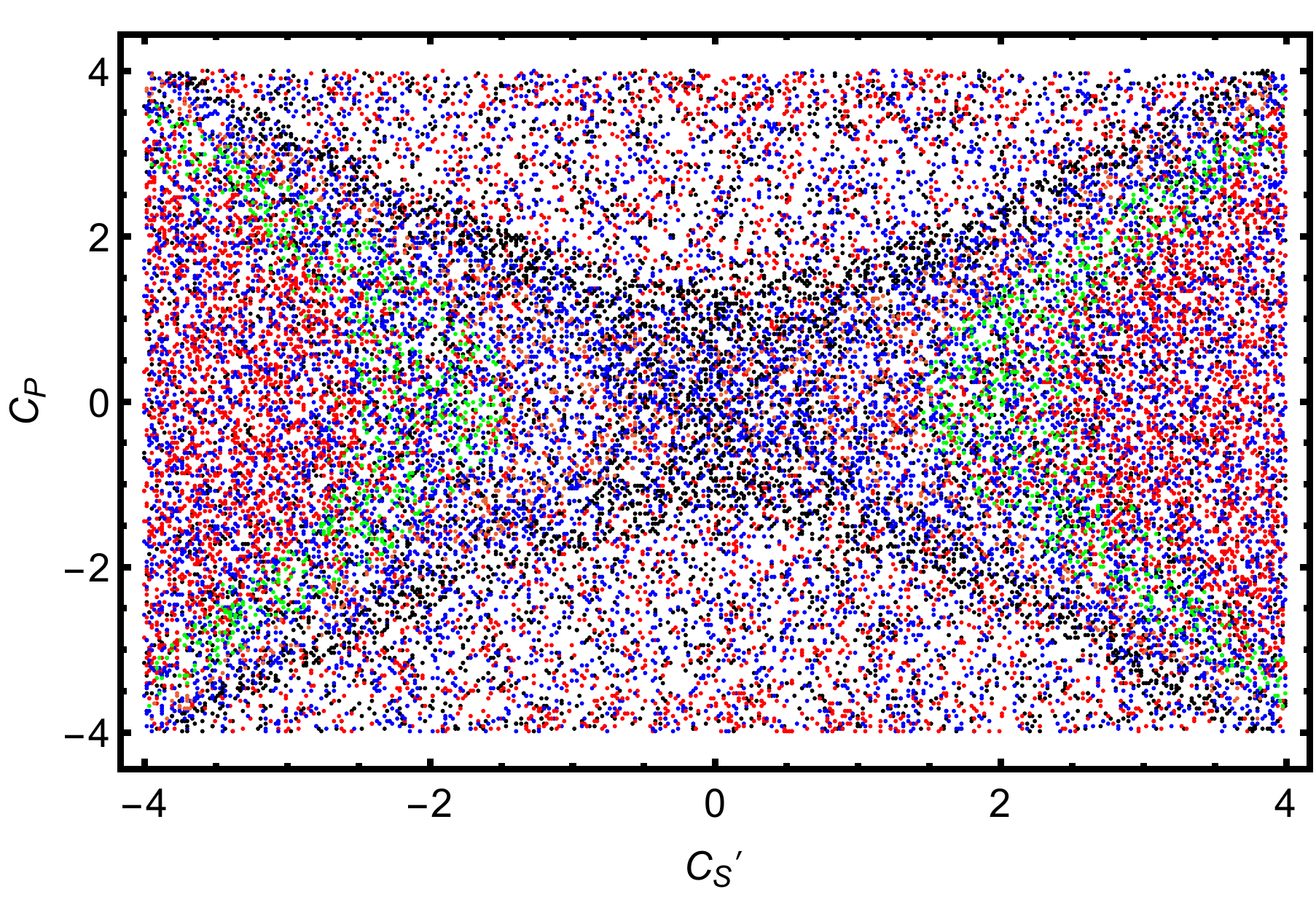} 
			\end{tabular}
		\end{center}
		\caption{$(C_S^{\prime},C_P)$ is the most favorable pair of Wilson coefficients. Green, Black, red and blue colors denote  $\frac{d\mathcal{B}}{ds},\; F_L, \; A_{FB}^{\ell}$ and $A_{FB}^{\Lambda}$ respectively. }\label{MFpair}
	\end{figure}

	We have extracted the most favorable pair of new WC's which is $(C_S^{\prime},C_P)$ as shown in Fig. {\ref{MFpair}}. This pair satisfy individual observables $\frac{d\mathcal{B}}{ds},\; F_L, \; A_{FB}^{\ell}$ and $A_{FB}^{\Lambda}$ in large-recoil bin $s\in [0.1,2]$ GeV$^2$ and low-recoil bins $s\in [15,16]$ GeV$^2$, $[16,18]$ GeV$^2$ and $[18,20]$ GeV$^2$. It means that it can satisfy all experimental data available for $\Lambda_b$ decay observables respecting the $B-$physics constraints. Density of plots in Fig. {\ref{MFpair}} shows how the respective parametric space of $(C_S^{\prime},C_P)$ is favorable by these decay observables.
	
\section{Summary and Conclusion\label{conclusion}}
	Here, our investigation on NP is not restricted to operators that already appear in the SM; i.e., the four-fermion ones built out of $V-A$ $b \to s$ and vector/axial leptonic currents - but actually also considers (pseudo)scalar and tensor operators, and the $V+A$ combination for the $b \to s$ current. In literature it is known as the model independent approach. By taking the most general weak interaction Hamiltonian, we have discussed the impact of new $VA$, $SP$ and tensor $T$ couplings on above mentioned physical observables in $\Lambda_b \to \Lambda (\to p\pi^-)\mu^{+}\mu^{-}$ decay. Most of them are the ratios of angular coefficients and hence show little dependence on the uncertainties involved in the calculation of hadronic FFs therefore, in future these observables can serve as a tool to look for the imprints of currents which are beyond the SM physics.

	
	First of all we have plotted $d\mathcal{B}/ds$, $A_{FB}^{\ell}$, $A_{FB}^{\Lambda}$,  $A_{FB}^{\ell \Lambda}$, $F_{L,T}$, $Y_{2,\;3sc,\;4sc}$, $\mathcal{P}_{3,\; 8,\; 9}$ and $\alpha^{(\prime)}_{i}$ where, $i = \theta_\ell,\; \theta_{\Lambda},\; \xi,\; L, \; U$ by considering the non-zero mass for the final state leptons. We deduce that in case of  $\mu$ as final state lepton the mass effects are negligible at low-recoil for all the observables - but mildly effected some observables, such as $F_T$, $\alpha_{\theta_{\ell}}$, $Y_2$, $Y_{4sc}$ and $\mathcal{P}_9$ in high-recoil region. Therefore, based on our analysis it can be inferred that one can safely ignore the muon mass terms in the expressions of the lepton helicity fractions as was done in \cite{Das:2018sms}. By using the following ranges of new WCs that are allowed by the $B-$physics data with the global fit sign suggestions
     \begin{eqnarray}
     C_V &=& [-1.61,-1], \hspace{1cm} C_V^{\prime}=0, \hspace{1cm} C_A=1, \hspace{1cm} C_A^{\prime}=-0.4 \notag \\
     C_S^{(\prime)} &=& C_P^{(\prime)}=[-4,4], \hspace{0.7cm} C_T=0.72, \hspace{0.4cm} C_{T5}=0.2, \label{Csranges}
     \end{eqnarray}
  	we have the following findings:
  	\begin{itemize}
  		\item There is a mismatch between the SM predictions and the LHCb data of $d\mathcal{B}/ds$ particularly in high $s$ region. Only new $VA$ couplings in three low $s$ bins, $[0.1,2]$ GeV$^2$, $[2,4]$GeV$^2$ and $[6,8]$GeV$^2$ are able to accommodate this data.
  		\item The SM values of $F_L$ and $A^\Lambda_{FB}$ fall within the error bars of the LHCb data in all bins except in $s\in[16,18]$ GeV$^2$ where $VA$, $SP$ and $T$ couplings are also unable to accommodate the data.
  		\item The data of $A^\ell_{FB}$ deviates from the SM values in high $s$ region and only $SP$ couplings show some promising effect to accommodate it.
  	\end{itemize}

	It is also observed that the zero-crossing of observables shift towards high $s$ region when new $VA$ couplings are introduced in addition to the SM WCs which is not the case for the $SP$ couplings. In case of $\alpha_{\theta_{\ell}},\alpha_U$ and $\alpha_L$ when we include $VA$ couplings their values are modified slightly for the combination $C_V=-1.34,C_A^{\prime}=-0.4$, $C_V^{\prime}=C_A=0$. However, the data of $A_{FB}^{\ell}$, particularly in high $s$ region favors the new $VA$ couplings in comparison to the SM couplings alone. Now compared to $VA$ and $T$ couplings, the constraints are less stringent on $SP$ WCs therefore, their influence on above discussed observables is more prominent. Also, the data of $d\mathcal{B}/ds$, $A_{FB}^{\ell}$ and $A_{FB}^{\Lambda}$ in some high $s$ bins favor the $SP$ couplings. In case of the WCs corresponding to the tensor currents the value of $A_{FB}^{\Lambda}$ in high $s$ region come closer to the experimental value which is neither the case for the SM nor for any other NP couplings. Hence, one can deduce that there is not a single new coupling among $VA$, $SP$ and $T$ operators that can simultaneously accommodate the whole available LHCb data of these four observables in all $s$ bins.
	
	To overcome this difficulty, we have also examined the impact of $C_V$, $C_S$, $C_S^{\prime}$, $C_P$ and $C_P^{\prime}$ by considering these couplings in pairs $(C_X , C_Y)$ where $X, Y = V, S, S^{\prime}, P, P^{\prime}$. The goal was to explore their allowed parametric space to see whether the combinations of these new couplings satisfy the available LHCb data for all the four observables in large-recoil bin $s\in [0.1,2]$ GeV$^2$ and in low-recoil bins $s\in[15,16],\; [16,18]$ GeV$^2$ and $s\in[18,20]$ GeV$^2$ or not. We observed that at large-recoil $s \in [0.1,2]$ GeV$^2$ region, the measured values of $d\mathcal{B}/ds$, $F_L$, $A_{FB}^{\ell}$ and $A_{FB}^{\Lambda}$ could be justified simultaneously by taking the combinations given in Eq. (\ref{combleft}) while the ranges of WCs mentioned in Eq. (\ref{Csranges}) remain unchanged. At low-recoil, for the bin $s\in [15,16]$ GeV$^2$, these combinations also accommodate LHCb data for all four observables simultaneously but the ranges of $SP$ WCs mentioned in Eq. (\ref{Csranges}) are further constrained while the ranges of $VA$ WCs remain the same. It reflects that these two bins can provide a good opportunity to search for the NP when more accurate data will be available from the Run 3 of the LHC. In the $s \in [16,18]$ GeV$^2$ bin, the measurements of all observables cannot be accommodated by any of the combinations of NP WCs. In this region, only a few combinations satisfy the measured ranges of $F_L$ and $A_{FB}^{\Lambda}$ simultaneously but none of them satisfy $A_{FB}^{\ell}$. In last bin of low-recoil region; i.e.,  $s\in[18,20]$ GeV$^2$ we tried to accommodate the data of $A_{FB}^{\ell}$, $A_{FB}^{\Lambda}$ and $d\mathcal{B}/ds$ simultaneously and found that it is still possible for the several combinations of NP WCs. However, doing so for these three observables we got more severe constraints on the the $SP$ WCs as compared to the one in $s \in [15,16]$ GeV$^2$ bin while the range of $VA$ WCs remains the same as allowed by the $B$-physics data. Finally, by using the allowed parametric space of these WCs constrained by the data of above mentioned observables in $\Lambda_b$ decays, we predicted the values of these four observables in their corresponding bins and find that they could be potentially measured in future at the LHCb and other planned experiments. 
	
	It has already been discussed that in this study we have calculated the expressions of the lepton helicity fractions in the presence of lepton mass therefore, this analysis can be easily extended to the $\Lambda_b \to \Lambda \tau^{+}\tau^{-}$ decays. Doing so, we found that the non-zero mass of tauons significantly modify the values of different physical observables both in the SM as well as in the presence of new WCs. Finally, by considering the uncertainties involved in the FFs and other input parameters, we have calculated the values of all the 19 observables for $\Lambda_b \to \Lambda (\to p\pi^-)\tau^{+}\tau^{-}$ decay.

\newpage
\appendix
\section*{Appendix}{\label{App}}
\subsection*{ Helicity amplitudes for hardronic part}
It is well known fact that the helicity formalism provide a convenient way to find the projections of $\Lambda_b \to \Lambda$ matrix elements on the direction of the polarization of virtual gauge boson \cite{Gutsche:2013pp,Boer:2014kda,Yan:2019tgn}. In case of vector currents, we can write
\begin{equation}
H_V^t(s_{\Lambda_b},\; s_{\Lambda})=\varepsilon^{\mu*}_t \langle \Lambda(P_{\Lambda},s_{\Lambda})\vert \bar{s}\gamma^{\mu}b \vert \Lambda(P_{\Lambda_b},s_{\Lambda_b})  \rangle \label{HelV}
\end{equation}
where $\epsilon^{\mu*}_t$ denotes the time-like polarization of the virtual gauge boson and $s_{\Lambda_b}$ and $s_{\Lambda}$ are the spin-projections of initial and final state baryons on the $z-$axis in their rest frames, respectively. Using Eq. (\ref{HelV1}) and the kinematical relations defined in \cite{Das:2018sms,Gutsche:2013pp} along with $\varepsilon^{\mu}_t = \frac{1}{\sqrt{s}}\left(q_0,\;0,\;0,\;-\left|\vec{q}\right|\right)$, the non-zero helicity components for time-like polarization from Eq. (\ref{HelV1}) read
\begin{eqnarray}
H^t_V(+1/2,+1/2)=H^t_V(-1/2,-1/2)=f_V^t(s)\frac{m_{\Lambda_b}-m_{\Lambda}}{\sqrt{s}}\sqrt{s_+}.\label{HelVNZ}
\end{eqnarray}
In case of longitudinal polarization $\varepsilon^{\mu*}_0 = \frac{1}{\sqrt{s}}\left(\left|\vec{q}\right|,\;0,\;0,\;-q_0\right)$, the corresponding helicity amplitude becomes
\begin{equation}
H_V^0(s_{\Lambda_b},\; s_{\Lambda})=\varepsilon^{\mu*}_0 \langle \Lambda(P_{\Lambda},s_{\Lambda})\vert \bar{s}\gamma^{\mu}b \vert \Lambda(P_{\Lambda_b},s_{\Lambda_b})  \rangle \label{LV1}
\end{equation}
and using Eq. (\ref{HelV1}), the non-zero longitudinal components for vector current will take the form
\begin{equation}
H^0_V(+1/2,+1/2) = H^0_V(-1/2,-1/2) = f_V^0(s)\frac{m_{\Lambda_b}+m_{\Lambda}}{\sqrt{s}}\sqrt{s_-}.\label{LHelVNZ}
\end{equation}
with $s_{-} = (m_{\Lambda_b}-m_{\Lambda})^2-s$.
Likewise, for the transverse polarization $\varepsilon^{\mu*}_{\pm} = \frac{1}{\sqrt{2}}\left(0,\;\pm1,\;i,\;-q_0\right)$
\begin{equation}
H_V^{\pm}(s_{\Lambda_b},\; s_{\Lambda})=\varepsilon^{\mu*}_{\pm} \langle \Lambda(P_{\Lambda},s_{\Lambda})\vert \bar{s}\gamma^{\mu}b \vert \Lambda(P_{\Lambda_b},s_{\Lambda_b})  \rangle,
\end{equation}
the corresponding non-zero helicity components are
\begin{equation}
H^{+}_V({-1/2,+1/2}) = H^-_V(+1/2,-1/2)=-f_V^{\perp}(s)\sqrt{2s_-}\label{THelVNZ}.
\end{equation}

In case of axial-vector currents, from Eq. (\ref{HelA1}) the corresponding non-zero components for time-like, longitudinal and transverse polarizations of virtual boson are
\begin{eqnarray}
H^t_A(+1/2,+1/2) &=& -H^t_A(-1/2,-1/2) = f_A^t(s)\frac{m_{\Lambda_b}+m_{\Lambda}}{\sqrt{s}}\sqrt{s_-},\label{HAT2}\\
H^0_A(+1/2,+1/2) &=&-H^0_A(-1/2,-1/2)=f_A^0(s)\frac{m_{\Lambda_b}-m_{\Lambda}}{\sqrt{s}}\sqrt{s_+},\label{HAL2}\\
H^{+}_A(-1/2,+1/2) &=& -H^-_A(+1/2,-1/2)=-f_A^{\perp}(s)\sqrt{2s_+}.\label{THA2}
\end{eqnarray}

For the dipole operators $i\bar{s}q_\nu \sigma^{\mu\nu}b$ and  $i\bar{s}q_\nu \sigma^{\mu\nu}\gamma_5 b$, the respective transition matrix elements are defined in Eq. (\ref{HelT1}) and in this particular case the corresponding non-zero helicity components for different polarizations of virtual boson are
\begin{eqnarray}
H^0_T(+1/2,+1/2) &=& H^0_T(-1/2,-1/2) = -f_T^0(s)\sqrt{ss_-} \\
H^0_{T_5}(+1/2,+1/2)&=& H^0_{T_5}({-1/2,-1/2})=f_{T_5}^0(s)\sqrt{s s_+}.
\end{eqnarray}
The helicity amplitude corresponding to the tensor current i.e., $\bar{s}\sigma^{\mu\nu}b$ becomes
 \begin{eqnarray}
H_{T^\prime}^{m,n}(s_{\Lambda_b,}s_{\Lambda})= \varepsilon^{\mu*}_m \epsilon^{\mu*}_n \langle \Lambda(P_{\Lambda},s_{\Lambda})\vert \bar{s}i\sigma^{\mu\nu} b \vert \Lambda(P_{\Lambda_b},s_{\Lambda_b})  \rangle , \label{HT1}
\end{eqnarray}
where $m,n=t,0,\pm$. Using the expression of $\langle \Lambda\left(P_{\Lambda},\;s_{\Lambda}\right)\left|\bar{s}i \sigma^{\mu\nu} b\right| \Lambda_b\left(P_{\Lambda_b},\;s_{\Lambda_b}\right)\rangle$ from \cite{Das:2018sms} (c.f. Eq. (C.7)), the non-zero components for virtual bosons's time-like, longitudinal, transverse and the possible combination of these polarization becomes
\begin{eqnarray}
H^{0t}_{T^\prime}(+1/2,+1/2) &=& H^{0t}_{T^\prime}(-1/2,-1/2) = -f_{T}^0(s) \sqrt{s_-} \label{HTNZ1} \\
H^{+t}_{T^\prime}(-1/2,+1/2) &=& H^{-t}_{T^\prime}(+1/2,-1/2) = f_{T_5}^{\perp}(s) \frac{m_{\Lambda_b}+m_{\Lambda}}{\sqrt{s}} \sqrt{2s_-} \label{HTNZ2} \\
H^{+0}_{T^\prime}(-1/2,+1/2) &=& H^{-0}_{T^\prime}(+1/2,-1/2) = f_{T_5}^{\perp}(s) \frac{m_{\Lambda_b}-m_{\Lambda}}{\sqrt{s}} \sqrt{2s_+}\label{HTNZ3}\\
H^{+-}_{T^\prime}(+1/2,+1/2) &=& -H^{+-}_{T^\prime}(-1/2,-1/2) = -f_{T_5}^{0}(s) \sqrt{s_+} \label{HTNZ4}
\end{eqnarray}
The remaining components can be obtained by using the relation $H^{m,n}_{T^\prime}(s_{\Lambda_b},s_{\Lambda})=-H^{n,m}_{T^\prime}(s_{\Lambda_b},s_{\Lambda})$.

In case of the scalar and pseduo-scalar currents, the corresponding helicity amplitudes along with their non-zero components can be obtained from Eqs. (\ref{S1}) and (\ref{S2}) and these are
\begin{eqnarray}
H^t_S(+1/2,+1/2) &=& H^t_S(-1/2,-1/2) = f_V^t(s)\frac{m_{\Lambda_b}-m_{\Lambda}}{m_b} \sqrt{s_+},\label{HS1}\\
H^t_P(+1/2,+1/2) &=& -H^t_P(-1/2,-1/2) = -f_A^t(s)\frac{m_{\Lambda_b}-m_{\Lambda}}{m_b} \sqrt{s_-}\label{HSP1}.
\end{eqnarray}

\subsection{Helicity amplitudes for lepton-part}
Just like the hadronic part, the leptonic helicity amplitudes the scalar $(S)$, pseudo-scalar $(P)$, vector $(V)$ and axial-vector $(A)$ currents having non-zero contribution are \cite{Yan:2019tgn}
\begin{eqnarray}
L_S(+1/2,+1/2) &=& -L_S(-1/2,-1/2)= s\beta_{\ell}, \quad\quad\quad\quad\quad\quad\quad  L_P(+1/2,+1/2)=L_P(-1/2,-1/2)=-s , \notag \\
L_V^{\pm}(+1/2,+1/2)&=& L_V^{\mp}(-1/2,-1/2)=\mp \sqrt{2} m_{\ell}\sin{\theta_{\ell}},\quad\quad\quad \; L_V^{0}(+1/2,+1/2)= -L_V^{0}(-1/2,-1/2)=- 2m_{\ell} \cos\theta_{\ell}, \notag \\
L_V^{0}(+1/2,-1/2 )&=& L_V^{0}(-1/2,+1/2) =s\sin\theta_{\ell},\quad\quad\quad\quad\quad\quad \; L_V^{+}(+1/2,-1/2)= -L_V^{-}(-1/2,+1/2)= \frac{s}{\sqrt{2}}(1-\cos\theta_{\ell}), \notag \\
L_V^{-}(+1/2,-1/2) &=& -L_V^{+}(-1/2,+1/2)= \frac{s}{\sqrt{2}}(1+\cos\theta_{\ell}), \quad\quad L_A^t(+1/2,+1/2)=L_A^t(-1/2,-1/2)=-2m_{\ell}, \notag \\
L_A^{+}(+1/2,-1/2) &=& L_A^{-}(-1/2,+1/2)=\frac{sv}{\sqrt{2}}(1-\cos\theta_{\ell}) , \quad\quad\quad L_A^{-}(+1/2,-1/2)=L_A^{+}(-1/2,+1/2)=\frac{sv}{\sqrt{2}}(1+\cos\theta_{\ell}), \notag \\
L_A^{0}(+1/2,-1/2) &=& -L_A^{0}(-1/2,+1/2)= sv\sin\theta_{\ell}.
\end{eqnarray}
Similarly, from the tensor current $\sigma_{\mu\nu}(1\mp \gamma_5)$ the non-zero leptonic helicity amplitudes are
\begin{eqnarray}
L_{T^{\prime}}^{t,\pm}(+1/2,+1/2) &=& -L_{T^{\prime}}^{t,\pm}(-1/2,-1/2)=\mp\frac{s}{\sqrt{2}}\sin\theta_{\ell},\; L_{T^{\prime}}^{t,+}(-1/2,+1/2) = -L_{T^{\prime}}^{t,-}(+1/2,-1/2) = \sqrt{2}(1-\cos\theta_{\ell})m_{\ell} \notag \\
L_{T^{\prime}}^{t,-}(-1/2,+1/2) &=& -L_{T^{\prime}}^{t,+}(+1/2,-1/2) = \sqrt{2}m_{\ell}(1+\cos\theta_{\ell}), \; L_{T^{\prime}}^{t,0}(+1/2,+1/2) = -L_{T^{\prime}}^{t,0}(-1/2,-1/2)=- s \cos\theta_{\ell} \notag \\
L_{T^{\prime}}^{+,-}(+1/2,+1/2) &=& L_{T^{\prime}}^{+,-}(-1/2,-1/2) = -s \cos\theta_{\ell}\beta_{\ell},  \; L_{T^{\prime}}^{\pm,0}(+1/2,+1/2) = L_{T^{\prime}}^{\pm,0}(-1/2,-1/2) =-\frac{s}{\sqrt{2}}\sin\theta_{\ell}\beta_{\ell}, \label{lpart} \\
L_{T^{\prime}}^{t,0}(-1/2,+1/2) &=& L_{T^{\prime}}^{t,0}(+1/2,-1/2) =2m_{\ell}\sin\theta_{\ell},\; L_{T5^{\prime}}^{t,\pm}(+1/2,+1/2) = -L_{T5^{\prime}}^{t,\pm}(-1/2,-1/2) = \pm \frac{s}{\sqrt{2}}\sin\theta_{\ell}\beta_{\ell}\notag\\
L_{T5^{\prime}}^{t,0}(+1/2,+1/2) &=& L_{T5^{\prime}}^{t,0}(-1/2,-1/2)= s \cos\theta_{\ell}\beta_{\ell}, \; L_{T5^{\prime}}^{+,-}(+1/2,+1/2) = -L_{T5^{\prime}}^{+,-}(-1/2,-1/2)= s \cos\theta_{\ell} ,\notag \\
L_{T5^{\prime}}^{-,+}(+1/2,-1/2) &=& L_{T5^{\prime}}^{-,+}(-1/2,+1/2)= 2m_{\ell} \sin\theta_{\ell}, \; L_{T5^{\prime}}^{+,0}(+1/2,-1/2) = -L_{T5^{\prime}}^{-,0}(-1/2,+1/2)= {\sqrt{2}} (1+\cos\theta_{\ell})m_{\ell}, \notag \\
L_{T5^{\prime}}^{\pm,0}(+1/2,+1/2) &=& -L_{T5^{\prime}}^{\pm,0}(-1/2,-1/2)= \frac{s}{\sqrt{2}} \sin\theta_{\ell},\; L_{T5^{\prime}}^{-,0}(+1/2,-1/2) = L_{T5^{\prime}}^{+,0}(-1/2,+1/2)= -{\sqrt{2}} (1- \cos\theta_{\ell})m_{\ell}\notag .
\end{eqnarray}
It can be seen that by setting the lepton mass $m_{\ell}$ to be zero, we can obtain the relations given in \cite{Das:2018sms} .

In terms of the hadronic and leptonic currents, the square of amplitudes corresponding to different currents can be assembled as
\begin{eqnarray}
\vert M_{VA}\vert^2 &=& \frac{1}{4}\sum_{s_{\Lambda_b},s_{\Lambda}} \sum_{s_{\Lambda}^{\prime}} \sum_{s_{\ell_1},s_{\ell_2}} \sum_{m,n} \sum_{m^{\prime},n^{\prime}} H^m(s_{\Lambda_b},s_{\Lambda}) H^{n*}(s_{\Lambda_b},s_{\Lambda^{\prime}}) g_{m,m^{\prime}} g_{n,n^{\prime}} L^{m^{\prime}}(s_{\ell_1},s_{\ell_2}) L^{n^{\prime^*}}(s_{\ell_1},s_{\ell_2}) \Gamma^{\prime}(s_{\Lambda},s_{\Lambda}^{\prime}),\notag
\end{eqnarray}
where $\lambda= (m_{\Lambda_b}^2-m_{\Lambda}^2-s)^2+4sm_{\Lambda}^2$ and $L^{m^{\prime},\;n^{\prime}}$ are the helicity amplitudes for the leptonic part given in Eq. (\ref{lpart}). Likewise
\begin{eqnarray}
\vert M_{SP}\vert^2 &=&\frac{1}{4}\sum_{s_{\Lambda_b},s_{\Lambda}} \sum_{s_{\Lambda}^{\prime}} \sum_{s_{\ell_1},s_{\ell_2}}  H(s_{\Lambda_b},s_{\Lambda}) H^*(s_{\Lambda_b},s_{\Lambda^{\prime}}) L(s_{\ell_1},s_{\ell_2}) L^*(s_{\ell_1},s_{\ell_2}) \Gamma^{\prime}(s_{\Lambda},s_{\Lambda}^{\prime})\notag\\
\vert M_{T^{\prime}}\vert^2 &=& \sum_{s_{\Lambda_b},s_{\Lambda}} \sum_{s_{\Lambda}^{\prime}} \sum_{s_{\ell_1},s_{\ell_2}} H^{mn}(s_{\Lambda_b},s_{\Lambda}) H^{rs*}(s_{\Lambda_b},s_{\Lambda^{\prime}}) g_{mm^{\prime}} g_{nn^{\prime}}
g_{rr^{\prime}} g_{ss^{\prime}} L^{m^{\prime}n^{\prime}}(s_{\ell_1},s_{\ell_2}) L^{r^{\prime}s^{\prime}*}(s_{\ell_1},s_{\ell_2}) \Gamma^{\prime}(s_{\Lambda},s_{\Lambda}^{\prime}) \notag\\
M_{VA}M^*_{SP}+h.c.&=& \frac{1}{4}\sum_{s_{\Lambda_b},s_{\Lambda}} \sum_{s_{\Lambda}^{\prime}} \sum_{s_{\ell_1},s_{\ell_2}} \left[  H^m(s_{\Lambda_b},s_{\Lambda}) H^*(s_{\Lambda_b},s_{\Lambda^{\prime}}) L^{m^{\prime}}(s_{\ell_1},s_{\ell_2}) L^*(s_{\ell_1},s_{\ell_2})+h.c. \right] \Gamma^{\prime}(s_{\Lambda},s_{\Lambda}^{\prime})\notag\\
M_{VA}M^*_{T^{\prime}}+h.c.&=&\frac{1}{2}\sum_{s_{\Lambda_b},s_{\Lambda}} \sum_{s_{\Lambda}^{\prime}} \sum_{s_{\ell_1},s_{\ell_2}} \left[ H^m(s_{\Lambda_b},s_{\Lambda}) H^{*rs}(s_{\Lambda_b},s_{\Lambda^{\prime}})  L^{m^{\prime}}(s_{\ell_1},s_{\ell_2}) L^{r^{\prime}s^{\prime}*}(s_{\ell_1},s_{\ell_2}) +h.c. \right] \Gamma^{\prime}(s_{\Lambda},s_{\Lambda}^{\prime})\notag\\
M_{SP}M^*_{T^{\prime}}+h.c. &=& \frac{1}{2}\sum_{s_{\Lambda_b},s_{\Lambda}} \sum_{s_{\Lambda}^{\prime}} \sum_{s_{\ell_1},s_{\ell_2}} \left[ H(s_{\Lambda_b},s_{\Lambda})  H^{mn*}(s_{\Lambda_b},s_{\Lambda^{\prime}}) L(s_{\ell_1},s_{\ell_2}) L^{r^{\prime}s^{\prime}*}(s_{\ell_1},s_{\ell_2}) + h.c. \right] \Gamma^{\prime}(s_{\Lambda},s_{\Lambda}^{\prime})
\end{eqnarray}
where summation over the repeated indices is understood.

The various angular coefficients appearing in Eq. (\ref{eq6}) are defined as
\begin{eqnarray}
K_{1ss} &=& \vert \widetilde{C}_{9}^+ \vert^2 H_{1V}^{0,+}+ \vert \widetilde{C}_{9}^- \vert^2 H_{1A}^{0,+} + \vert C_{7}^+ \vert^2 H_{1T}^{0,+} +\vert C_{7}^- \vert^2 H_{1T5}^{0,+} + \vert \widetilde{C}_{10}^+ \vert^2 H_V^{t,0,+} + \vert \widetilde{C}_{10}^- \vert^2 H_A^{t,0,+} \notag \\
&+& \Re\left[C_7^+ C_9^{+*}\right] H_{4(V,T)} + \Re\left[C_7^- C_9^{-*}\right] H_{4(A,T5)}, \notag \\
K_{1cc} &=& \vert \widetilde{C}_{9}^+ \vert^2 H_{2V}^{0,+}+ \vert \widetilde{C}_{9}^- \vert^2 H_{2A}^{0,+} + \vert \widetilde{C}_{10}^+\vert^2 H_{3V}^{t,+} + \vert \widetilde{C}_{10}^- \vert^2 H_{3A}^{t,+} + \left( \vert C_{7}^+\vert^2+\vert C_{7}^-\vert^2 \right) H_{2T}^{0,+}  \notag \\
&+& 2 \Re\left[C_7^+ C_9^{+*}\right] H_{5(V,T)}+ 2 \Re\left[C_7^- C_9^{-*}\right] H_{5(A,T5)}, \notag \\
K_{2ss} &=& \alpha \Re \left[ \widetilde{C}_{10}^+ \widetilde{C}_{10}^{-*} \right] \left(2 \beta^2 H_{A,V}^{t,t}(+1/2,+1/2) + 2 v^2 H_{A,V}^{0,0}(+1/2,+1/2)+v^2H_{A,V}^{+,+}(-1/2,+1/2) \right) \notag \\
&+& \alpha \Re \left[ \widetilde{C}_{9}^+ \widetilde{C}_{9}^{-*} \right] H_{4(A,V)}+ \alpha \Re \left[ C_{7}^+ C_{7}^{-*}\right] H_{4(T,T5)} + \alpha \Re \left[ C_{7}^+ \widetilde{C}_{9}^{-*}\right] H_{4(A,T)} +\alpha \Re \left[ C_{7}^- \widetilde{C}_{9}^{+*}\right] H_{4(V,T5)}, \notag \\
K_{2cc} &=& 2\alpha \Re \left[ \widetilde{C}_{9}^+ \widetilde{C}_{9}^{-*} \right]H_{5(A,V)}+2\alpha \Re \left[ \widetilde{C}_{10}^+ \widetilde{C}_{10}^{-*} \right] \left( \beta^2 H_{A,V}^{t,t}(+1/2,+1/2) +v^2H_{A,V}^{+,+}(-1/2,+1/2) \right) \notag \\
&+& 2\alpha \left( \Re \left[ C_{7}^+ C_{7}^{-*} \right] H_{5(T,T5)}+\Re \left[ C_{7}^+ \widetilde{C}_{9}^{-*} \right] H_{5(A,T)}+\Re \left[ C_{7}^- \widetilde{C}_{9}^{+*} \right] H_{5(V,T5)} \right), \notag \\
K_{1c} &=& -2v \left(\Re \left[ C_{7}^+ \widetilde{C}_{10}^{-*} \right] H_{A,T}^{+,+}(-1/2,+1/2) +\Re \left[ C_{7}^- \widetilde{C}_{10}^{+*} \right] H_{V,T5}^{+,+}(-1/2,+1/2) \right) \notag \\
&-& 2v \left[ \Re \left( \widetilde{C}_{9}^+ \widetilde{C}_{10}^{-*} \right] H_{A,V}^{+,+}(-1/2,+1/2)+\Re \left[ \widetilde{C}_{9}^- \widetilde{C}_{10}^{+*} \right] H_{A,V}^{+,+}(-1/2,+1/2) \right), \notag \\
K_{2c} &=& -2v\alpha \left(\Re \left[ C_{7}^+ \widetilde{C}_{10}^{+*} \right] H_{V,T}^{+,+}(-1/2,+1/2)+\Re \left[ C_{7}^- \widetilde{C}_{10}^{-*} \right] H_{A,T5}^{+,+}(-1/2,+1/2) \right) \notag \\
&-& 2v\alpha \left(\Re \left[ \widetilde{C}_{9}^+ \widetilde{C}_{10}^{+*} \right] \vert H_{V}^{+}(-1/2,+1/2)\vert^2+\Re \left[ \widetilde{C}_{9}^- \widetilde{C}_{10}^{-*} \right] \vert H_{A}^{+}(-1/2,+1/2)\vert^2 \right) \notag \\
K_{3s} &=& \sqrt{2}v\alpha \left( Im\left[ C_{7}^+ \widetilde{C}_{10}^{-*} \right] H^R_{6(A,T)}+ \Im\left[ C_{7}^- \widetilde{C}_{10}^{+*} \right] H^R_{6(V,T5)} +\Im\left[ \widetilde{C}_{9}^+ \widetilde{C}_{10}^{-*} \right] H^R_{6(A,V)} +\Im\left[ \widetilde{C}_{10}^+ \widetilde{C}_{9}^{-*} \right] H^R_{6(A,V)} \right), \notag \\
K_{4s} &=& \sqrt{2}v\alpha \left( Re\left[ C_{7}^+ \widetilde{C}_{10}^{+*} \right] H^R_{6(T,V)}  +2 \Re\left[ \widetilde{C}_{9}^+ \widetilde{C}_{10}^{+*}\right] H_V^0(+1/2,+1/2)H_V^{+*}(-1/2,-1/2)  \right) \notag \\
&-& \sqrt{2}v\alpha \left( Re\left[ C_{7}^- \widetilde{C}_{10}^{-*} \right] H^R_{6(A,T5)} +2 \Re\left[ \widetilde{C}_{9}^- \widetilde{C}_{10}^{-*}\right] H_A^0(+1/2,+1/2)H_A^{+*}(-1/2,-1/2)  \right), \notag
\end{eqnarray}
\begin{eqnarray}
K_{3sc} &=& \sqrt{2}v^2 \alpha \left( -\Im\left[ \widetilde{C}_{9}^+ C_{7}^{+*} \right] H^L_{6(V,T)}+ \Im\left[ \widetilde{C}_{9}^- C_{7}^{-*} \right] H^L_{6(A,T5)} \right), \notag \\
K_{4sc} &=& -\sqrt{2}v^2 \alpha \left( \Re \left[ C_{7}^+ C_{7}^{-*} \right] H^L_{6(T,T5)} +\Re \left[ \widetilde{C}_{9}^+ C_{7}^{-*} \right] H^L_{6(T5,V)} \right) \notag \\
&+& \sqrt{2}v^2 \alpha\left( \Re \left[ \widetilde{C}_{9}^+ \widetilde{C}_{9}^{-*} \right] H^L_{6(A,V)} +\Re \left[ \widetilde{C}_{9}^- C_{7}^{+*} \right] H^L_{6(A,T)} \right),
\end{eqnarray}
where $\beta=\frac{2m_{\ell}^2}{\sqrt{s}}$, $v^{\prime}=\sqrt{1+\beta^2}$ and
\begin{eqnarray}
H_{x,y}^{m,n}(s_{\Lambda_b},s_{\Lambda}) &=& H_x^m(s_{\Lambda_b},s_{\Lambda}) H_y^{n*}(s_{\Lambda_b},s_{\Lambda}), \; H_{1x}^{0,+} = \vert H_x^0(+1/2,+1/2) \vert^2+\frac{1}{2}v^{\prime^2}  \vert H_x^+(-1/2,+1/2) \vert \notag \\
H_{2x}^{0,+} &=& \beta^2 \vert H_{x}^0(+1/2,+1/2) \vert^2+  \vert H_x^+(-1/2,+1/2) \vert^2 ,\; H_{3x}^{t,+} = \beta^2 \vert H_{x}^t(+1/2,+1/2) \vert^2+ v^2 \vert H_x^+(-1/2,+1/2) \vert^2 \notag \\
H_{4(x,y)} &=& 2  H_{x,y}^{0,0}(+1/2,+1/2) + v^{\prime^2} H_{x,y}^{+,+}(-1/2,+1/2),\; H_{5(x,y)} = \beta^2  H_{x,y}^{0,0}(+1/2,+1/2) + H_{x,y}^{+,+}(-1/2,+1/2) \notag \\
H_{6(x,y)}^{R,L} &=& H_x^0(+1/2,+1/2) H_y^{+*}(-1/2,+1/2) \pm H_x^+(-1/2,+1/2) H_y^{0*}(+1/2,+1/2),\notag\\
H_x^{t,0,+} &=& \beta^2 \vert H_x^t(+1/2,+1/2) \vert^2+ v^2 \vert H_x^0(+1/2,+1/2) \vert^2 +\frac{1}{2}v^{2}  \vert H_x^+(-1/2,+1/2) \vert^2
\end{eqnarray}
with $x,y=V,A,T,T5$.

The square of amplitudes corresponding to $SP$ and $T^\prime$  operators, that are absent in the SM are
\begin{eqnarray}
\vert M_{SP} \vert^2 &=& \left[ (v^2\vert C_S^+ \vert^2 +\vert C_P^+ \vert^2) \vert H_S(+1/2,+1/2) \vert^2 +(v^2\vert C_S^- \vert^2 +\vert C_P^- \vert^2) \vert H_P(+1/2,+1/2) \vert^2 \right] \cos\theta_{\Lambda} \notag \\
&+& 2\alpha \Re\left[C_P^+C_P^{-*}\right]H_S(+1/2,+1/2) H_P^*(+1/2,+1/2) +2\alpha \Re\left[C_S^+C_S^{-*}\right]H_s(+1/2,+1/2) H_P^*(+1/2,+1/2)\notag\\
\vert M_{T^{\prime}} \vert^2 &=& 8\left[\vert C_T \vert^2 H_{7T^{\prime}}+\vert C_{T5} \vert^2 H_{8T^{\prime}}\right]\sin^2\theta_{\ell} + 16 \left[\vert C_T \vert^2 H_{T^{\prime}}+\vert C_{T5} \vert^2 H_{10,T^{\prime}}\right]\cos^2\theta_{\ell} \notag \\
&+& 32\alpha \Re\left[C_{T5}C_T^* \right] H_{11,T^{\prime}} \sin^2\theta_{\ell}\cos\theta_{\Lambda} + 64\alpha \Re\left[C_{T5}C_T^* \right] H_{12,T^{\prime}} \cos^2\theta_{\ell}\cos\theta_{\Lambda} -32 \sqrt{2} v^2 \alpha \Re\left[C_{T5}C_T^* \right] H_{13,T^{\prime}} \cos\theta_{\ell}\cos\theta_{\Lambda}\notag\\
\label{ST-Cont}
\end{eqnarray}
where
\begin{eqnarray}
H_{7,T^{\prime}} &=& v^{\prime^2} \vert H_{T^{\prime}}^{t,-}(+1/2,-1/2) \vert^2 +v^2 \vert H_{T^{\prime}}^{0,-}(+1/2,-1/2) \vert^2 +\beta^2 \vert H_{T^{\prime}}^{+,-}(+1/2,+1/2) \vert^2 \notag \\
H_{8,T^{\prime}} &=& v^{2} \vert H_{T^{\prime}}^{t,-}(+1/2,-1/2) \vert^2 +v^{\prime^2} \vert H_{T^{\prime}}^{0,-}(+1/2,-1/2) \vert^2 +\beta^2 \vert H_{T^{\prime}}^{+,-}(+1/2,+1/2) \vert^2 \notag \\
H_{9,T^{\prime}} &=& v^2 \vert H_{T^{\prime}}^{+,-}(+1/2,+1/2) \vert^2+ \vert H_{T^{\prime}}^{t,0}(+1/2,+1/2) \vert^2 +\beta^2  \vert H_{T^{\prime}}^{t,-}(-1/2,+1/2) \vert^2 \notag \\
H_{10,T^{\prime}} &=& \vert H_{T^{\prime}}^{+,-}(+1/2,+1/2) \vert^2+v^2 \vert H_{T^{\prime}}^{t,0}(+1/2,+1/2) \vert^2 +\beta^2  \vert H_{T^{\prime}}^{0,-}(-1/2,+1/2) \vert^2 \notag \\
H_{11,T^{\prime}} &=& \beta^2 H_{T^\prime}^{+,-}(+1/2,+1/2) (H_{T^{\prime}}^{t,0}(+1/2,+1/2))^*-H_{T^{\prime}}^{0,-}(+1/2,-1/2)(H_{T^{\prime}}^{t,-}(+1/2,-1/2))^* \notag \\
H_{12,T^{\prime}} &=& \left(1-\frac{\beta^2}{2} \right)  H_{T^\prime}^{+,-}(+1/2,+1/2) (H_{T^{\prime}}^{t,0}(+1/2,+1/2))^*- \frac{\beta^2}{2} H_{T^{\prime}}^{0,-}(+1/2,-1/2)(H_{T^{\prime}}^{t,-}(+1/2,-1/2))^* \notag \\
H_{13,T^{\prime}} &=& H_{T^\prime}^{+,-}(+1/2,+1/2) (H_{T^{\prime}}^{t,-}(+1/2,-1/2))^*+ H_{T^{\prime}}^{0,-}(+1/2,-1/2)(H_{T^{\prime}}^{t,0}(+1/2,+1/2))^*
\end{eqnarray}

\end{document}